\documentclass[longauth]{aa} 

\usepackage{graphicx}
\usepackage{txfonts}

\def\degree{$^{\circ}~$}

\usepackage{color} 
\usepackage{pbox} 
\usepackage{amsmath} 

\begin{document} 
   \title{Are fractured cliffs the source of cometary dust jets ?\\insights from OSIRIS/Rosetta at 67P}

\author{
J.-B. Vincent\inst{\ref{mps}}
\and N. Oklay\inst{\ref{mps}}
\and M. Pajola\inst{\ref{padova-spa}}  
\and S. Höfner\inst{\ref{mps}, \ref{igep}}
\and H. Sierks\inst{\ref{mps}}
\and X. Hu\inst{\ref{mps}, \ref{igep}}
\and C. Barbieri\inst{\ref{padova-phy}}
\and P. L. Lamy\inst{\ref{lam}}
\and R. Rodrigo\inst{\ref{csic}, \ref{issi}}
\and D. Koschny\inst{\ref{estec}}
\and H. Rickman\inst{\ref{uppsala}, \ref{pas}}
\and H. U. Keller\inst{\ref{igep}}
\and M. F. A'Hearn\inst{\ref{umd}, \ref{gauss}, \ref{mps}} 
\and M. A. Barucci\inst{\ref{lesia}}
\and J.-L. Bertaux\inst{\ref{latmos}}
\and I. Bertini\inst{\ref{padova-spa}}
\and S. Besse\inst{\ref{estec}}
\and D. Bodewits\inst{\ref{umd}}
\and G. Cremonese\inst{\ref{inaf-pad}} 
\and V. Da Deppo\inst{\ref{padova-lux}}
\and B. Davidsson\inst{\ref{uppsala}}
\and S. Debei\inst{\ref{padova-ind}}
\and M. De Cecco\inst{\ref{uni-trento}}
\and M. R. El-Maarry\inst{\ref{bern}} 
\and S. Fornasier\inst{\ref{lesia}} 
\and M. Fulle\inst{\ref{inaf-tri}}
\and O. Groussin\inst{\ref{lam}} 
\and P. J. Guti\'errez\inst{\ref{iaa}}
\and P. Guti\'errez-Marquez\inst{\ref{mps}} 
\and C. G\"uttler\inst{\ref{mps}} 
\and M. Hofmann\inst{\ref{mps}}
\and S. F. Hviid\inst{\ref{dlr}} 
\and W.-H. Ip\inst{\ref{ncu}} 
\and L. Jorda\inst{\ref{lam}} 
\and J. Knollenberg\inst{\ref{dlr}}
\and G. Kovacs\inst{\ref{mps}, \ref{gabor}} 
\and J.-R. Kramm\inst{\ref{mps}} 
\and E. K\"uhrt\inst{\ref{dlr}} 
\and M. K\"uppers\inst{\ref{esac}}
\and L. M. Lara\inst{\ref{iaa}}
\and M. Lazzarin\inst{\ref{padova-phy}}
\and Z.-Y. Lin \inst{\ref{ncu}}
\and J. J. Lopez Moreno \inst{\ref{iaa}}
\and S. Lowry\inst{\ref{kent}} 
\and F. Marzari\inst{\ref{padova-phy}}
\and M. Massironi\inst{\ref{padova-spa}}
\and F. Moreno\inst{\ref{iaa}} 
\and S. Mottola\inst{\ref{dlr}} 
\and G. Naletto\inst{\ref{padova-inf}, \ref{padova-spa}, \ref{padova-lux}} 
\and F. Preusker\inst{\ref{dlr}} 
\and F. Scholten\inst{\ref{dlr}} 
\and X. Shi\inst{\ref{mps}}
\and N. Thomas\inst{\ref{bern}}
\and I. Toth\inst{\ref{hungary}, \ref{lam}} 
\and C. Tubiana\inst{\ref{mps}}
}

\institute{
Max-Planck Institut fuer Sonnensystemforschung, Justus-von-Liebig-Weg, 3 37077 Goettingen, Germany \email{vincent@mps.mpg.de}\label{mps}
\and Centro di Ateneo di Studi ed Attivitá Spaziali "Giuseppe Colombo" (CISAS), University of Padova, Via Venezia 15, 35131 Padova, Italy \label{padova-spa}
\and Institute for Geophysics and Extraterrestrial Physics, TU Braunschweig, 38106 Braunschweig, Germany \label{igep}
\and Department of Physics and Astronomy "G. Galilei", University of Padova, Vic. Osservatorio 3, 35122 Padova, Italy \label{padova-phy}
\and Aix Marseille Universit\'e, CNRS, LAM (Laboratoire d'Astro-physique de Marseille) UMR 7326, 13388, Marseille, France \label{lam}
\and Centro de Astrobiologia (INTA-CSIC), European Space Agency (ESA), European Space Astronomy Centre (ESAC), P.O. Box 78, E-28691 Villanueva de la Canada, Madrid, Spain \label{csic} 
\and International Space Science Institute, Hallerstrasse 6, 3012 Bern, Switzerland \label{issi} 
\and Research and Scientific Support Department, European Space Agency, 2201 Noordwijk, The Netherlands  \label{estec} 
\and Department of Physics and Astronomy, Uppsala University, Box 516, 75120 Uppsala, Sweden \label{uppsala}
\and PAS Space Research Center, Bartycka 18A, 00716 Warszawa, Poland \label{pas} 
\and Department for Astronomy, University of Maryland, College Park, MD 20742-2421, USA \label{umd} 
\and Gauss Professor Akademie der Wissenschaften zu Göttingen, 37077 Göttingen, Germany \label{gauss}
\and LESIA, Observatoire de Paris, CNRS, UPMC Univ Paris 06, Univ. Paris-Diderot, 5 Place J. Janssen, 92195 Meudon Pricipal Cedex, France \label{lesia} 
\and LATMOS, CNRS/UVSQ/IPSL, 11 Boulevard d'Alembert, 78280 Guyancourt, France  \label{latmos} 
\and INAF Osservatorio Astronomico di Padova, Vicolo dell'Osservatorio 5, 35122 Padova \label{inaf-pad}Italy
\and CNR-IFN UOS Padova LUXOR, Via Trasea 7, 35131 Padova, Italy \label{padova-lux}
\and Department of Industrial Engineering University of Padova Via Venezia, 1, 35131 Padova \label{padova-ind}
\and University of Trento, via Sommarive, 9, 38123 Trento, Italy \label{uni-trento}
\and Physikalisches Institut, Sidlerstrasse 5, University of Bern, CH-3012 Bern, Switzerland  \label{bern}
\and INAF - Osservatorio Astronomico di Trieste, via Tiepolo 11, 34143 Trieste, Italy \label{inaf-tri}
\and Instituto de Astrofisica de Andalucia-CSIC, Glorieta de la Astronomia, 18008 Granada,  Spain \label{iaa} 
\and Institute of Planetary Research, DLR, Rutherfordstrasse 2, 12489 Berlin, Germany \label{dlr} 
\and Institute for Space Science, National Central University, 32054 Chung-Li, Taiwan \label{ncu} 
\and Budapest University of Technology and Economics, Department of Mechatronics, Optics and Engineering Informatics, Muegyetem rkp 3, Budapest, Hungary H-1111 \label{gabor}
\and ESA/ESAC, PO Box 78, 28691 Villanueva de la Ca\~nada, Spain \label{esac} 
\and Centre for Astrophysics and Planetary Science, School of Physical Sciences, The University of Kent, Canterbury CT2 7NH, United Kingdom\label{kent}
\and Department of Information Engineering, University of Padova, Via Gradenigo 6/B, 35131 Padova, Italy \label{padova-inf}
\and Observatory of the Hungarian Academy of Sciences, PO Box 67, 1525 Budapest, Hungary \label{hungary}
}

\date{Accepted December 4, 2015}

 
  \abstract
{Dust jets, i.e. fuzzy collimated streams of cometary material arising from the nucleus, have been observed in-situ on all comets since the Giotto mission flew by comet 1P/Halley in 1986. Yet their formation mechanism remains unknown. Several solutions have been proposed, from localized physical mechanisms on the surface/sub-surface (see review in \cite{belton2010}) to purely dynamical processes involving the focusing of gas flows by the local topography \citep{crifo2002}. While the latter seems to be responsible for the larger features, high resolution imagery has shown that broad streams are composed of many smaller features (a few meters wide) that connect directly to the nucleus surface.}
   {We monitored these jets at high resolution and over several months to understand what are the physical processes driving their formation, and how this affects the surface.}
   {Using many images of the same areas with different viewing angles, we performed a 3-dimensional reconstruction of collimated jets, and linked them precisely to their sources on the nucleus.}
   {We show here observational evidence that the Northern hemisphere jets of comet 67P arise from areas with sharp topographic changes and describe the physical processes involved. We propose a model in which active cliffs are the main source of jet-like features, and therefore the regions eroding the fastest on comets. We suggest that this is a common mechanism taking place on all comets.}
   {}

   
   \titlerunning{comets, 67P, jet-like features, Rosetta}
   \authorrunning{J.-B. Vincent et al}

   \maketitle

\section{Introduction}

In March 1986, the Giotto mission flew by the nucleus of comet 1P/Halley. One of its many discoveries was the realization that cometary activity is not evenly distributed across the illuminated surface. Indeed, it appeared that most of the nucleus was dark and seemingly inactive, while a few percent of the surface gave rise to strong collimated flows of gas and dust, commonly referred to as "jets". In the following 30 years, 5 different cometary nuclei have been visited by space probes, one of them twice (9P/Tempel 1). Fly-by missions provided tremendous data, but by definition could only observe their target at a fixed heliocentric distance and for a short time. This led to a better description of cometary jets, but no definite answer as to what are the physical processes driving them, and what they mean for the nucleus surface evolution. Several solutions have been proposed, from localized physical mechanisms on the surface/sub-surface (see review in \cite{belton2010}) to purely dynamical processes involving the focusing of gas flows by the local topography \citep{keller94, crifo2002}. While the latter seems to be responsible for the larger features, high resolution imagery has shown that broad streams are composed of many smaller features (a few meters wide) that connect directly to the nucleus surface and can be associated to specific morphologies. 
In his review of mechanisms producing collimated outflows, \cite{belton2010} proposes a nomenclature of different types of jet-like activity. 
\textit{Type I} describes dust release dominated by the sublimation of H$_2$O through the porous mantle; \textit{Type II} is controlled by the localized and persistent effusion of super-volatiles from the interior; while \textit{Type III} is characterized by episodic releases of super-volatiles. 

\textit{Type I} jets do not appear to be associated to specific morphology and are generally broader and more diffuse than other features. On 67P this would be the case for the largest jet arising from Hapi region, the interface between the two lobes of the nucleus, since late July 2014. The source of this feature could not be associated to a particular terrain, other than the whole smooth surface of Hapi itself, which shows also a brighter and bluer average spectrum in visible and infrared \citep{sierks2015, fornasier2015, lara2015, lin2015, capaccioni2015}.
\textit{Type II} jets, also called \textit{filaments} display a much more collimated structure and have been traced back to specific regions of cometary nuclei. For instance on 81P/Wild 2 they could be associated to the walls of large, vertical sided pits \citep{brownlee2004, sekanina2004}. On 9P/Tempel 1, such jets were also linked to a ragged area bordering a large smooth terrain, and in particular to a the scarped edge of this terrain \citep{farnham2007}.
\textit{Type III} are more sporadic events, probably related to micro-outbursts or other explosive processes.

To progress on this topic, it is necessary to have a longer time coverage of activity, and high resolution data of the nucleus showing how the surface changes in active areas. ESA's Rosetta space probe is observing comet 67P/Churyumov-Gerasimenko (hereafter 67P) over 2 years of nominal mission, and this has allowed us to achieve a detailed characterization of the comet diurnal and seasonal evolution.

This paper describes how we have linked jets to active sources observed from arrival in August 2014 (3.6 AU) to equinox in May 2015 (1.6 AU). During that time the sub-solar latitude migrated from +42.4\degree to -5.7\degree, therefore the results we present here apply only to the Northern Summer on 67P. We will discuss in section \ref{sec:discussion} how this work can be extrapolated to other regions and different comets. 

Although the jets characterized in this paper would be described as \textit{Type II} in the above nomenclature, we will see that they are not necessarily related to the presence of super volatiles, and can very well be driven by water sublimation alone, if there is some morphologic control of the flow. In that sense they would be more like a small scale \textit{Type I} or a new type altogether.

\section{Methods}

\subsection{Observations}\label{sec:observations}
The OSIRIS instrument is composed of two elements: a narrow angle camera (NAC, FoV = 2.18 \degree) and a wide angle camera (WAC, FoV = 11.89 \degree), both equipped with sets of filters chosen to investigate the composition of nucleus and coma \citep{keller2007}. We typically monitored gas and dust activity with the WAC, about once every two weeks for heliocentric distances greater than 2 AU, and once per week afterward. The nominal sequence had a set of observations once per hour for a full comet rotation (12h). As both Rosetta and the comet were coming closer to Earth, the data volume available increased and we updated the observational sequence to one set of observations every 20 min, for 14h. This gives us a good coverage of the diurnal and seasonal evolution of the comet. 

Dust jet observations consist of 2 images in the visible spectrum (central wavelength 612.6 nm, bandwidth 9.8 nm), one with short exposure for the nucleus, one long exposure to see the faintest structures. Typically, the long exposure is 30 times greater than the short one; for instance at 3 AU we used exposure times of respectively 15 and 0.5 seconds.
In addition, we make use of the high dynamic range of the CCD (16 bits) to detect faint structures at very high resolution in the shadowed areas of the nucleus observed with the NAC orange filter (central wavelength 649.2 nm, bandwidth 84.5 nm). During the epoch considered in this work, the spacecraft orbited 67P at distances ranging from 10 to 200~km, mainly on a terminator orbit. Therefore we achieved an imaging resolution of 0.15~m to 3.7~m with the NAC; 0.8~m to 20~m with the WAC.

We define jets as fuzzy collimated dust coma structures seemingly arising from the nucleus. They are usually detected with no image processing in long exposure images. The brightness levels of short exposures often need to be stretched to emphasize the lowest few percents of their pixel values. Previous missions have used a logarithmic stretch to reveal jets (i.e. \citet{farnham2007}); as Rosetta flies much closer to its target than previous spacecrafts, and the OSIRIS camera have a larger dynamic range, a linear stretch is generally sufficient to reveal most coma structures.

Figures \ref{fig:STP017_GAS}, \ref{fig:STP030_AP}, and \ref{fig:STP051_RE} show typical monitoring sequences, and are well representative of the large numbers of jets detected routinely. We counted on average 20 jets at any given time, which is only a lower limit for the real number of dust features. Each jet is at least a few pixels wide. Images at higher resolution show that jets can always be resolved into thinner features, typically as small as a few pixels, and it is difficult to define what is the smallest jet size. Longer exposures and different image processing can help reveal fainter/smaller jets, but the stray light and general coma signal increase limit our effective resolution anyway. Therefore, we do not assume here that we can identify all sources at anytime, but simply that we have collected enough statistics to draw significant conclusions. We will come back to this problem in Section \ref{sec:fractures_sources} and discuss how our findings can describe what is the smallest diameter of a jet.

\subsection{Geometric inversion of local sources}\label{sec:inversion}
As seen in our observations, jet-like features are detected in most images and rotate with the nucleus. Similar studies on other comets have shown that often jets can be associated with specific locations on the surface, presenting a different composition and/or morphology than other areas. For instance some areas of comet 9P/Tempel 1 would remain active and produce jets even when far into the night, hinting at sublimation of material much more volatile than water ice \citep{feaga07, vincent10b}. The high resolution provided by OSIRIS allows in principle an accurate determination of active sources, although the process is not straight forward. Indeed, single images provide only 2-dimensional information and one need to combine several observations to reconstruct the true 3-dimensional structures we want to investigate. To achieve this we developed two independent techniques: \textit{blind} and \textit{direct} inversion, detailed in the following sections. These methods are very similar to those used by \cite{sekanina2004} and \cite{farnham2013} in their respective studies of the jets of comets 81P/Wild 2 and 9P/Tempel 1.

Another aspect of jet studies is the detailed investigation of the physics taking place in the jet itself, as cometary material arises from the surface and possibly fragments or sublimates. We focus here on the use of our inversions to understand better the sources themselves, describe their morphology, evolution, and the associated surface physical processes. Readers interested in photometric properties of dust jets from 67P are referred to \cite{lara2015} and \cite{lin2015}, or to \cite{vincent2013} for a numerical modeling of such features on a larger scale.

\subsubsection{Blind inversion}

This first algorithm considers each feature in the images as an isolated jet. Although it seems obvious that a jet can be tracked from one image to another, appearance can be deceiving and we do not impose this condition for the inversion. For each collimated structure, it is safe to assume that the dust is emitted in a plane defined by the observer (Rosetta) and the central line of the jet. A single image does not allow us to determine if the jet is inclined toward or away from the observer. In a second step we calculate the intersection between the jet-plane and a shape model of the nucleus, oriented beforehand to match the geometric conditions of our observations. We used a shape model reconstructed by photogrammetry \citep{preusker2015}; Spacecraft and comet attitudes and trajectories were retrieved with the SPICE library \citep{acton1996}. Each jet-plane/comet intersection defines a line of possible sources on the surface of the nucleus. For a same jet observed at different times this technique provides different set of source-lines but their intersection defines the unique location of the source. We refine the inversion by repeating this process for each jet and each image in the sequence.

This technique is very robust, because it does not make any assumptions on the jets before the inversion. However, when there are too many small jets close to each other the inversion results are often too noisy. Therefore we interpret the output as a probability map of source areas rather than the exact map of active spots. It is very good for tracking diurnal evolution on a regional scale.

\subsubsection{Direct inversion}

The direct inversion aims to achieve a tri-dimensional reconstruction of the jets close to the surface. It works in a similar way as the blind inversion, with the added assumption that we are able to identify the same jet from one image to an other, if possible separated by 10-30 \degree of nucleus rotation (20 min to 1h between observations). This ensures optimal conditions for stereo reconstruction. For each source we can define the plane Rosetta-jet at two different times; the intersection of these two planes in the comet frame describes in 3 dimensions the core line of the jet. This technique provides not only the source location, but also the initial direction of the jet. We can then use these parameters to simulate the jet at a different time, and compare with other images to assess the quality of the reconstruction. This approach is well suited for the sharpest features for which the main direction is easily measurable in our images. Figure \ref{fig:inversion} summarizes the technique.

\subsubsection{Error estimate}

The accuracy of both inversions depends on the resolution of our images and uncertainties on the spacecraft pointing. To estimate the latter we generate simulated images of the nucleus for every observation and compare with the real images. We usually get close to pixel perfect match. A larger source of error comes from the underlying assumption that jets extend all the way to the surface. In reality, stray light contamination prevents us actually seeing the first few pixels above the surface in jet exposures. In addition, to that, shadows cast by nearby topography can hide the source and first expansion zone of the jet. Depending on the resolution, this represents a blind area of a few meters to a few tens of meters in elevation. It is possible that interactions between different gas flows generates complex gas streams which merge and get collimated only a few meters above the surface, thus preventing any localization more precise than a few 10s of meters on the surface. We will see in section \ref{sec:discussion} that morphology and color variations minimize this uncertainty. 

Because most sources are active for about half a comet rotation, inversion of image pairs taken several hours apart should give us a subset of common areas. This is indeed the case, although we observe a slight dispersion of a few \degree in lat/lon (less than 100m uncertainty on the surface).

\section{Results}\label{sec:results}

We present here a summary of active sources for several different epochs, covering the time from August 2014 (3.6 AU) to May 2015 (1.6 AU), and sampling different heliocentric distances and sub solar latitudes. We selected a subset of sequences obtained at similar resolution and separated by a few months. As explained in Section \ref{sec:observations} we are not performing an inversion of all sequences and all jets observed, but rather try to identify a general pattern of how these sources are distributed spatially over the nucleus, and how this distribution evolves over time. It is important to understand that the following results are not a exhaustive catalog of all sources, but are statistically significant to describe the general behaviour of the activity arising from the Northern hemisphere of 67P.

This section and the following often refer to source regions by the official name used within the Rosetta community. For instance \textit{Hapi} is the smooth area between the two lobes, \textit{Hatmehit} is the large depression on the small lobe, \textit{Seth} is the pitted region on the big lobe, and so on. A complete description of these regions and associated morphological features has been published in \cite{thomas2015} and \cite{elmaarry2015a}.

\subsection{Sources in August-September 2014, 3.5 AU, sub-solar latitude +42.4 \degree}
\begin{figure}[h!]
\centering\
\includegraphics[width=\hsize]{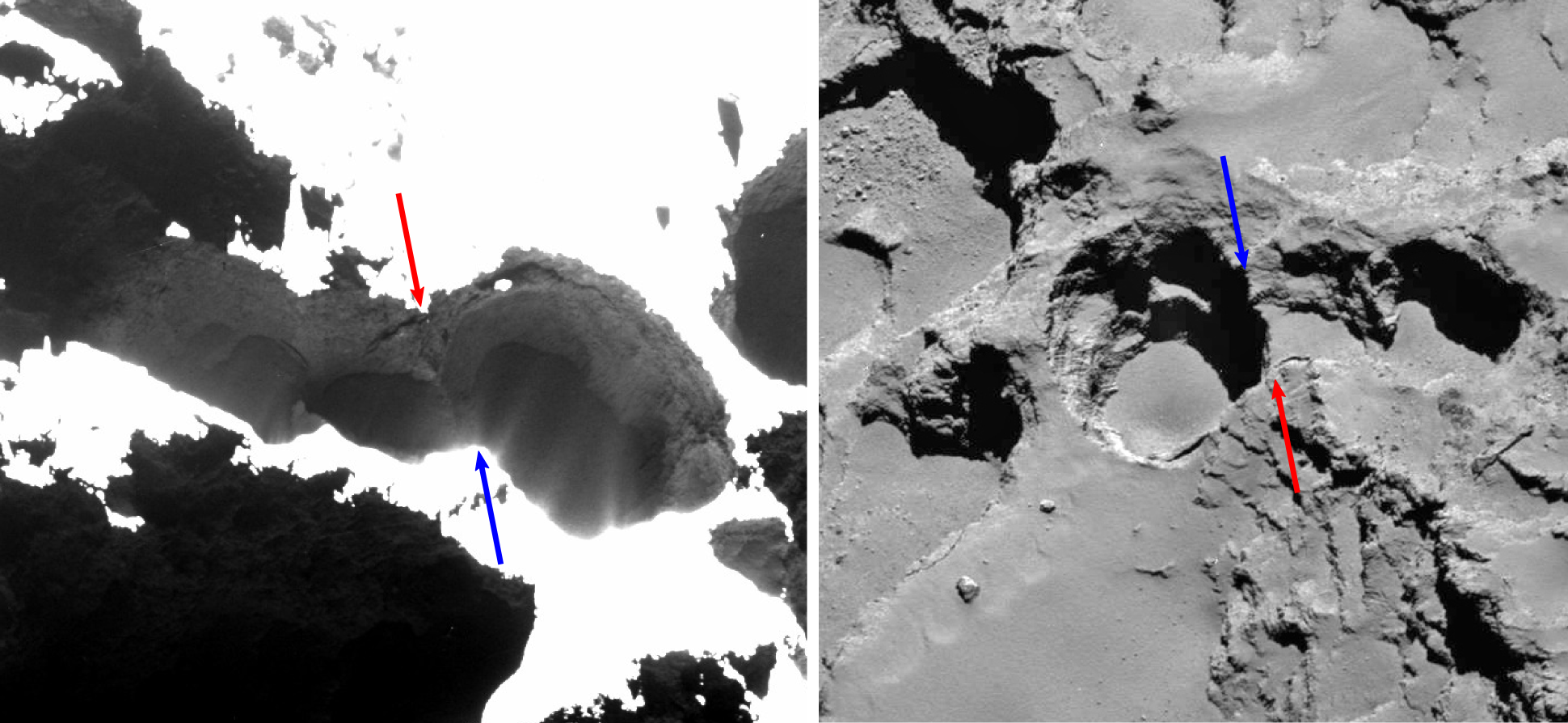}
\caption{Small jets arising from the fractured edge of a 200~m diameter pit. WAC image obtained from a distance of 9~km. Resolution: 0.73~m/px. Right panel shows the same pit from the opposite direction, exposing the morphology responsible for the activity seen in the left panel. For orientation, red and blue arrows indicate the same edges of the pit in both images. Image reference: WAC\_2014-10-20T08.15.50.752Z\_ID30\_1397549000\_F18}
\label{fig:jets_pits}
\end{figure}

OSIRIS started to resolve dust coma features at the end of July 2014, from a distance of 3000~km. These jets are described in details in \cite{sierks2015} and \cite{lara2015}, along with the inversion of their sources, localized in the Hapi region at the interface between the two lobes of the nucleus. This is consistent with ground based observations in the last two orbits, that reported the presence of an active region at +60 \degree northern latitude \citep{vincent2013}. As the spacecraft spiraled down to 30~km in August and September 2014, we refined this inversion and observed that the large jets were actually made of several smaller structures, strongly collimated (Figure \ref{fig:STP017_GAS}). These fine jets were linked to the "active pits" of Seth region, a set of deep circular depressions around latitude +60 \degree and longitude +210 \degree \citep{vincent2015b}, the walls of alcoves neighbouring Hapi, and some outcrops in Hapi. Figure \ref{fig:jets_pits} shows a high resolution view of one source. While in our inversions it seems like one large jet arises from the pit, one can see that this jet is actually composed of several features, much smaller. The pit appeared to be constantly active over a full comet rotation, but high-resolution images have shown that the active area within the pit was rotating with the illumination during the course of a comet day, the inner walls of the pit being sequentially illuminated \citep{vincent2015b}.
A map of other sources active at that time is given in Figure \ref{fig:maps}.

\subsection{Sources in November-December 2014, 2.8 AU, sub-solar latitude +36.2 \degree}
\begin{figure}[h!]
\centering\
\includegraphics[width=\hsize]{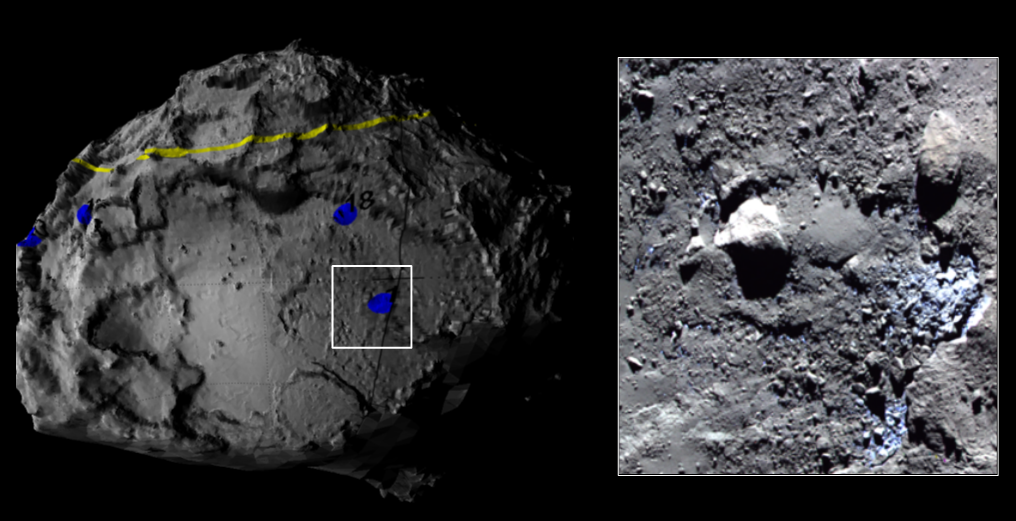}
\caption{Left panel: Map of sources active in December 2015, projected on the 3D shape model of 67P. Right: Close-view of an RGB color composite showing one of the sources with a typical bluer hue, corresponding to the location of the white box in the left panel. The color image reflects the variegation of the terrain in August 2014, when this area was not yet active. Distance from the comet center: 43~km, resolution: 0.75 m/px. The large block in the center is about 100~m wide. Reference image: NAC\_2014-09-05T06.35.55.557Z\_ID30\_1397549300\_F22.}
\label{fig:source_imh_blue}
\end{figure}

From October to November 2014, Rosetta's orbit was designed to ensure the best characterization of potential landing sites, the final selection, and landing. Therefore the spacecraft went down to 10~km mostly pointing at the surface. This gave us very high resolution images of the nucleus, but limited data on the jets as the nucleus was always larger than even the WAC field of view.
Immediately after landing, Rosetta retreated to 30~km and we acquired the first 20 min cadence sequence dedicated to the study of dust jets. It was also executed from a sub spacecraft latitude far in the Southern hemisphere (57 degree South). This configuration is ideal for jet monitoring because it provides us a view of all local times in a single frame, and the nucleus rotating in front of the camera gives the stereo-graphic views without the need for different spacecraft pointing. From a distance of 29 km we achieved 3 m/px resolution with the WAC.
We found that many more sources were active at that time (at least 10 jets per image) and we could trace them down to the surface (Figure \ref{fig:maps}). We consistently found sources on high slopes, like cliffs and walls of alcoves and pits. There is very little evidence for activity arising from smooth, flat areas, apart from the larger scale activity in Hapi. 
The Hapi activity was associated to terrains displaying bluer spectral slopes; when comparing our sources with published color maps by OSIRIS \citep{fornasier2015, oklay2015} or VIRTIS \cite{capaccioni2015}, it seems to be true also for the smaller sources. It is particularly interesting to see that new jets arise from areas identified as bluer in early images August 2014) but not yet active at that time. Figure \ref{fig:source_imh_blue} shows an example of such source.
Figure \ref{fig:maps} summarizes all inverted sources.

Active spots are typically observed closer to the equator than in September 2014, actually within 30 \degree of the sub-solar latitude. Inverted sources are listed in Figure \ref{fig:maps}.
We also observed that sources switch on and off within 20 min of crossing the terminator, this duration being an upper limit constrained by the cadence of our observations. This is particularly striking for sources in Seth and Hapi which experience very different diurnal illumination. They are typically illuminated only for 3 consecutive hours before entering 3 hours of night because of the strong self shadowing introduced by the large concavity of the nucleus. Hence, they experience two day/night cycles per nucleus rotation. 

\subsection{Sources in April 2015, 2 AU, sub-solar latitude +9.8 \degree}\label{sec:apr}
\begin{figure}[h!]
\centering\
\includegraphics[width=\hsize]{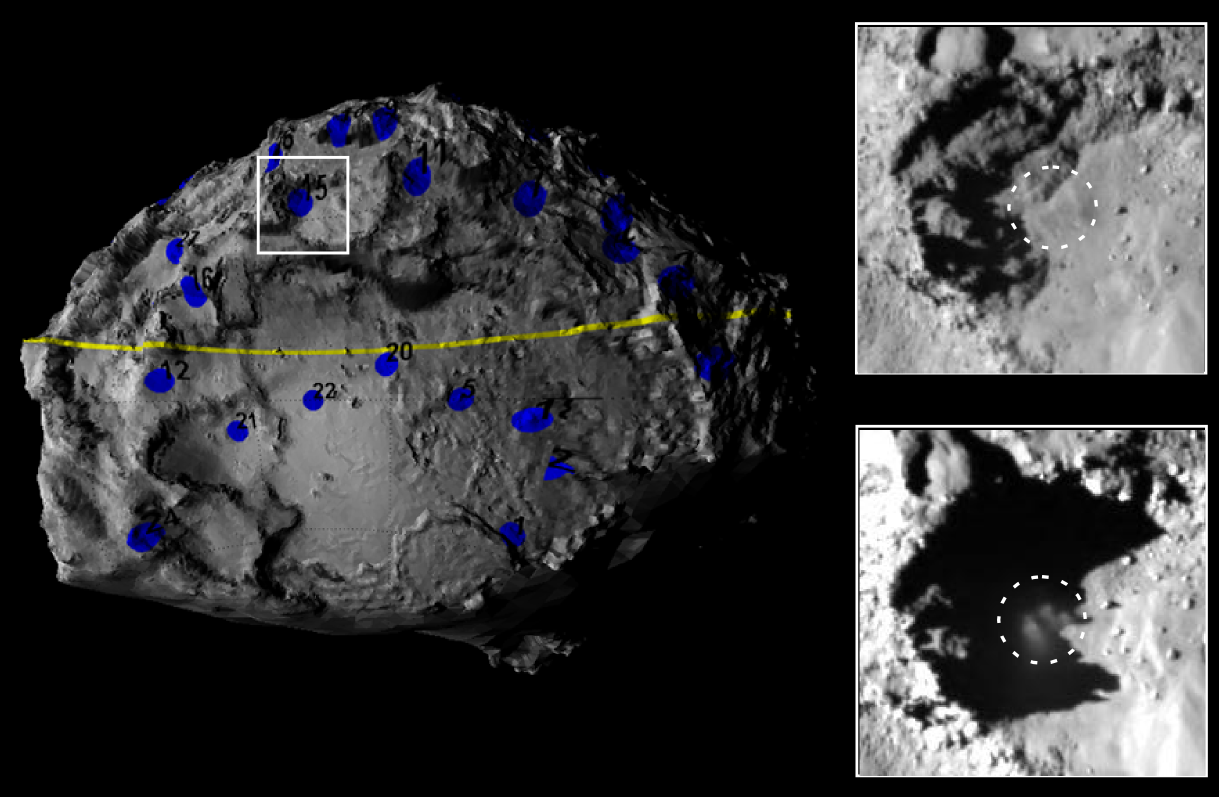}
\caption{Left panel: Map of sources active in April 2015, projected on the 3D shape model of 67P. Right: Close-view of one of the sources where a small jet is distinctly seen arising from the mass wasted area, even if it had passed the evening terminator since about 1/2h. Distance from the comet center: 91~km, resolution: 1.65 m/px. The circled area is about 80~m wide. Image references: NAC\_2015-04-25T13.22.43.406Z\_ID30\_1397549001\_F22 and NAC\_2015-04-25T14.02.42.662Z\_ID30\_1397549001\_F22.}
\label{fig:source_imh_cliff}
\end{figure}

The same trend in spatial distribution and behaviour of sources is confirmed with later sequences. Our next inversion of a sequence acquired in March 2015 shows again that sources lie close to cliffs, and their average latitude has migrated South, following the Sun. We start to observe new regions becoming active close to the equator in Ma'at and Imhotep regions (Figure \ref{fig:maps}).
As we get closer to the Sun, we also start to observe different types of activity, such as outbursts that may have originated from the night side \citep{knollenberg2015}. There are also indications of a few jets arising from more dusty areas and sustained for a while after sunset due to the increasing lag in surface cool down \citep{shi2015}. Figure \ref{fig:source_imh_cliff} shows an example of such jet, with the source clearly active while in the shadow.
As for the previous epochs, Figure \ref{fig:maps} summarizes all inverted sources.

\section{Discussion}\label{sec:discussion}

\subsection{Seasonal/Diurnal effects}
The results of our inversions show a strong correlation between active sources and sub-solar latitude. Jets footprints have been steadily migrating southward since August 2014, always remaining in a latitude band centered on the sub-solar point. This is perfectly consistent with what had been reported for the previous orbits, from ground based observations \citep{lara05, lara11, tozzi11, vincent2013}. We expect the active sources to keep following the same seasonal evolution. From perihelion (August 2015) to the next equinox (March 2016) most of the activity should arise from southern latitudes.

On a diurnal time scale, jets wake up and decay correlates with the illumination, with very little lag, at least until end of March 2015. It indicates that the sublimating ices responsible for lifting up the dust we observe must be distributed within the thermal skin depth of the surface (typically 1~cm, \citet{gulkis2015}), therefore preferentially in terrains poorly insulated. This is consistent with the fact that we do not see jets arising from smooth, dust-covered areas during the time period prior to March, 2015. The dust layer is indeed a good thermal insulator which prevents heat from reaching volatiles trapped underneath, if the layer thickness is larger than the diurnal thermal skin depth. On the contrary, cliffs and other vertical features do not have dust on their surface; it falls down with the gravity. Hence, for similar illumination conditions, volatiles trapped in cliff walls will receive more heat than ices laying under a dust layer.

As discussed in section \ref{sec:apr}, jet behaviour is somewhat different as we get closer to the Sun. Starting from the end of March 2015, we have seen more jets arising from smooth surfaces, and an increase in time-lag of jet switch-off after crossing the terminator. This is attributed to the increase in energy received from the Sun, which allows volatiles previously insulated to reach their sublimation temperature. \cite{shi2015} describes this type of jets in details.

\subsection{Fractured cliffs as potential sources of collimated jets}
The local surface topography and physical properties may allow or prevent the formation of a jet. But activity also exerts a feedback on the source terrain. Indeed, as jets arise, they carry dust and gas away from the surface and reshape the area. 
To understand better the link between topography, jet formation, and nucleus erosion, we looked at the detailed morphology of the active sources identified above and found a common set of features: when resolved down to their thinnest component, most jets observed in the Northern hemisphere of 67P always seem to arise from fractured cliffs, irrespective of their location on the nucleus. 
These active walls present many signs of ongoing erosion. Large debris fields can be observed below the cliffs, interpreted as blocks falling down from the wall. Cliffs upper edges display mass wasting features, with the upper dust layer seemingly flowing down as the edge of the cliff collapses. These granular flows expose underneath fractured terrains. These fractures may be surfacial, or indicate that cracks on the cliff wall propagate inward.

\subsection{Consequence for surface evolution}
We interpret this morphology as a the signature of a multi steps activity mechanism:
\begin{enumerate}
\item{Cliffs are first fractured by mechanical or thermal processes \cite{elmaarry2015b}}
\item{Fractures propagate into a matrix of dust and ices}
\item{Cracks allow the diurnal heat wave to penetrate deeper into the surface, reaching volatiles otherwise insulated}
\item{As the ices sublimate, the gas expands and escapes through the fractures. They may act as nozzles, effectively accelerating and focusing the gas flows until they reach a pressure sufficient to tore off dust particles from the fracture walls and form the dust jets we observe}
\item{The combination of continued cracking, expanding gas flows, and removal of ices by sublimation, weakens the structural integrity of the cliff, leading to collapse of the wall and mass wasting on the cliff table}
\item{The cliff continues to retreat until all volatiles are exhausted or if the topography has become too shallow to prevent the formation of an insulating dust mantle from the fallback.}
\end{enumerate}

In addition to the small jets emitted by the cliff wall, we also suggest that most of the talus material is also active. As blocks are detached from the wall and fall at the cliff foot, they expose their previously hidden surface, enriched in volatiles. This is clearly seen in multispectral images, which consistently shows bluer material at the bottom of cliffs (e.g. Figure \ref{fig:source_imh_blue}. We will come back to this in section \ref{sec:bright_mat}.

\begin{figure}[h!]
\centering
\includegraphics[width=\hsize]{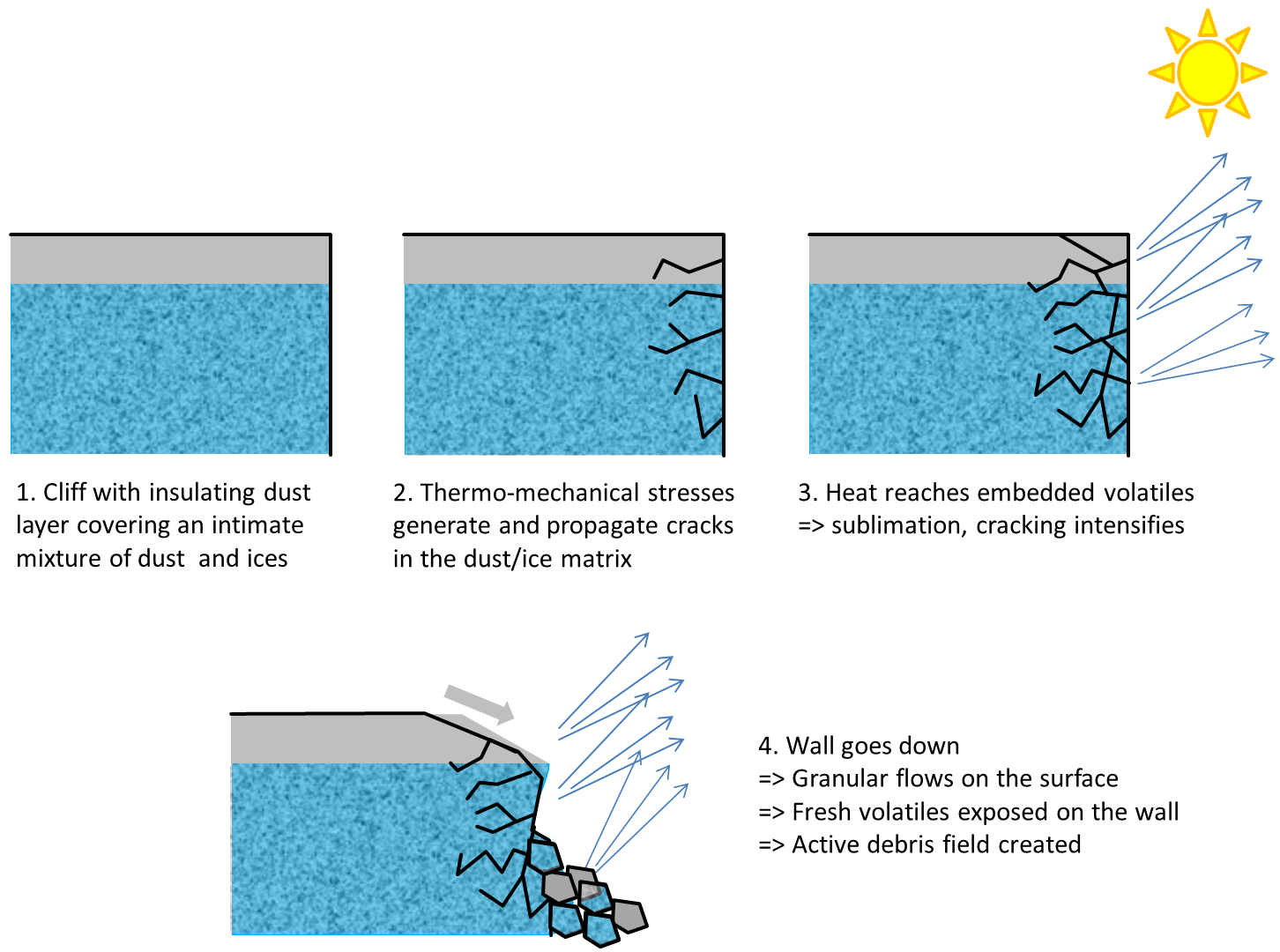}
\caption{Scenario of surface evolution describing how the regressive erosion of cliffs can expose volatiles and lead to localized activity.}
\label{fig:scenario}
\end{figure}

Figure \ref{fig:scenario} shows a sketch of this scenario, and the following sections present the observational evidence supporting our arguments.

\subsubsection{Cliff collapse}

A simple theoretical relation describes the maximum height of a cliff on a given body \citep{melosh2011}:
\begin{equation}
H_{max} = \frac{2c}{\rho g} \tan(45+\Phi /2) \quad \textrm{(meters)}
\end{equation}
with $c$ the cohesion, $\rho$ the material density, $g$ the body gravity, and $\Phi$ the angle of internal friction, or maximum angle of repose.

Because this formula does not account for intrinsic weaknesses in the material, it tends to predict cliffs slightly higher than in reality. It gives nonetheless the right order of magnitude. 

The physical parameters in this formula are taken from the literature. We chose a cohesion equivalent to the shear strength of the material (at most 50 Pa, \citet{vincent2015b, groussin2015}). Our calculation of the gravity accounts for the concave shape, and centrifugal force on the surface. With a mass of $1.10^{13}~kg$ and a density of $470~kg.m^{-3}$m \citep{sierks2015} we obtain a effective gravity of $2-8.10^{-4}~m.s^{-2}$, typically $4.10^{-4}~m.s^{-2}$ at the cliffs foot. The angle of internal friction is assumed to be at most 30 \degree, as this seems to be the steepest slope of granular flows on comets and asteroids (see \ref{sec:gflows}).

Applying this formula to comets, with the assumptions given above, predicts cliffs taller than 450 m, with an typical height of 920 m, which on 67P is observed only for the Hathor cliff (900~m).

The fact that all cliffs are smaller than this theoretical upper limit supports our choice of parameters. Our model shows that, given the right conditions, it is possible to grow or preserve cliffs much larger that what we observe. We interpret the fact that most cliffs are smaller than 100 m as an erosional effect. Even if the cliff is structurally strong enough to support itself, the combination of local gas flows and thermo-mechanical stresses will greatly reduce the cohesion of the material. As active eroding processes and intrinsic weaknesses of the material (for instance pre-existing faults) are not taken into account in the simple model, they may very well explain our observations.

\begin{figure}[h!]
\centering
\includegraphics[width=\hsize]{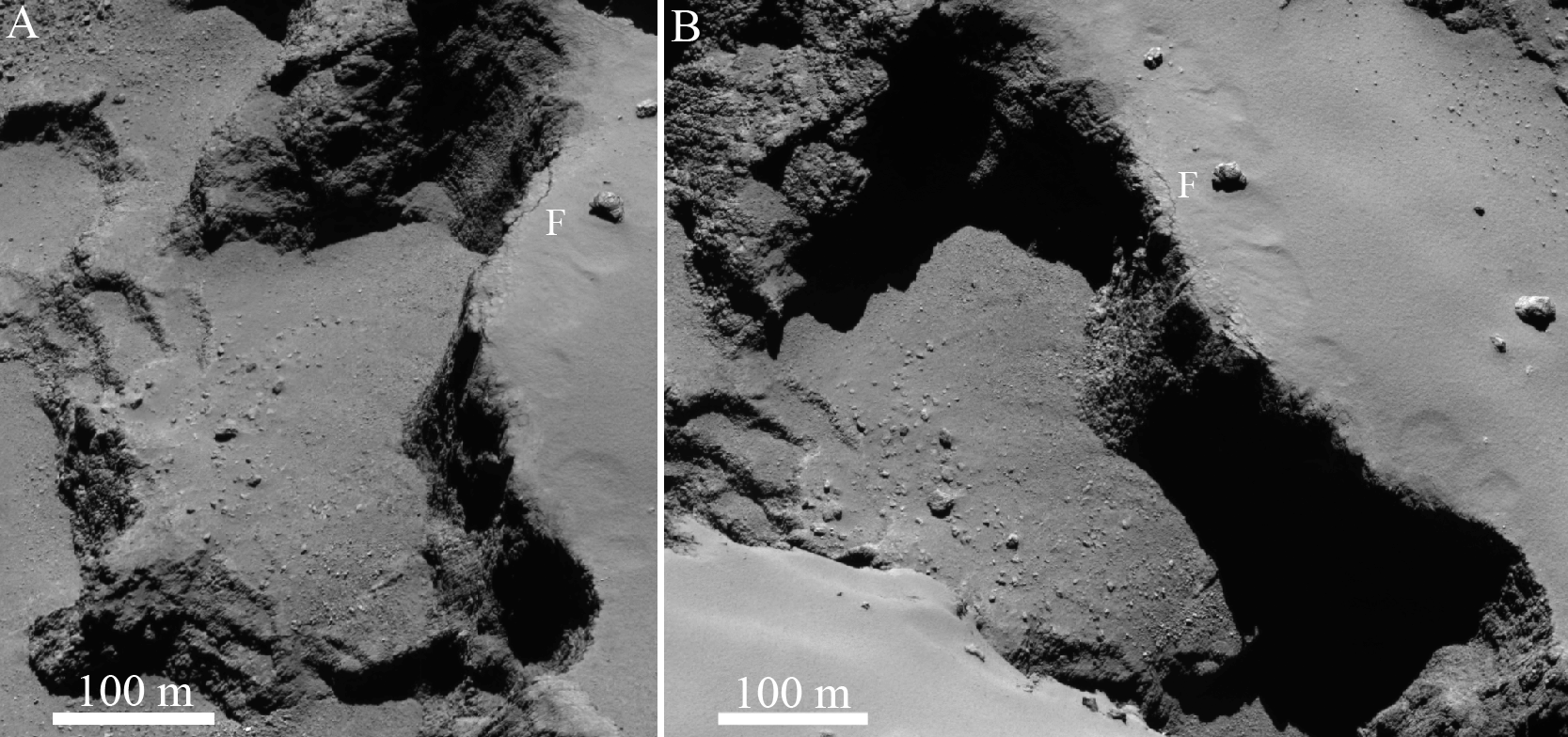}
\caption{The two views, A: NAC\_2014-09-21T02.34.51.832Z\_F41 and B: NAC\_2014-09-22T02.34.52.186Z\_F41 used in this work. F marks an 80~m long fracture present of the wall cliff. Image were acquired from a distance of 30~km and have a resolution of 0.5~m/px}
\label{fig:x}
\end{figure}

Indeed, high resolution images show that most cliffs display several types of fracturing patterns: Polygonal features on the walls, characteristics of sublimation and thermal stresses; larger features of the cliff edge likely leading to the slumping of material, for instance the 80~m long fracture marked in Figure \ref{fig:x}. A review of fracturing on the nucleus of 67P is available in \cite{elmaarry2015b}.

The fact that cliffs are collapsing is strongly supported by the observation that most cliffs present fallen debris at their foot. We decided to quantify this observation by performing a thorough boulder counting of the area in the proximity and below one of the major active cliffs located at the border between Seth and Hapi regions, at the edge of the area also known as Landing Site Candidate A (http://www.esa.int/Our\_Activities/Space\_Science/Rosetta/ Rosetta\_Landing\_site\_search\_narrows).

We made use of two OSIRIS NAC images obtained on September 21 at 02:34:51UT and on September 22 at 02:34:52 UT. Both images have a similar resolution, i.e. 0.48~m/px for the first one, and 0.49~m/px for the second one, being taken at 27.58~km from the comet nucleus center and 28.07~km, respectively. As we can see in Figure \ref{fig:x}, the two views, A and B, present unique observation geometries which give us the opportunity to analyze the boulder deposits located below the wall/cliff of Site A. Moreover, the high resolution provided by such images allows the unambiguous identification of boulders bigger than 1.5~m. This value corresponds to the 3 pixels sampling rule, chosen to minimize the likelihood of false detection \citep{nyquist1928}. 

\begin{figure}[h!]
\centering
\includegraphics[width=\hsize]{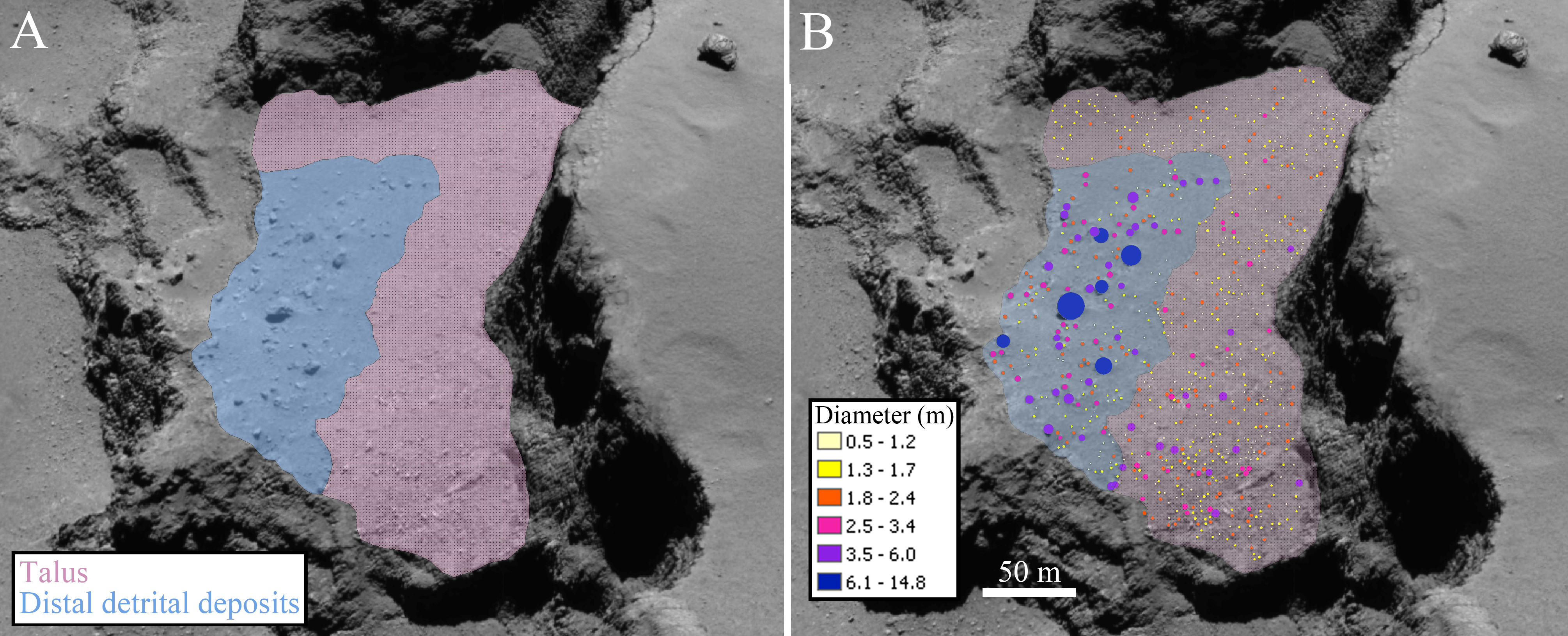}
\caption{Boulder statistics below one of the active cliffs. Panel A: the two units, i.e the talus and the distal detrital deposits, identified for this work. Panel B: The location of the detected boulders, subdivided in different diameters. The difference in sizes between boulders located below the cliff and those far from it is evident.}
\label{fig:y}
\end{figure}

In Figure \ref{fig:y} we show the reference image (A) used in this work and the two main units we have identified. Image (B) in the same figure shows the boulder counts for each unit. The talus unit is the area that is directly below the cliff, while the distal detrital deposits is located about 100~m far from it. 
The division between the talus and the distal detrital deposit is based on the different geomorphological textures they show. The talus deposit is constituted by smaller boulders typically located below or in close proximity to the overlying cliff. This finer material shows an homogeneous texture that is not recognizable in the distal detrital deposit. On the contrary, the distal detrital deposit lays further out from the cliff, showing bigger boulders with a more heterogeneous and articulated texture.

The total boulder counting of the area used the technique presented in \cite{pajola2015}, Figure 7. By manually identifying the boulders through the ArcGIS software, we measured their position on the surface of the comet, and assuming their shapes as circumcircles, we derived their maximum length, i.e. the diameter, and the corresponding area. 

We identified a total number of 730 boulders, 657 of them satisfying the 3 pixels sampling rule. The diameter of the other 73 boulders can be estimated from their cast shadows but since their statistics is not complete they cannot be considered in this analysis. In Figure \ref{fig:y} C the location of such boulders is showed together with a size color-coded distribution. The number of boulders bigger than 1.5~m is 469 for the talus deposit and 188 for the distal detrital deposits.

\begin{figure}[h!]
\centering
\includegraphics[width=\hsize]{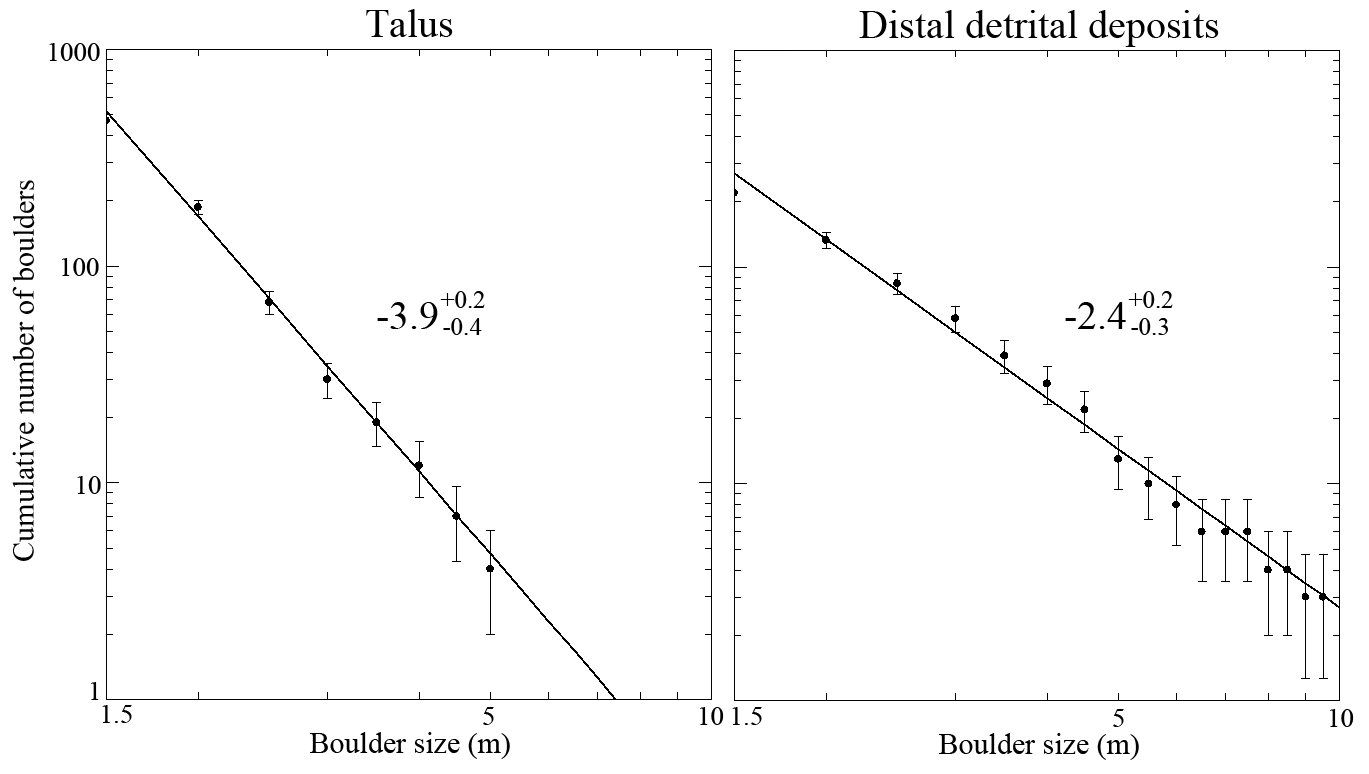}
\caption{ Cumulative number of boulders computed for both the Talus and the Distal detrital deposits. The bin size here is 0.5~m, i.e. the resolution per pixel derived from the OSIRIS NAC images used.}
\label{fig:z}
\end{figure}

A clear dichotomy can be observed between the sizes of the boulders that belong to the talus area and those that constitute the distal deposits (Figure \ref{fig:y}, panel C). Indeed, while the boulders directly below the cliff hardly reach 4~m in diameter, those that are further out from it are frequently above this dimension, reaching in one case a value of 15~m. 
In order to quantify the power-law index of the boulder distributions, we plotted the cumulative number of boulders versus their size. The obtained results for both the talus and the distal detrital deposits are presented in Figure \ref{fig:z}.

By fitting the plots of \ref{fig:z}, we obtained two different power-law index for the two considered areas: -3.9 +0.2/-0.4 for the talus deposit, and -2.4 +0.2/-0.3 for the distal detrital deposits. When comparing such two values, a steeper power-law index, as observed in the talus deposit, means that there is an increase of the population of smaller-size boulders with respect to the bigger ones. On the contrary, a greater number of bigger boulders in the distal detrital deposits coupled with less smaller boulders, has the main result to lower the power slope index derived from the cumulative distribution. 

These numbers should be compared to the global boulder size distribution on 67P presented in \cite{pajola2015} for blocks greater than 7~m, as well as the boulder size distribution of debris at the bottom of active pits shown in \cite{vincent2015b}. These papers have shown that a power law index of -3.9 falls inside the -3.5 to -4 range suggested for gravitational events triggered by sublimation and/or thermal fracturing causing regressive erosion. Despite differences in sizes, this index seems to confirm that the talus deposit we see below the cliff may well be the result of regressive erosion favored by sublimation from the cliff itself. 
Such talus does not present bigger blocks as the detrital deposit does, because its margins are continuously refurbished by blocks and grains from the nearby cliff, as is the case for other pits and alcoves in Seth region. On the contrary, the distal detrital deposits are generally characterized by bigger boulders, with respect to the talus ones, and a coarse presence of blocks smaller than 3 m. This is confirmed by the lower, -2.4 index, and a lower cumulative number of identified blocks, 188 versus 469. 

As we have not yet detected a new boulder in these areas, it is not clear whether the larger blocks in the distal detrital deposit are actually related to the cliff erosion or if they originate from a completely different process. Even in the low gravity of the comet, it seems unlikely that a >10~m diameter block could travel >100~m distance without breaking into smaller elements.

\subsubsection{Granular flows}\label{sec:gflows}
\begin{figure}[h!]
\centering
\includegraphics[width=\hsize]{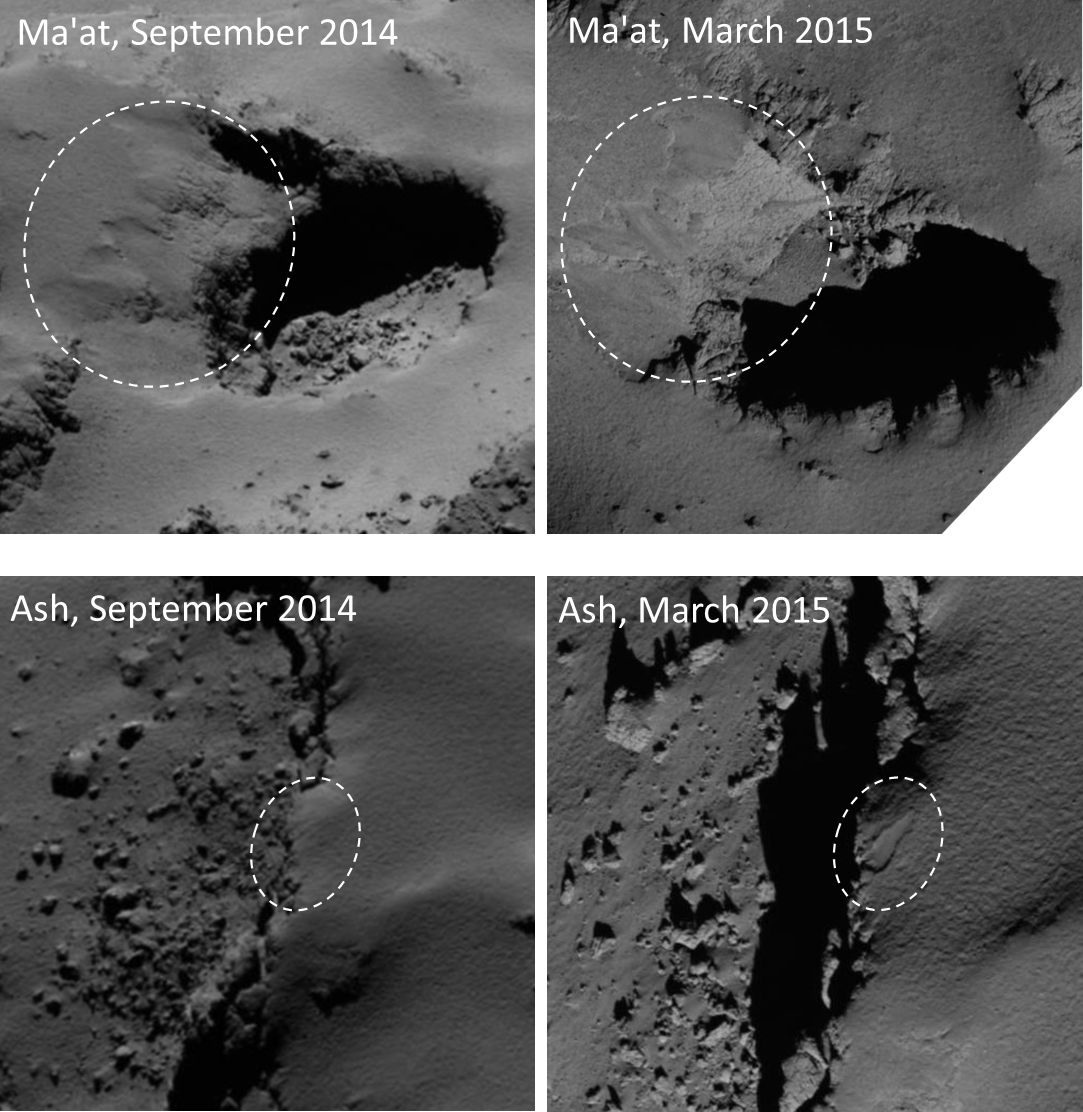}
\caption{Evolving flows, observed in September 2014 and March 2015. Top panel: flows from Ma'at regions between two of the active pits have changed; their outline is different and they seem to have expanded laterally. Bottom panel: new flow in Ash region, following the partial collapse of one active cliff. Image references for top panels: NAC\_2014-09-12T01.33.04.375Z\_ID30\_1397549400\_F22 \& NAC\_2015-03-28T16.18.50.365Z\_ID10\_1397549000\_F82, and bottom panels: NAC\_2014-09-10T18.34.00.345Z\_ID30\_1397549000\_F22 \& NAC\_2015-03-28T14.19.49.405Z\_ID10\_1397549000\_F82).}
\label{fig:flows}
\end{figure}

Mass wasting is observed close to the top edge of most active cliffs, with a flow directed along the local slope. Some flows are visible in Figure \ref{fig:x} (arched shapes in the dust layer on the cliff plateau). Figure \ref{fig:flows} shows a few more examples of such avalanches at different places on the nucleus. 

We calculated the slopes of this terrains taking into account local gravity and centrifuge force. The gravity of the body is calculated on the polyhedral shape of the nucleus \citep{preusker2015}, assuming a constant density of $~500 kg.m^{-3}$ (value from \cite{sierks2015} updated to the latest estimate of the comet volume). We find that slopes of granular flows vary from 20 to 30 \degree. These slopes define the maximum angle of repose for cometary material, slightly below what is measured for granular material on rocky bodies (typical value ~30 \degree).

It is not obvious from our images whether these features are material shuffled around by local gas release, or "dry" granular not triggered by a gas flow. We tend to favour the latter explanation because of the general good agreement between flow direction, gravity, and size distribution of debris. In-place reorganizing of dust would affect only the smaller grains as the gas flows involved cannot push meter-size debris down the cliffs. 

We propose that granular flows are triggered by the collapse of cliffs. We do observe the edges of cliffs slumping down in many places, and losing blocks from erosion. At some point the dust mantle covering the cliff top is no longer supported on its edge and will start to flow down the local gravity vector (Figure \ref{fig:scenario}). We have observed a few examples of cliff edges about to collapse; with blocks appearing to be almost separated from the cliff by a large crack. These blocks typically extend 10~m from the cliff edge, a size comparable to the maximum size of blocks at the cliffs foot. We propose this scale as the maximum distance fractures can propagate inside the cliff before triggering the collapse.

If a cliff is active, we expect this process to be ongoing. The sublimation and release of volatiles from the cliff will further its weakening and generate new collapses, with local retreats of 10~m or more. Granular flows and other mass wasting previously formed will be reactivated or modified subsequently and we should be able to observe their evolution. Collapsing cliffs will also form new flows. This is challenging to measure because many changes are at the limit of our detection capabilities but we have observed evolving flows in a few areas, consistent with the model. See examples in Figure \ref{fig:flows}.

\subsubsection{Exposure of bright material}\label{sec:bright_mat}

The sources associated to activity, more exactly cliffs and collapsed wall features, are consistently associated to specific variegation. Edges of cliffs are a few percent brighter and bluer than inactive surfaces. This is true in both the Visible \citep{fornasier2015, oklay2015} and Infrared range of the spectra \citep{capaccioni2015}. 

In addition, more than a hundred very bright spots have been observed on the comet, most of them appearing as clusters in the debris field of active cliffs. They are interpreted as water ice exposed on the surface of boulders dislocated from the cliff during collapse \citep{pommerol2015}. 

We use this observational fact to reevaluate partially what the source of jets actually is. As explained in section \ref{sec:inversion}, our jet observations often do not sample the first few 10s of meters above the surface and the link jet-cliffs is only an extrapolation of the jet direction measured above this blind spot. It is very possible that the real active spot extends beyond the cliff itself, and includes the debris at the cliff foot, as well as the bright material newly exposed in mass wasted areas. The gas flows arising from all these features will compete against each other and the topography, leading to a focused jet. This can explain why dust jets starting from vertical surfaces seem to bend upward rapidly and finally expand radially away from the nucleus. Although it can be that the jet direction is fully controlled by morphological features, such as fractures, but it is more likely that the jet is pushed upwards by gas arising from the debris field at the bottom of the cliff. This case is compatible with the work of several authors who studied the trajectories of gas flows around complex topographic features \citep{crifo2002, skorov2006, vincent12}.

Can the variegation of the talus like the one seen in Figure \ref{fig:source_imh_blue} be purely explained by airfall material ejected from the cliff ? The bluer debris field typically extend 100 to 150~m from the cliffs. Simple ballistic considerations show that this distance is reached by material ejected at a speed of at most $15 cm.s^{-1}$, 10 times less than the escape velocity ($1 m.s^{-1}$), and 50 times less than the typical velocity of dust particles ejected from the surface ($3.75 m.s^{-1}$, \cite{rotundi2015}). Because we see the jets material escaping from these areas, it is unlikely that a significant fallback takes place, and it seems more probable than the variegation is due to intrinsic properties of the fallen blocks. As a significant fraction of their surface was not exposed to the Sun when they were still attached to the cliff, it seems logical that they contain more volatile material than the surrounding terrains.

\subsection{Theoretical model}\label{sec:fractures_sources}

Most jets from the Northern hemisphere of 67P arise from areas with sharp topography, in particular cliffs. We have discussed observational evidence for regressive erosion which may be triggered by the activity. The following paragraphs will review how these typical morphologies can help sustaining the activity.

If fractures act as nozzle for the gas flows, effectively leading to the formation of jets, it means that they also define the smallest possible jets. All missions to comets have resolved jets as small as the spatial scale of their observations, pushing down the lower limit but not fixing it. The good correlation between jets and fractured terrains on one side, smallest jet diameter and fracture width in these areas on the other, suggest that the smallest jets we observe are constrained by the fracture dimensions, a couple of meters wide at most. Note that because jets are fuzzy, and more than one fracture is active in any area, jets often appear to arise from a few fractures rather than one.
The idea of jets arising from cracks, holes, or other morphological features has been extensively discussed in several papers, for instance \cite{belton2013}, \cite{brucksyal2013} for comets, or \cite{yeoh2015} who modeled such phenomena to explain the water plumes of Enceladus.

A detailed modeling of the gas dynamics and dust acceleration within the fractures requires complex simulations and will be the subject of a future study. Instead, we present here some physical considerations which approximate the real processes taking place in these fractures.

The main constraint on our problem is that gas flows in the jets must be able to lift the dust grains other instruments have collected around the comet. These grains are quite large for comet standards. Indeed, light scattering properties in ground based observations of 67P are dominated by $100~\mu m$ size particles and both GIADA and COSIMA on board Rosetta show a lack of micron size grains \citep{rotundi2015, schulz2015}. 
Dust velocities are measured by GIADA (Rotundi et al, 2015). The cumulative distribution shows a mean velocity equal to $3.75~m.s^{-1}$, $\simeq 4$ times the escape velocity. The average dust density is $1900~kg.m^{-3}$.

Once a particle is released, drag force and gravity are the main forces that will control its trajectory. Other effects like radiation pressure are negligible as long as the dust motion is still coupled to the gas, which is the case for the short time and distance scales we consider here.

Nucleus gravity is calculated from the OSIRIS volume reconstruction ($21.4~km^3$) and the Rosetta Radio Science Investigation mass estimate ($1 \times 10^{13}~kg$), both values published in \cite{sierks2015}. We assume homogeneous density at first order. The average mean radius is 1720 m, and the average surface gravity is $2.25 \times 10^{-4}~m.s^{-2}$.

We use a simple aerodynamic model to calculate the drag force, assuming spherical particles and a gas molecule free mean path much smaller than the dust diameter.
\begin{equation}\label{eq:drag}
F_d = \frac{1}{2} \rho_g v_r^2 C_d A \qquad (N)
\end{equation}

With $\rho_g$ the gas density, $v_r$ the relative velocity of the gas w.r.t. the particle, $A$ the cross section of the particle, and $C_d$ the drag coefficient($\simeq 1$ for roughly spherical grains). Because $v_{dust} << v_{gas}$ we can replace $v_r$ with $v_{gas}$ in the equation.\\

We do not consider complex situations with gas percolating through a porous dust layer. We rather assume a sublimating source inside a fracture, or at the bottom of a wall, and look at how the arising gas flow could tear out dust grains in its vicinity. 

Dust can only be lifted if the drag force overcomes gravity + forces holding the particle in place. These forces are not well characterized for cometary material but can be approximated using the surface tensile strength. \cite{thomas2015}, \cite{vincent2015b}, and \cite{groussin2015} have independently estimated the tensile strength of the surface upper layers in various regions, by studying the morphology of pits and overhangs and reached the same average value of $Y = 50~Pa$. 

We consider a spherical particle half embedded in the nucleus surface. The tensile force holding the particle in place is $F_t = Y \times area = 50 \times \pi r^2 \quad (N)$ and is typically 5 orders of magnitude larger than the weight of the particle. One could also consider cohesion or shear strength instead of tensile strength, but the order or magnitude is similar.

The drag force is a function of particle size + gas pressure and velocity. A gas expanding from a flat surface would have a mean velocity purely driven by thermal consideration. The Maxwellian distribution gives the root mean squared speed:
\begin{equation}
v_{gas} = \sqrt{\frac{3k_BT}{ m}} \qquad (m.s^{-1})
\end{equation} 
with $k_B$ being Boltzmann's constant, and $m$ the molecular mass of $H_2O~ (3.10^{-26}~kg)$.

The gas density is defined by: 
\begin{equation}
\rho_{gas} = P/(R T) \qquad (kg.m^{-3})
\end{equation}
with R = 8.31447 (constant for perfect gases)

The vapor pressure of ice can be defined as function of temperature, usually defined empirically. Reviews of various models and experimental measurements are available in \cite{andreas2007} and \cite{gundlach2011}. In this paper we use the empirical relation:
\begin{equation}
P_{sat}(T) = a_1 e^{-a_2/T} \qquad (Pa)
\end{equation}

with $a_1 = 3.23 \times 10^{12}~Pa$ and $a_2 = 6134.6~K$ \citep{gundlach2011}.

%

Using these equations we can calculate the drag force as a function of the temperature. Results for different grain sizes are summarized in Table \ref{tab:sublimation}.

\begin{table*}
\centering 
\begin{tabular}{c c c c c c c c} 
\hline\hline
T (K) & v ($m/s$) & Psat (Pa) & $\rho_{gas}$ ($kg/m^3$) & Fdrag (N) & Fdrag (N) & Fdrag (N) & Fdrag (N)\\
\hline
170 & 484.36 & 6.88E-04 & 5.10E-07 & 4.48E-14 & 4.48E-12 & 4.48E-10 & 4.48E-08 \\
180 & 498.40 & 5.10E-03 & 3.58E-06 & 3.33E-13 & 3.33E-11 & 3.33E-09 & 3.33E-07 \\
190 & 512.05 & 3.07E-02 & 2.04E-05 & 2.00E-12 & 2.00E-10 & 2.00E-08 & 2.00E-06 \\
200 & 525.36 & 1.54E-01 & 9.78E-05 & 1.01E-11 & 1.01E-09 & 1.01E-07 & 1.01E-05 \\
210 & 538.33 & 6.64E-01 & 4.02E-04 & 4.33E-11 & 4.33E-09 & 4.33E-07 & 4.33E-05 \\
220 & 551.00 & 2.51E+00 & 1.45E-03 & 1.63E-10 & 1.63E-08 & 1.63E-06 & 1.63E-04 \\
230 & 563.38 & 8.43E+00 & 4.68E-03 & 5.49E-10 & 5.49E-08 & 5.49E-06 & 5.49E-04 \\
250 & 587.37 & 7.12E+01 & 3.66E-02 & 4.64E-09 & 4.64E-07 & 4.64E-05 & 4.64E-03 \\
260 & 599.00 & 1.83E+02 & 9.05E-02 & 1.19E-08 & 1.19E-06 & 1.19E-04 & 1.19E-02 \\
270 & 610.41 & 4.38E+02 & 2.09E-01 & 2.86E-08 & 2.86E-06 & 2.86E-04 & 2.86E-02 \\
280 & 621.61 & 9.87E+02 & 4.55E-01 & 6.43E-08 & 6.43E-06 & 6.43E-04 & 6.43E-02 \\
290 & 632.61 & 2.10E+03 & 9.38E-01 & 1.37E-07 & 1.37E-05 & 1.37E-03 & 1.37E-01 \\
300 & 643.43 & 4.25E+03 & 1.84E+00 & 2.77E-07 & 2.77E-05 & 2.77E-03 & 2.77E-01 \\
\hline 
\multicolumn{4}{r}{Particle size} & $1~\mu m$ & $10~\mu m$ & $100~\mu m$ & $1000~\mu m$  \\
\multicolumn{4}{r}{Tensile force (N)} & 1.96E-11 & 1.96E-09	& 1.96E-07 & 1.96E-05 \\
\hline
\end{tabular}
\caption{Gas dynamics and drag force as a function of grain size and temperature}
\label{tab:sublimation}
\end{table*}

\begin{figure}[h!]
\centering\
\includegraphics[width=\hsize]{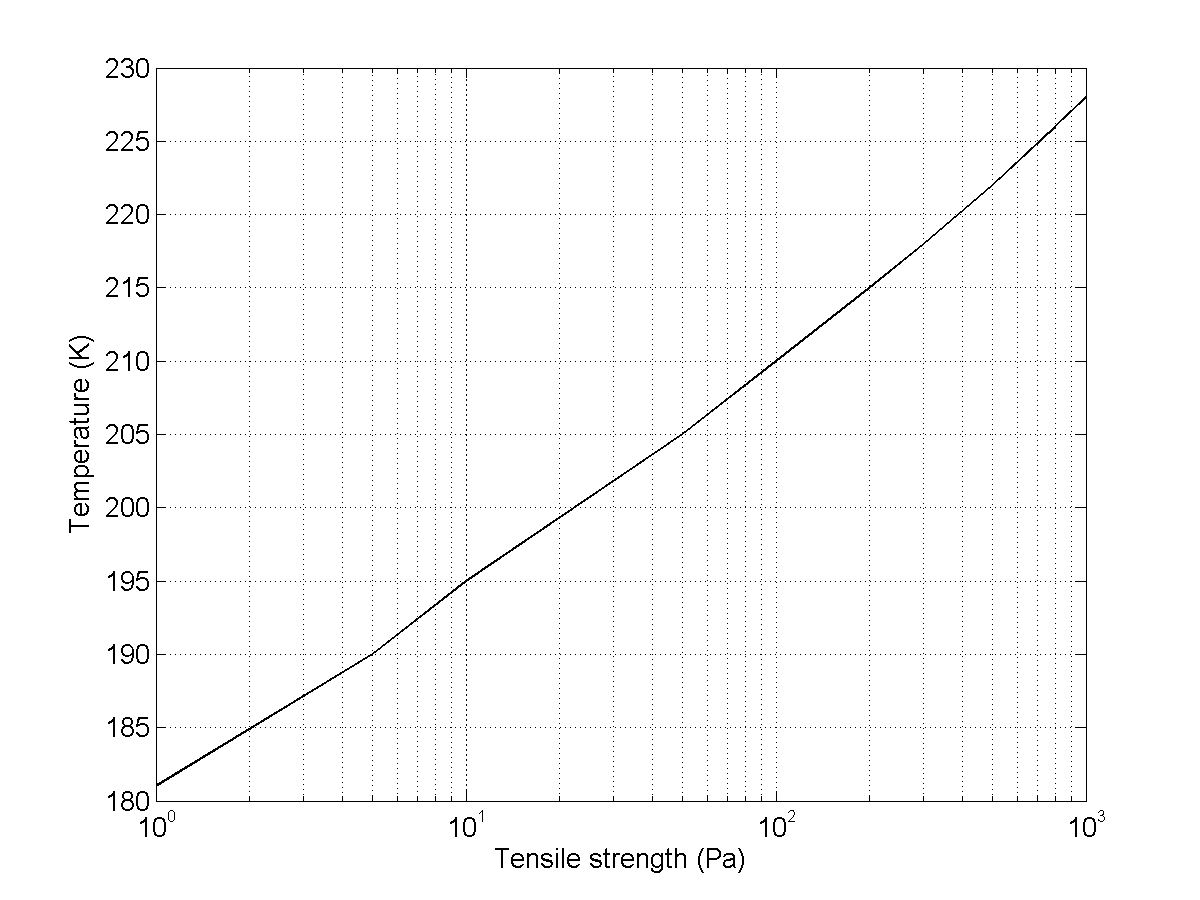}
\caption{Minimum sublimation temperature needed to produce a gas flow able to tear off dust grains half embedded in the nucleus, as a function of the tensile strength of the surface.}
\label{fig:lift_model}
\end{figure}

\begin{figure*}[h!]
\centering
\includegraphics[width=5.5cm]{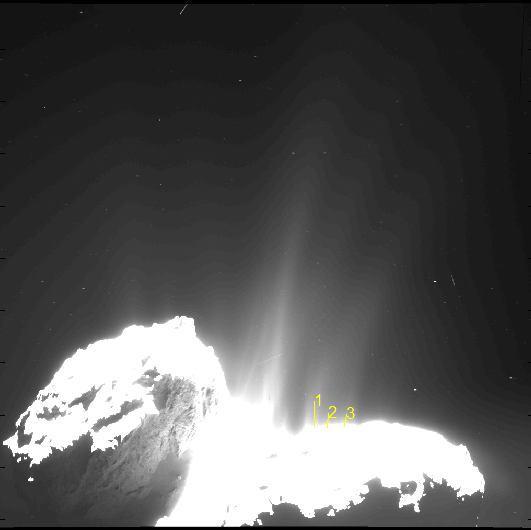}
\includegraphics[width=5.5cm]{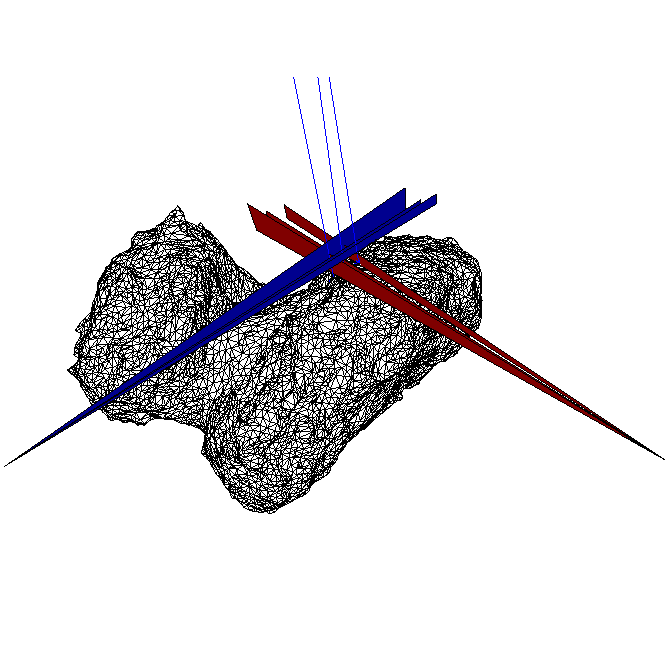}
\includegraphics[width=5.5cm]{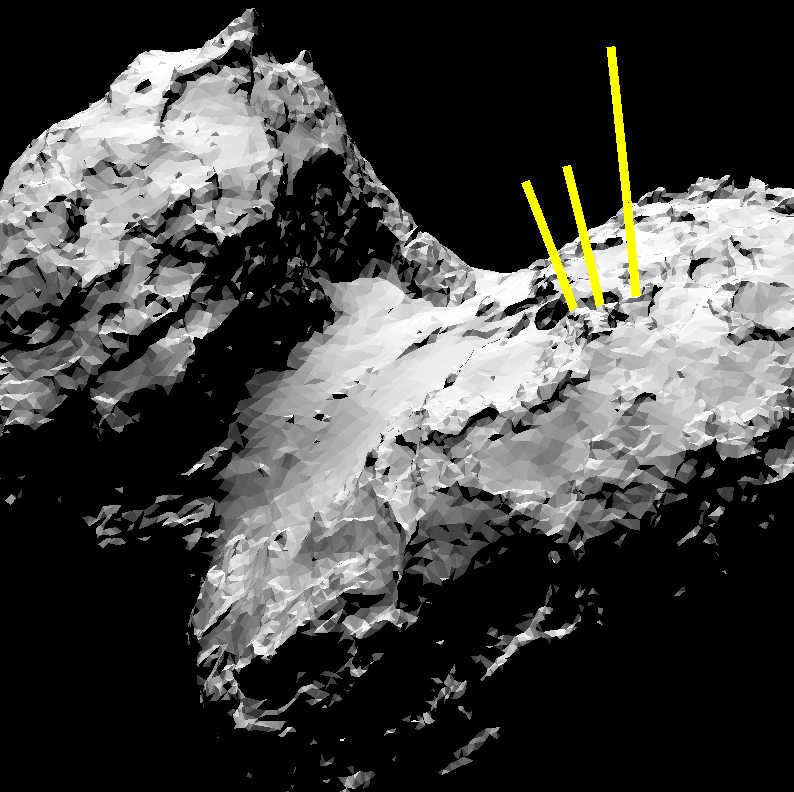}
\caption{Summary of the direct inversion technique. Left panel: jets are identified in two different images, only one is shown here, see Figure \ref{fig:STP017_GAS} for the full sequence. Middle panel: triangulation, for each image we know the precise position and attitude of the spacecraft. From this we define the planes of jets (spacecraft, line of sight, observed jet direction). The intersection between these planes defines the jet in three dimensions, and we calculate their intersection with the surface. Right panel: Knowing sources and directions, plus some assumptions on grains velocity, we can reconstruct the jets trajectories (yellow lines)}
\label{fig:inversion}
\end{figure*}

\begin{figure*}[h!]
\centering
\includegraphics[width=3.3cm]{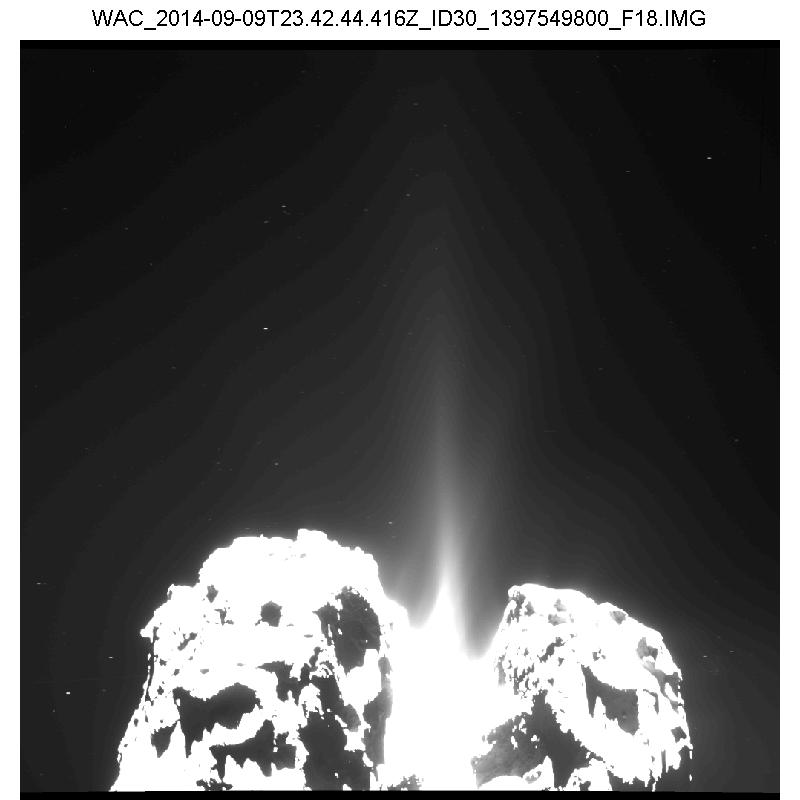}
\includegraphics[width=3.3cm]{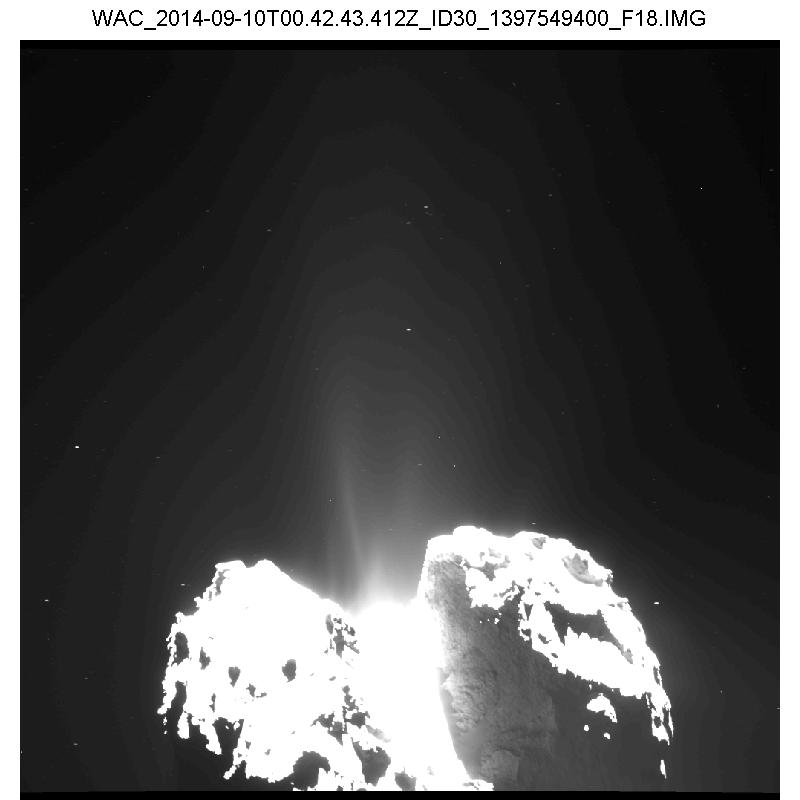}
\includegraphics[width=3.3cm]{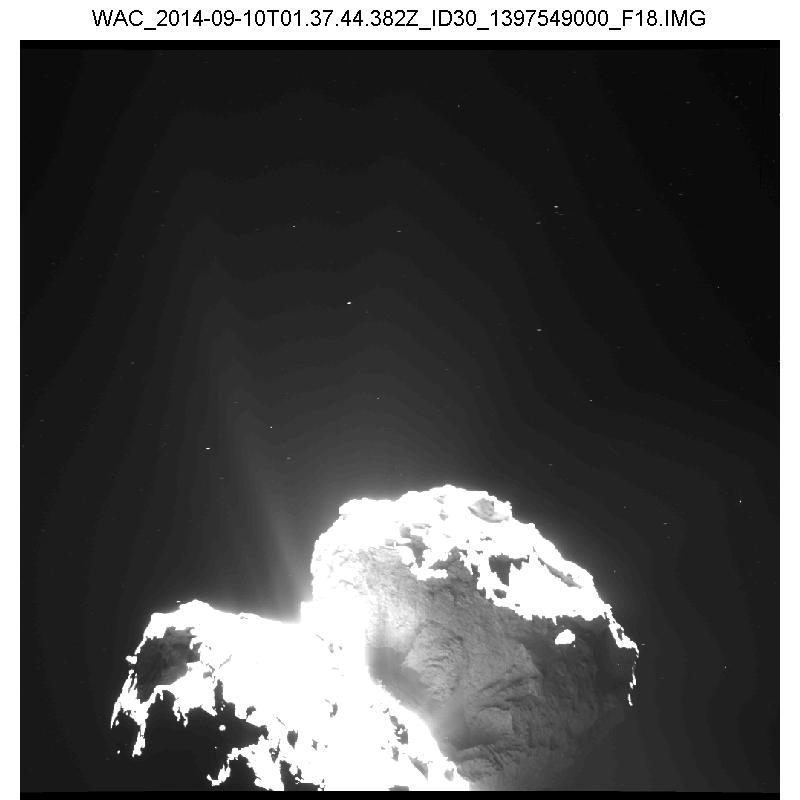}
\includegraphics[width=3.3cm]{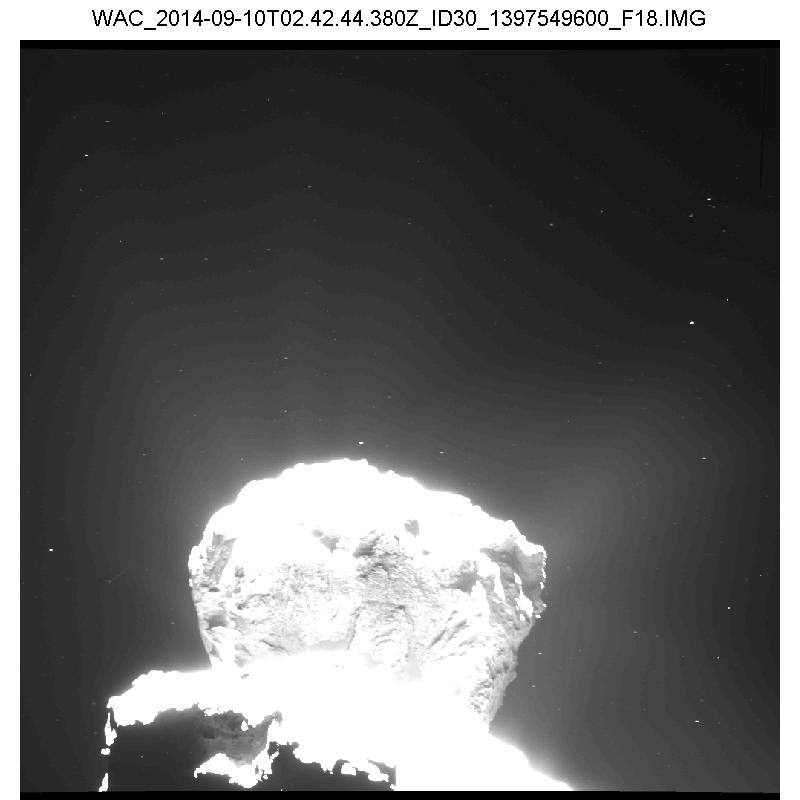}
\includegraphics[width=3.3cm]{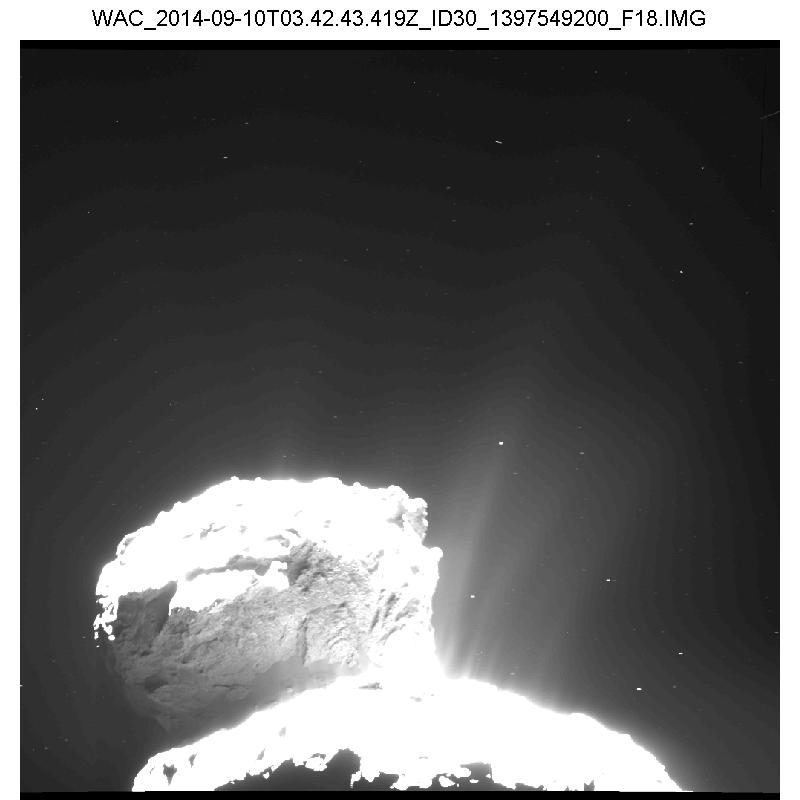}\\

\includegraphics[width=3.3cm]{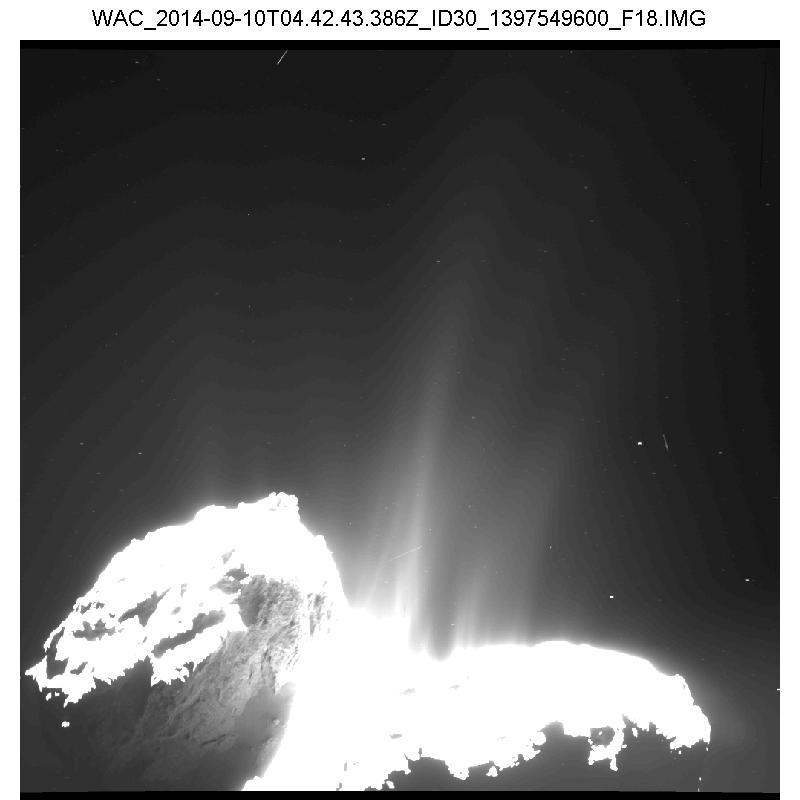}
\includegraphics[width=3.3cm]{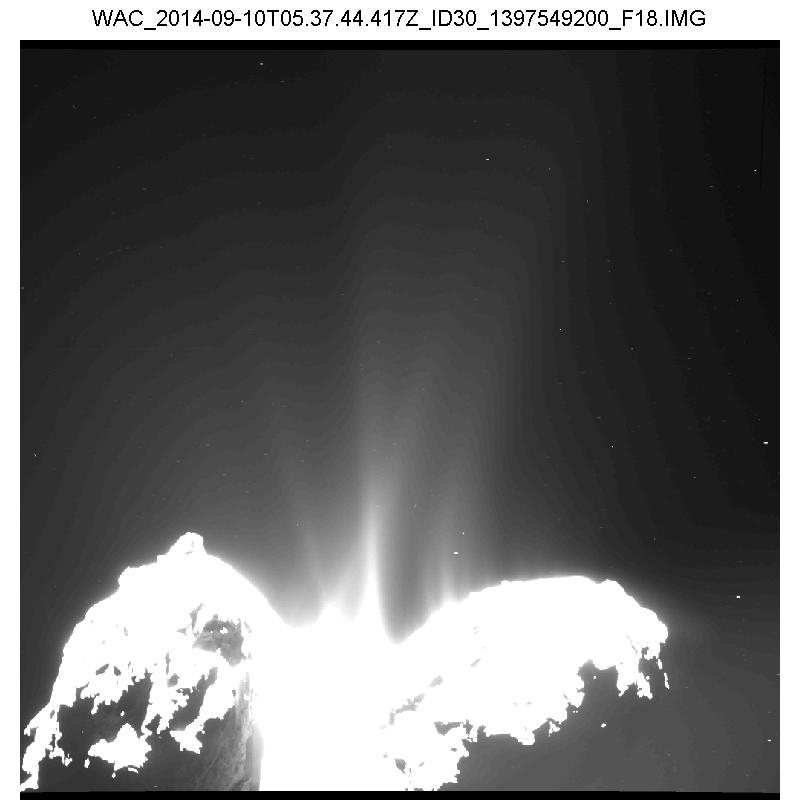}
\includegraphics[width=3.3cm]{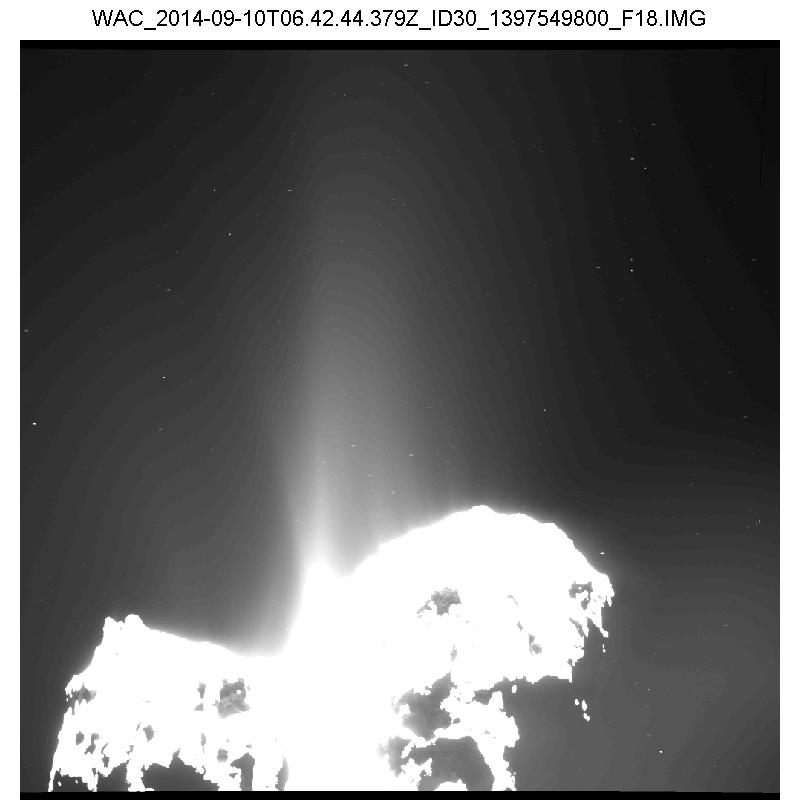}
\includegraphics[width=3.3cm]{STP017_GAS/b04_42.png}
\includegraphics[width=3.3cm]{STP017_GAS/b05_37.png}

\caption{10 WAC images from a sequence acquired on 10 September 2014, from a distance of 29 km and a sub-spacecraft latitude of +48 \degree. The sequence shows the diurnal variation of activity over 9 hours (72\%) of nucleus rotation. Every view is separated from the previous one by about 1h (30 \degree of nucleus rotation). Sun and comet North pole are pointing up in these images. Brightness levels are stretched linearly to emphasize the 5\% least bright pixel values.}
\label{fig:STP017_GAS}
\end{figure*}

\begin{figure*}[h!]
\centering
\includegraphics[width=3.3cm]{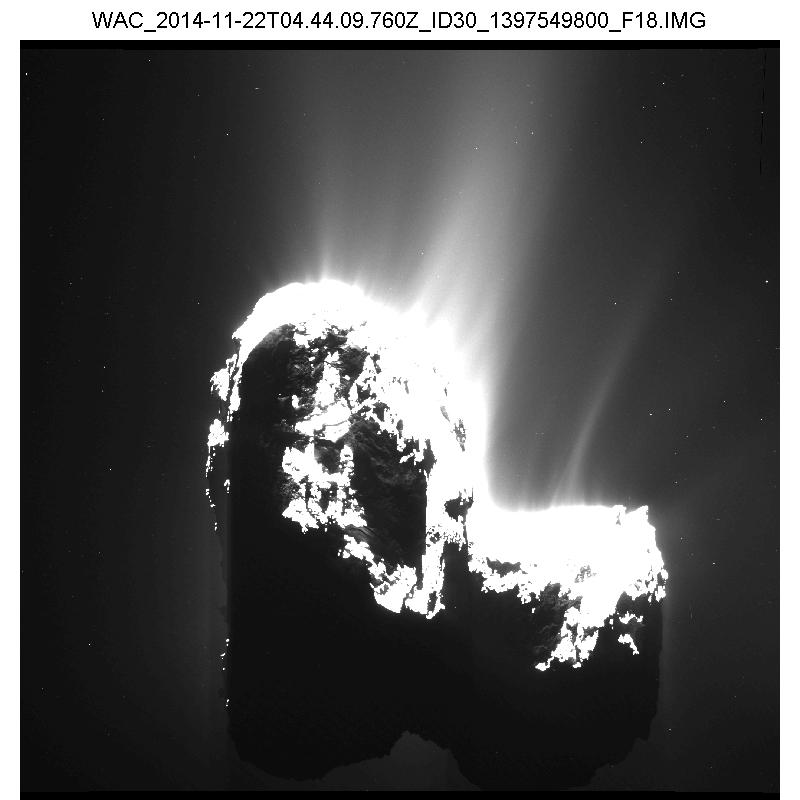}
\includegraphics[width=3.3cm]{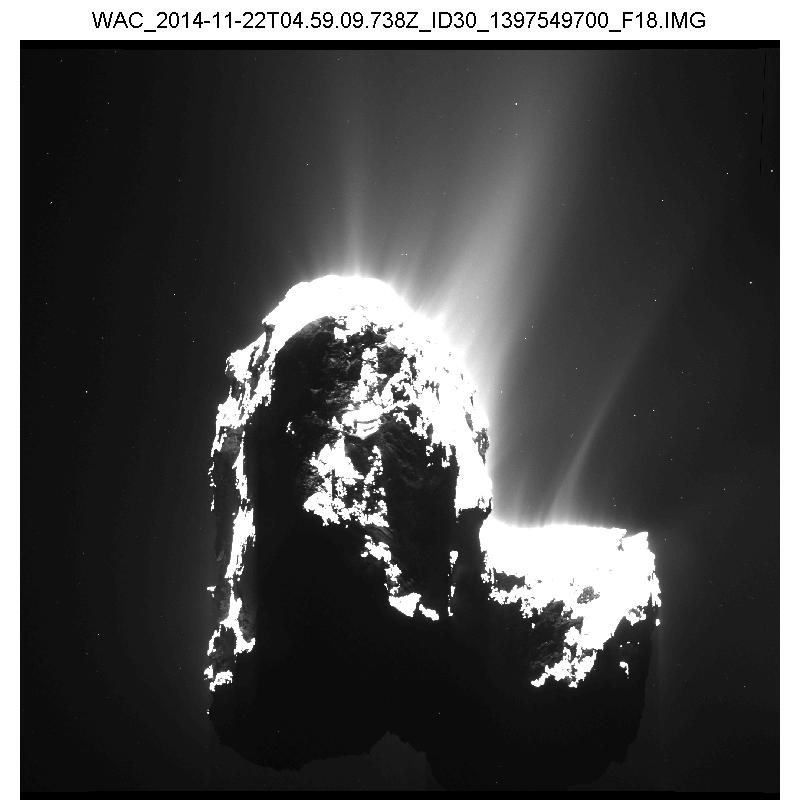}
\includegraphics[width=3.3cm]{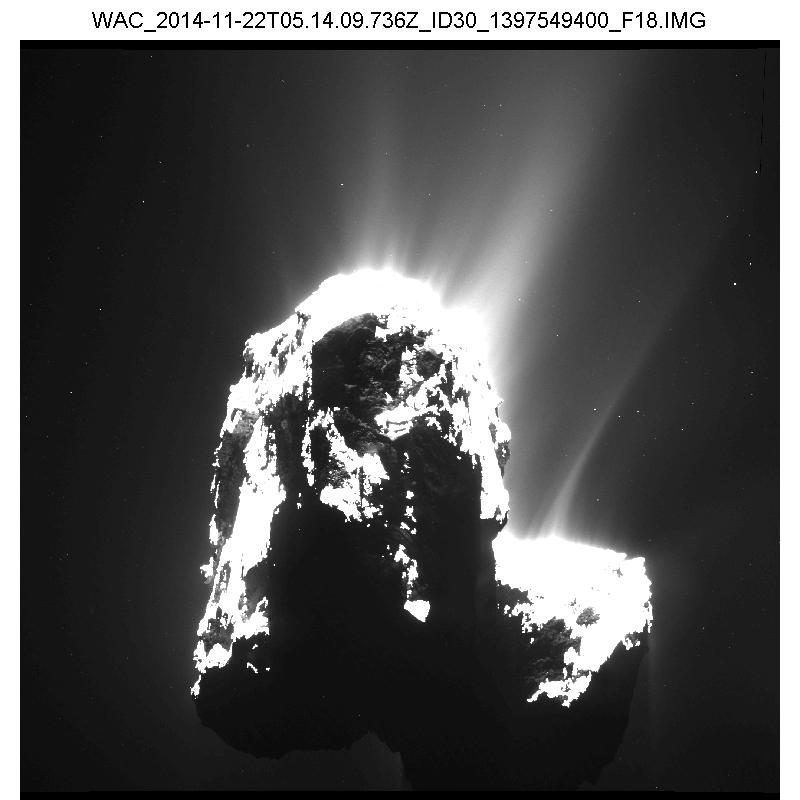}
\includegraphics[width=3.3cm]{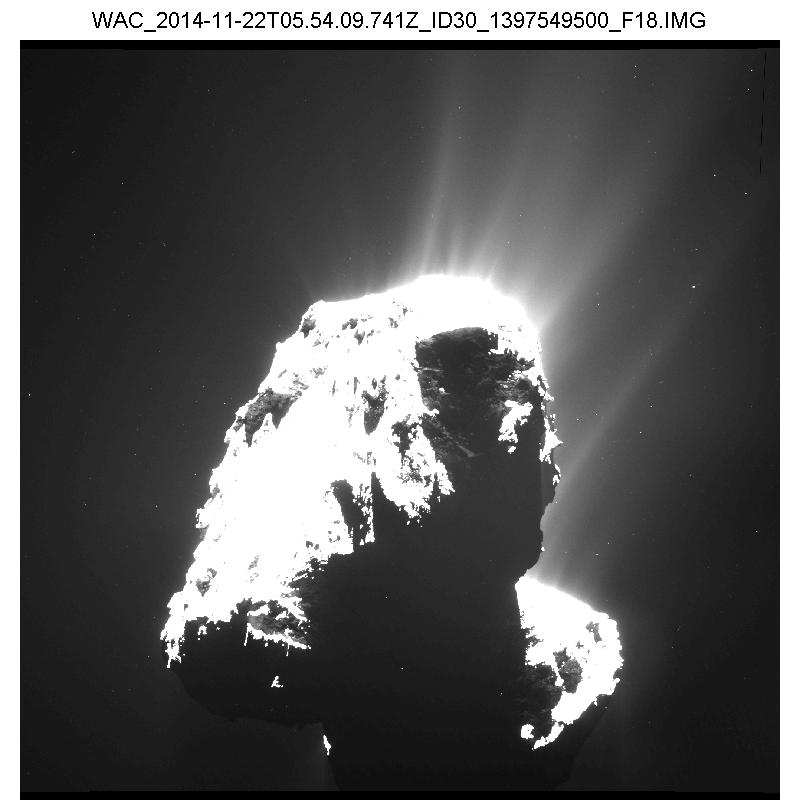}
\includegraphics[width=3.3cm]{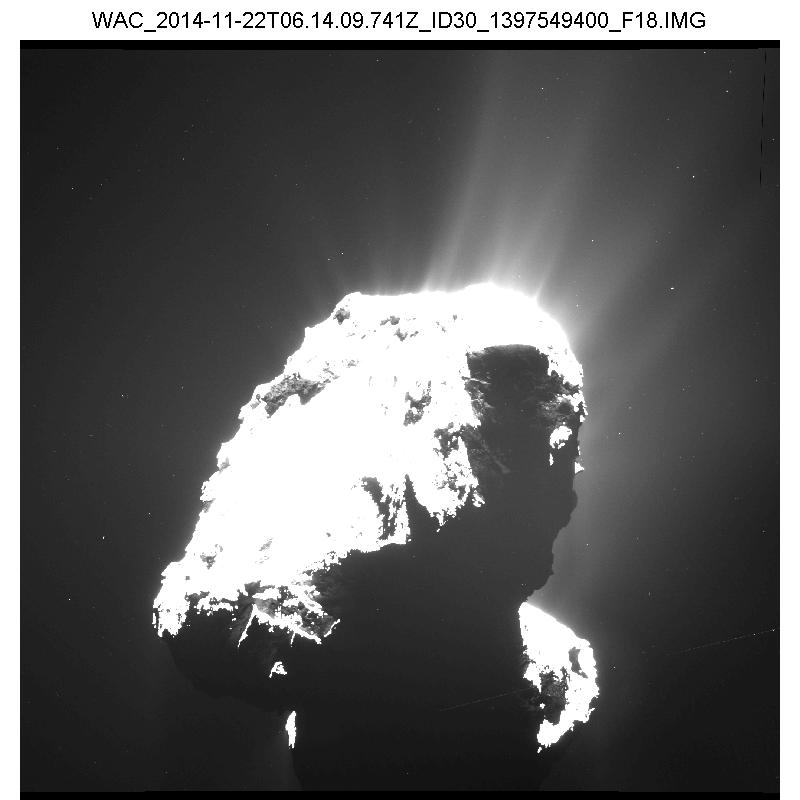}\\

\includegraphics[width=3.3cm]{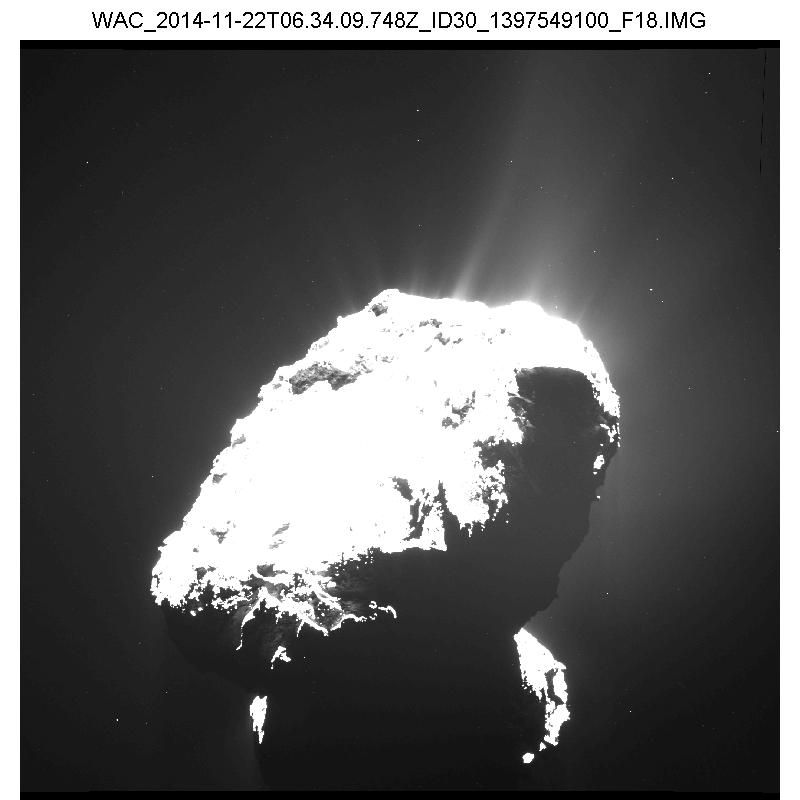}
\includegraphics[width=3.3cm]{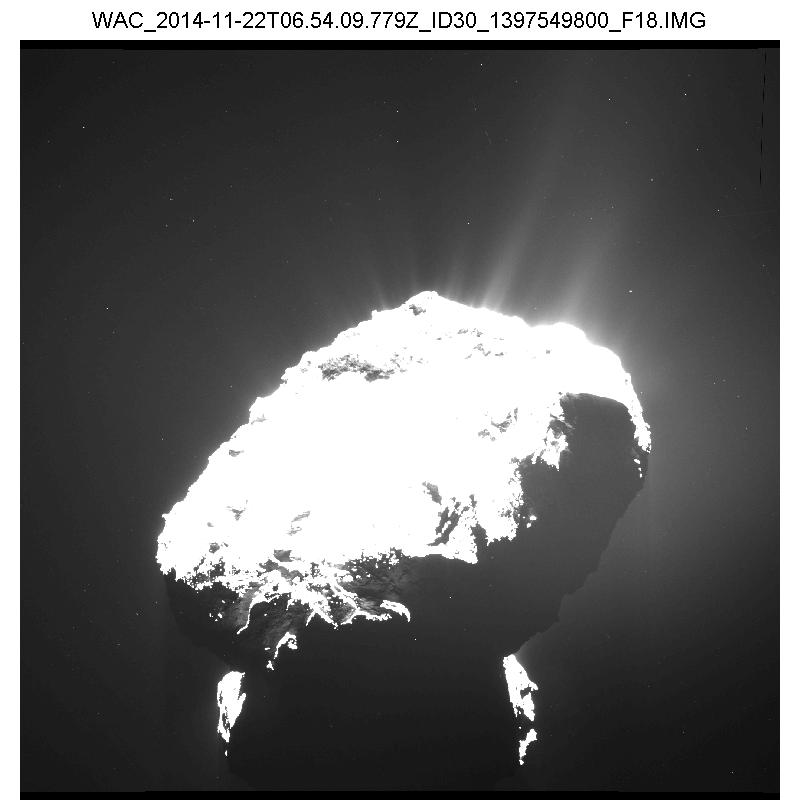}
\includegraphics[width=3.3cm]{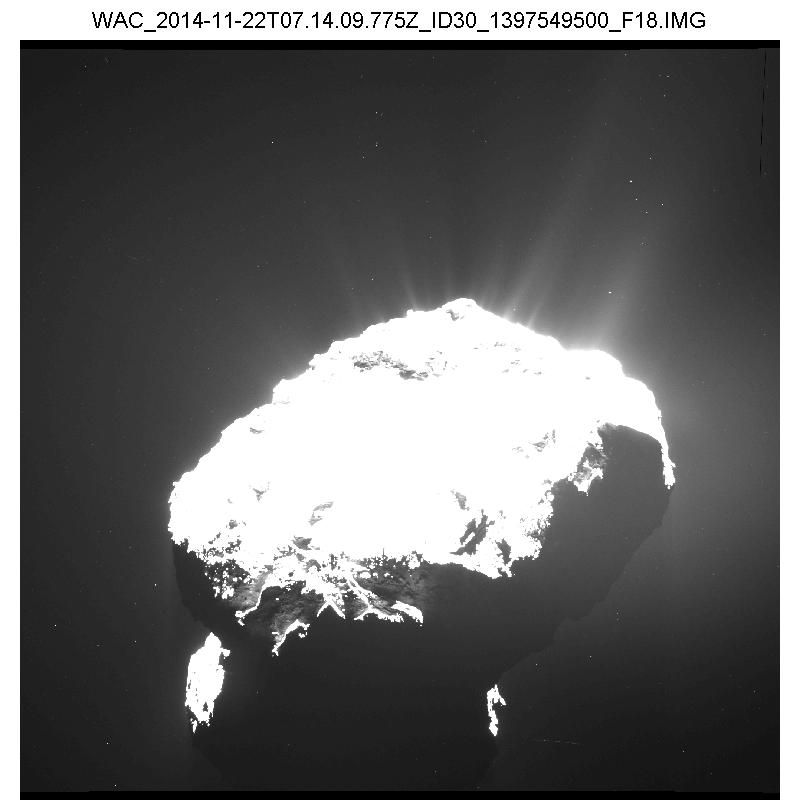}
\includegraphics[width=3.3cm]{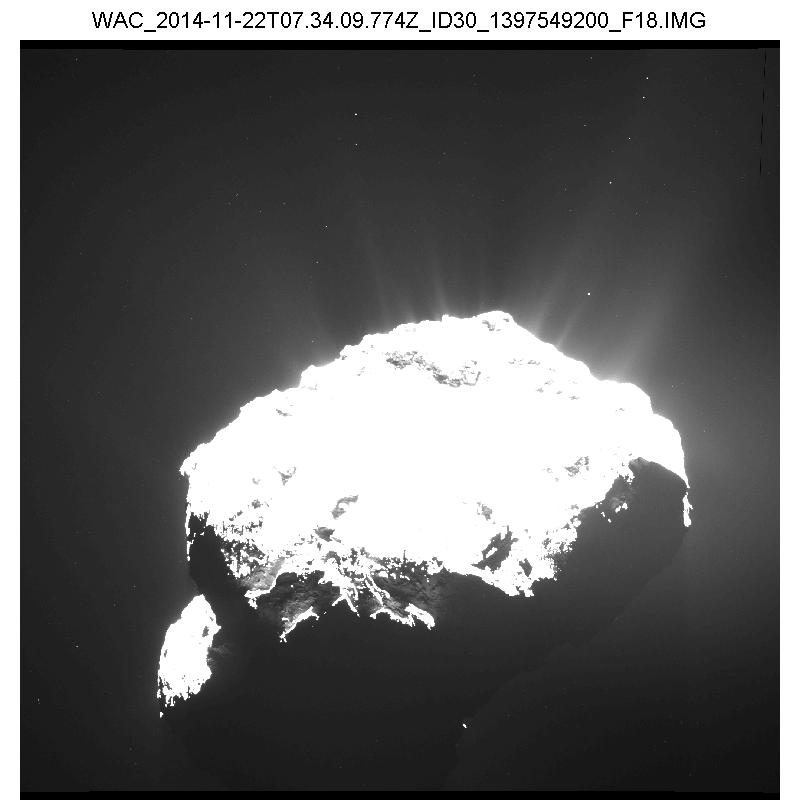}
\includegraphics[width=3.3cm]{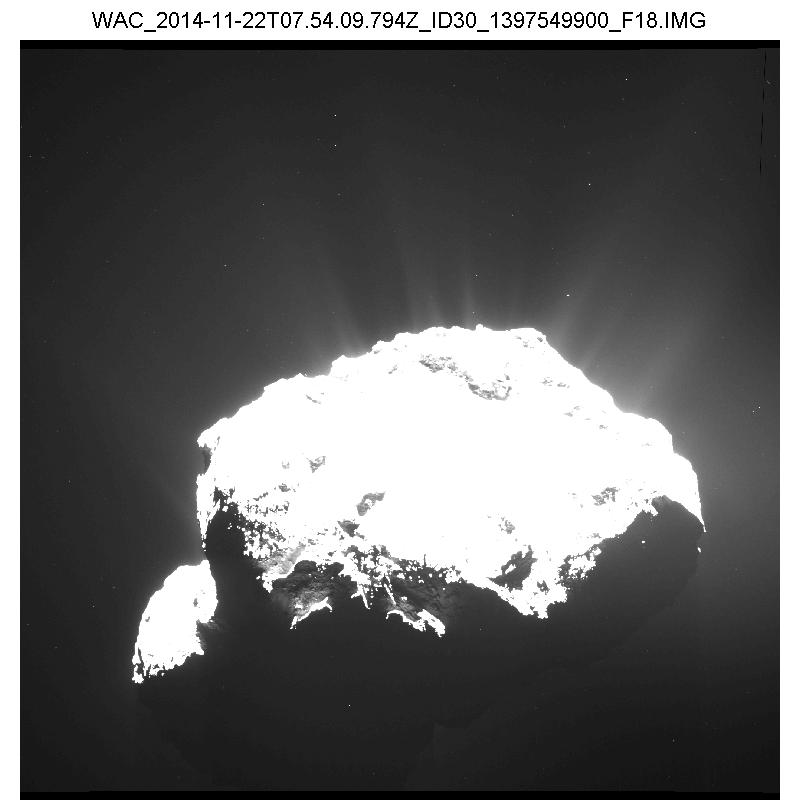}\\

\includegraphics[width=3.3cm]{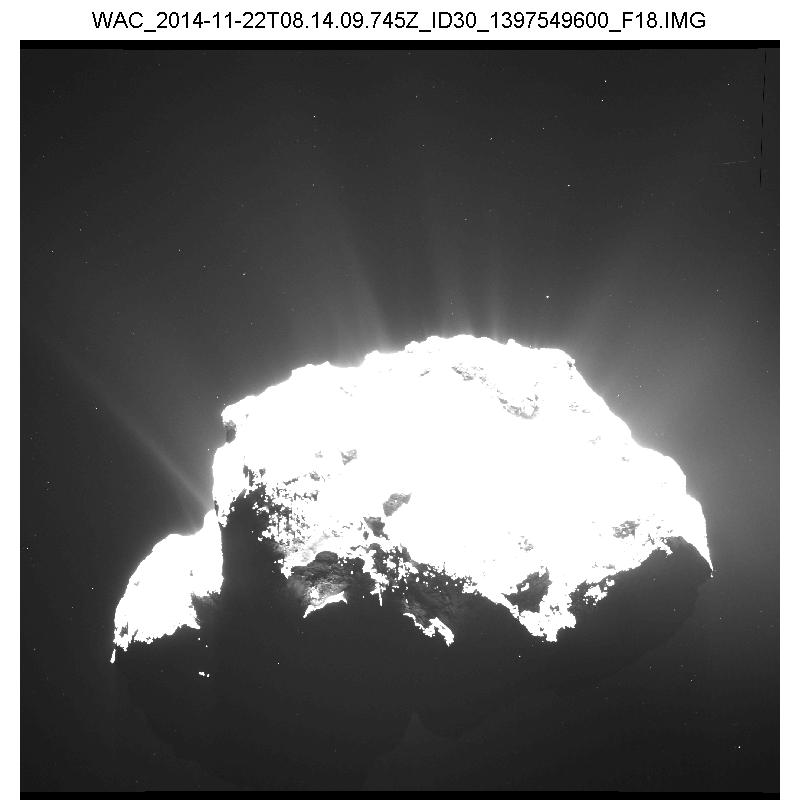}
\includegraphics[width=3.3cm]{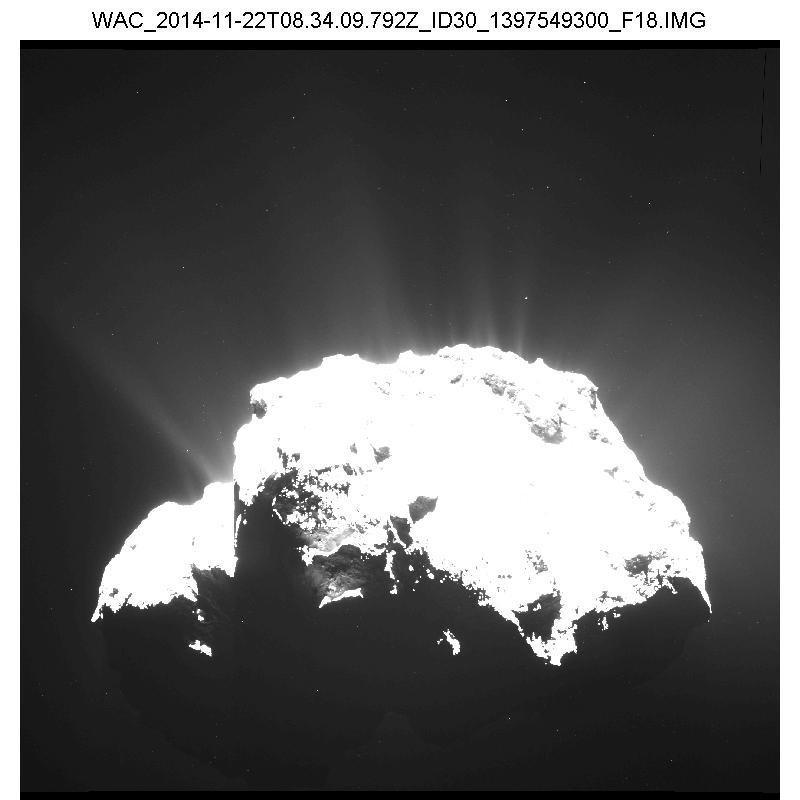}
\includegraphics[width=3.3cm]{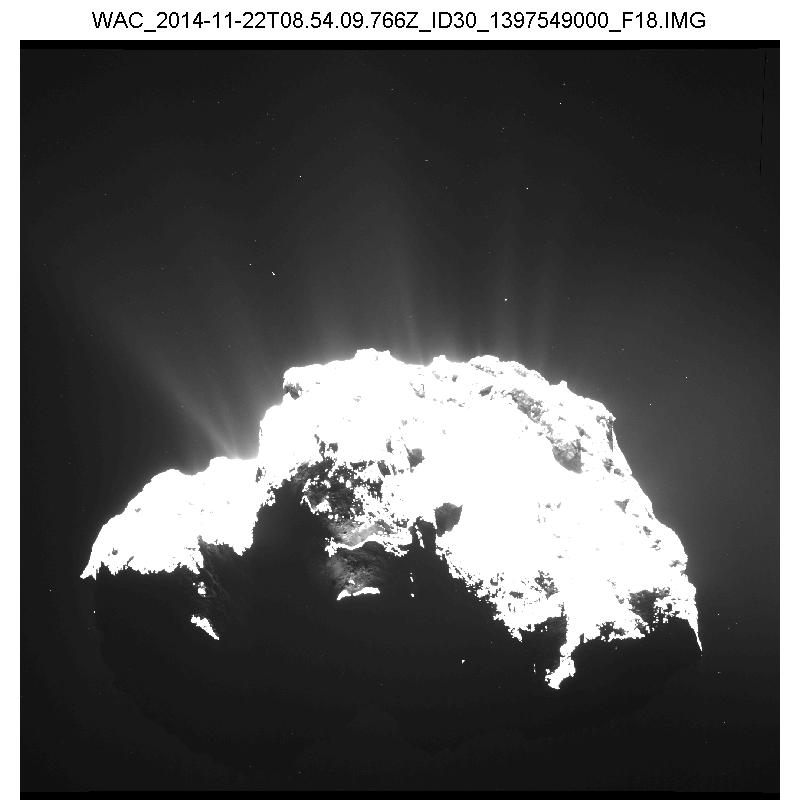}
\includegraphics[width=3.3cm]{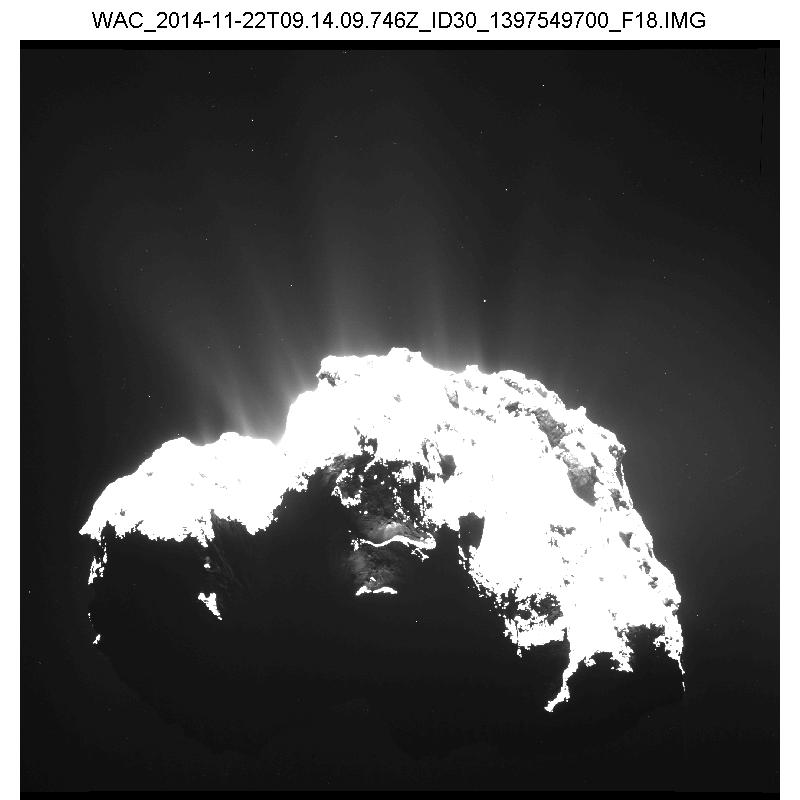}
\includegraphics[width=3.3cm]{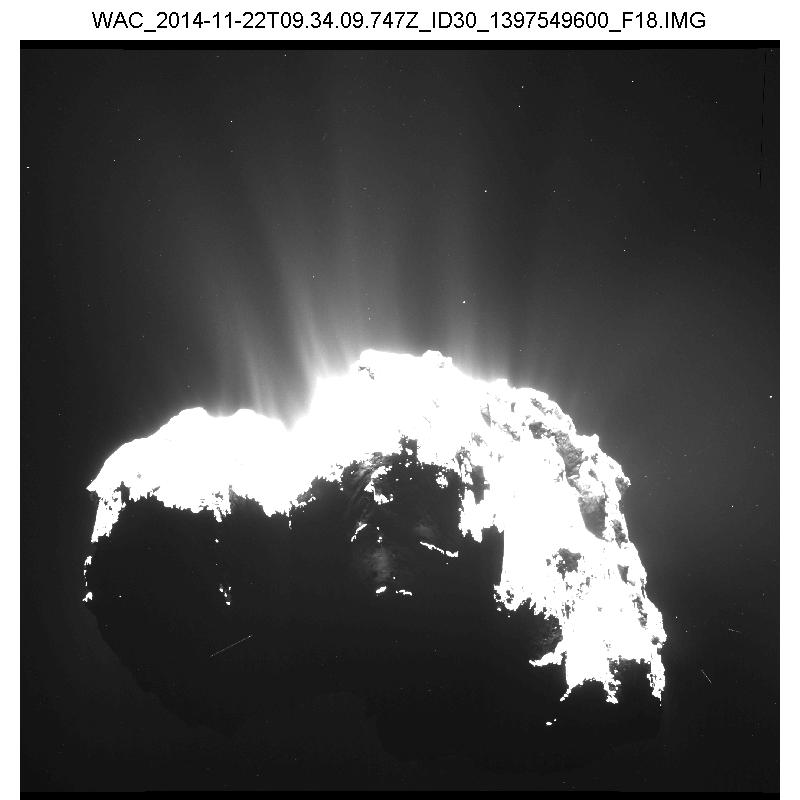}\\

\includegraphics[width=3.3cm]{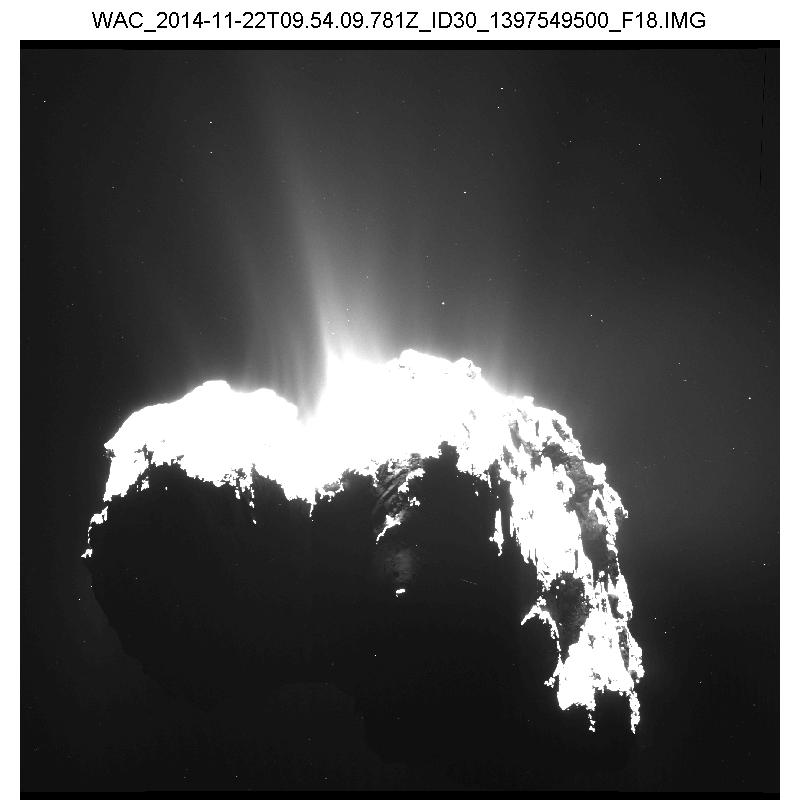}
\includegraphics[width=3.3cm]{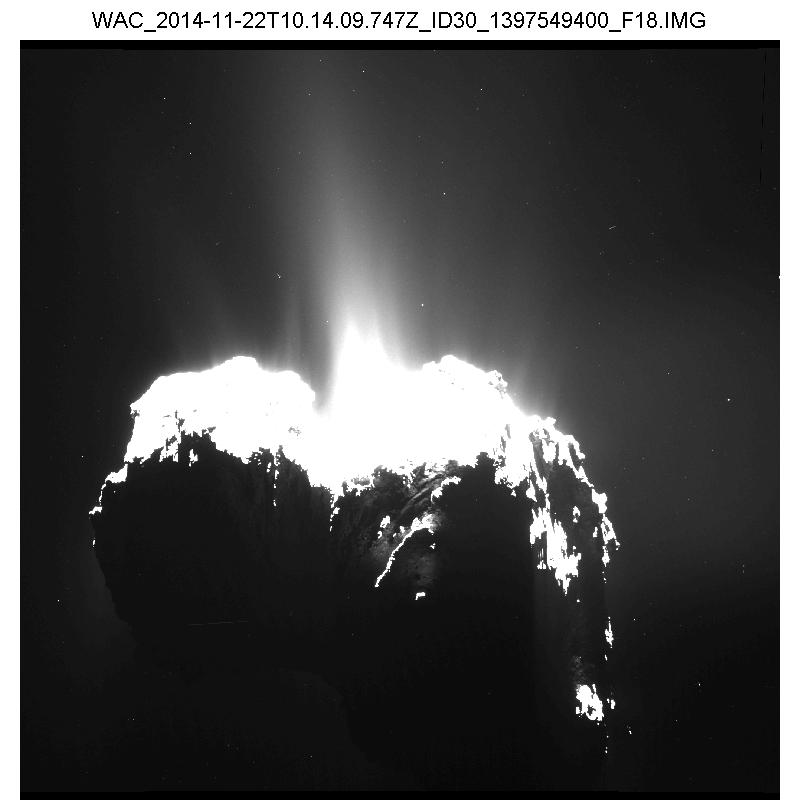}
\includegraphics[width=3.3cm]{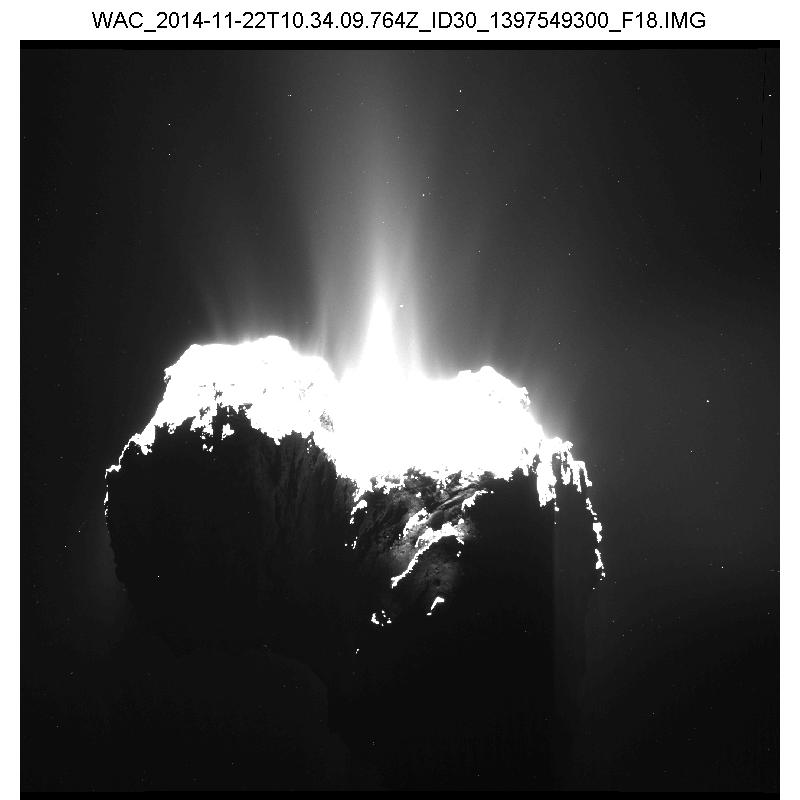}
\includegraphics[width=3.3cm]{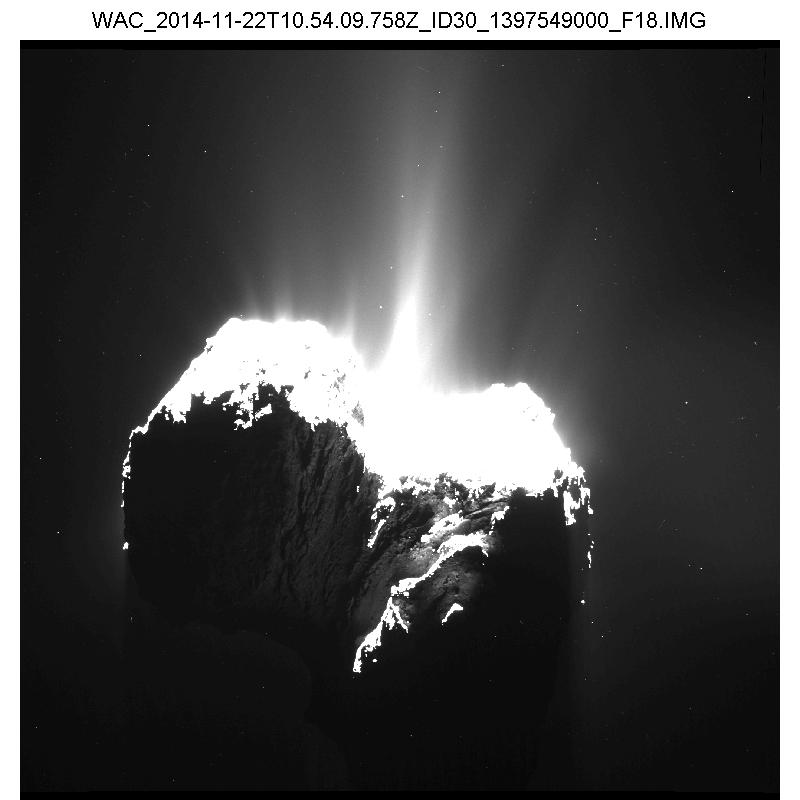}
\includegraphics[width=3.3cm]{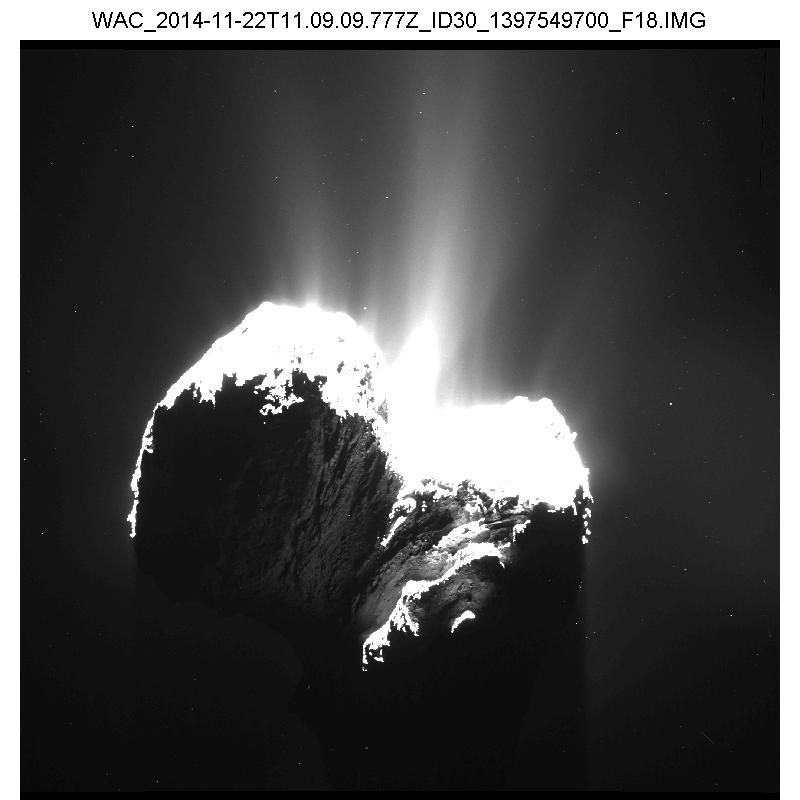}\\

\includegraphics[width=3.3cm]{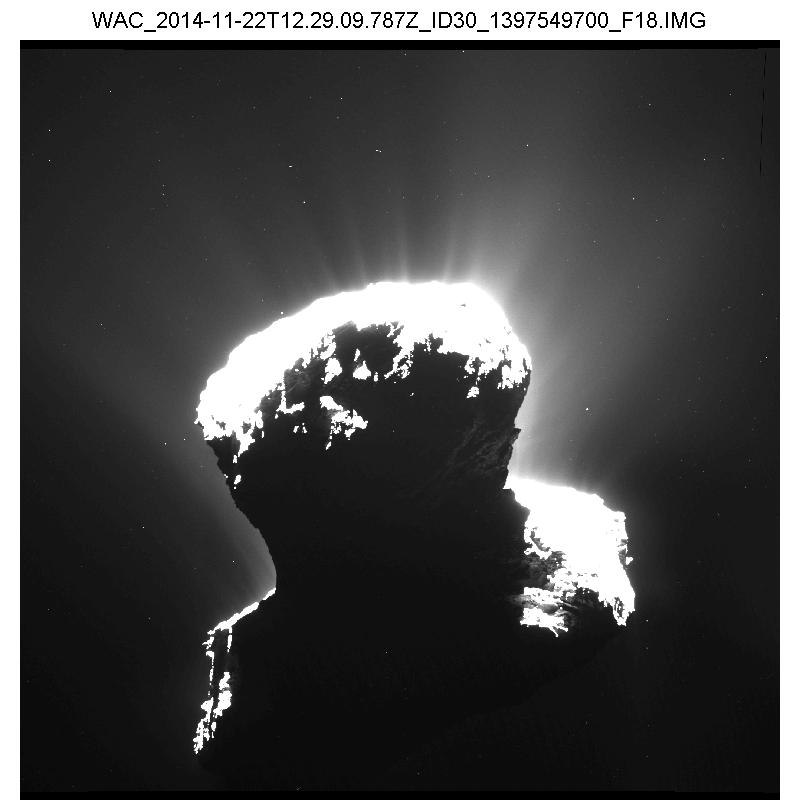}
\includegraphics[width=3.3cm]{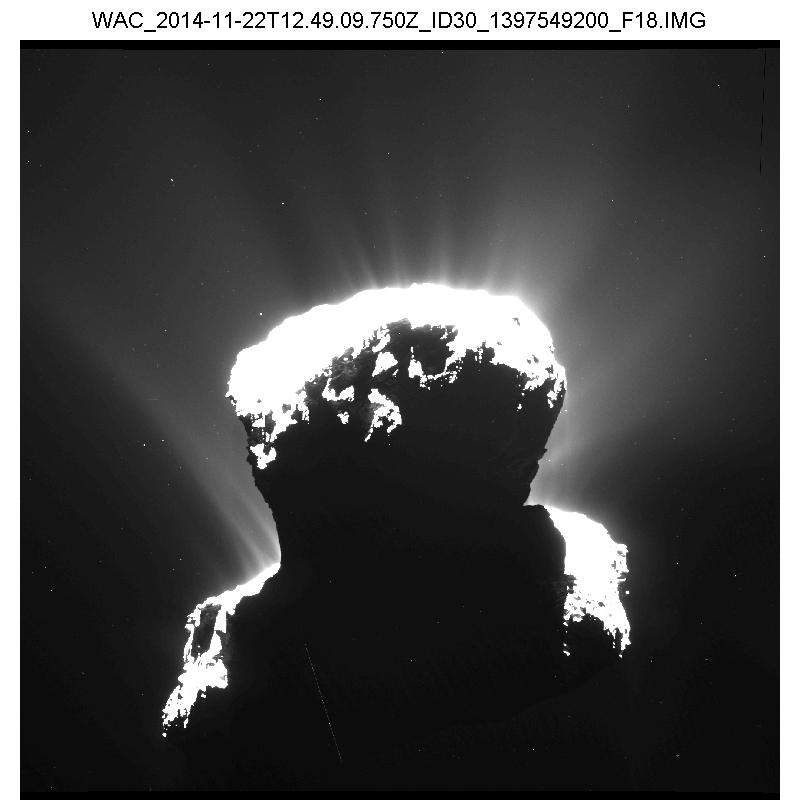}
\includegraphics[width=3.3cm]{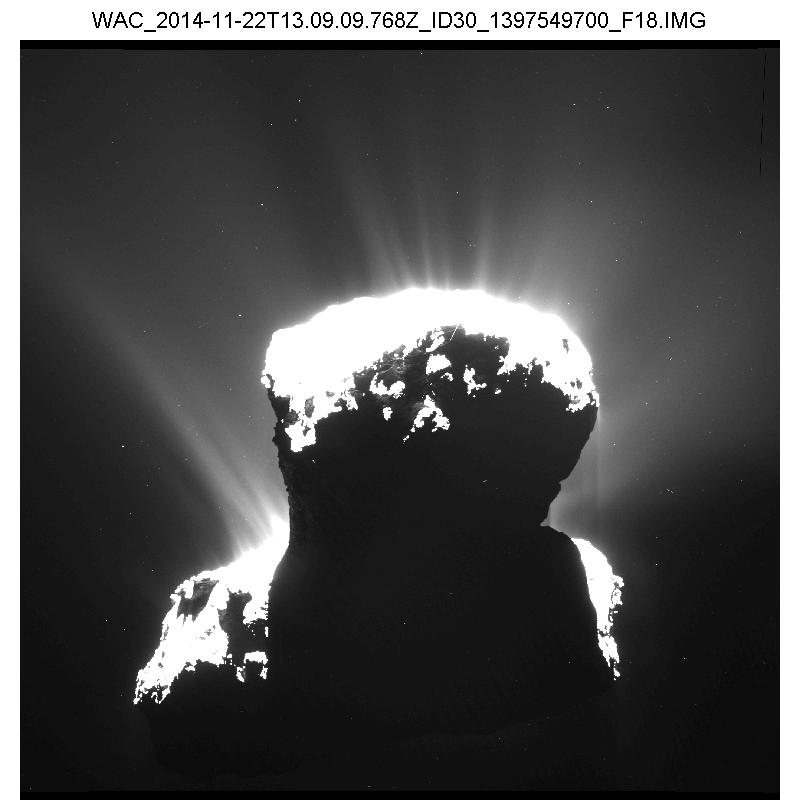}
\includegraphics[width=3.3cm]{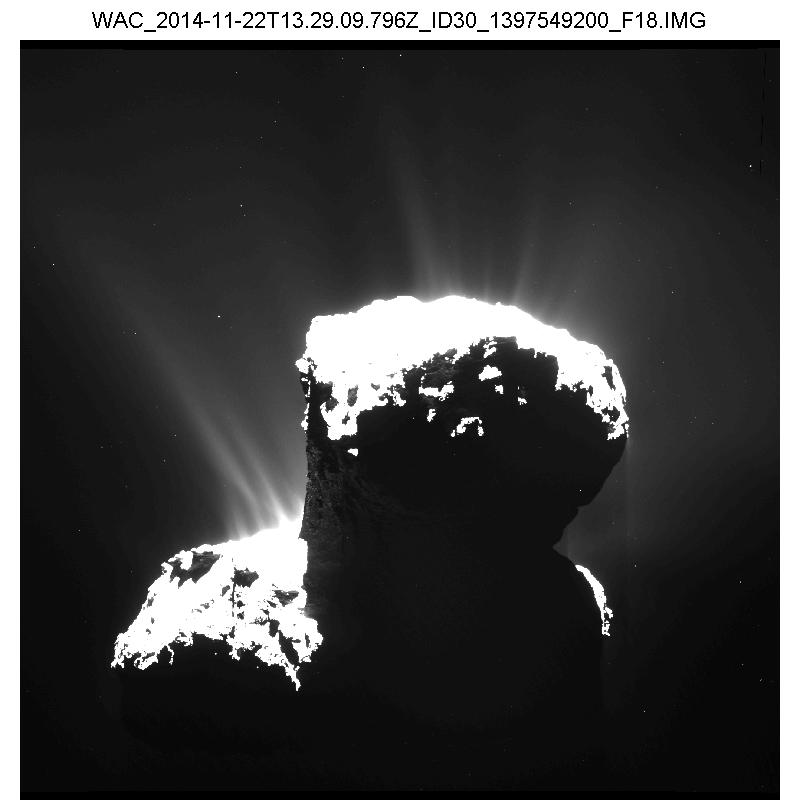}
\includegraphics[width=3.3cm]{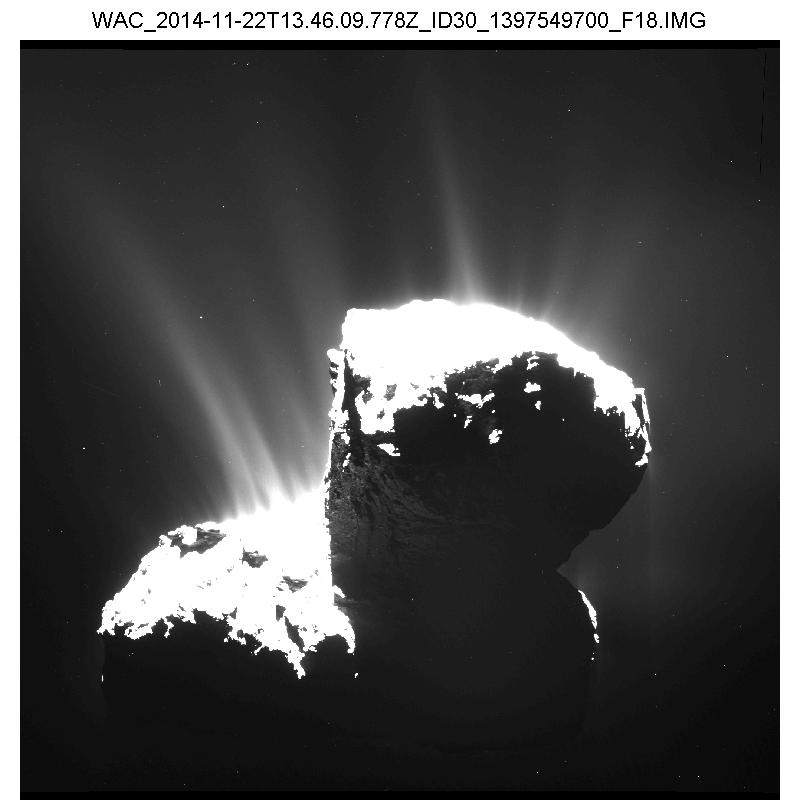}

\caption{25 WAC images from a sequence acquired on 22 November 2015, from a distance of 29 km and a sub-spacecraft latitude of -57 \degree. The sequence shows the diurnal variation of activity over 10.8 hours of rotation (88\% of the total period). Every view is separated from the previous one by about 20 minutes (10 \degree of nucleus rotation). There is a gap of 2 hours between the fourth and last rows and after the last image due to spacecraft navigation slots. Sun and comet North pole are pointing up in these images. Brightness levels are stretched linearly to emphasize the 5\% least bright pixel values.}
\label{fig:STP030_AP}
\end{figure*}

\begin{figure*}[h!]
\centering
\includegraphics[width=3.3cm]{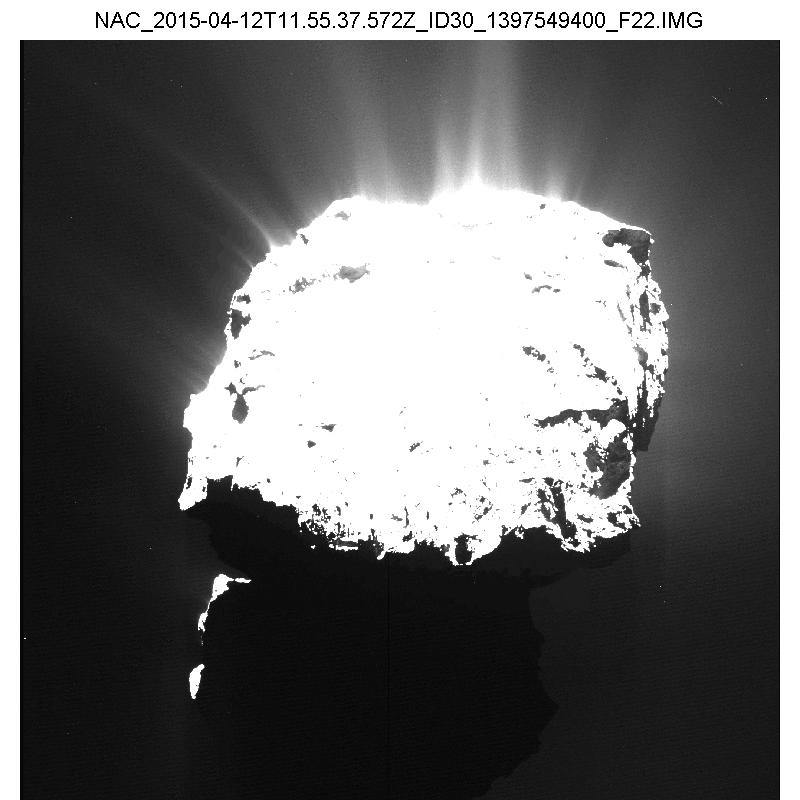}
\includegraphics[width=3.3cm]{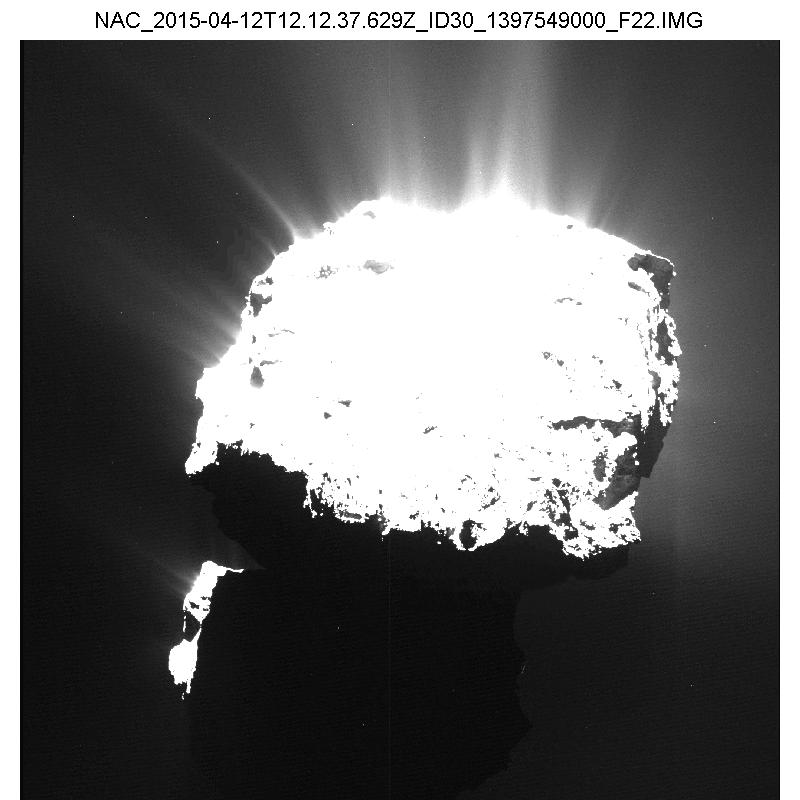}
\includegraphics[width=3.3cm]{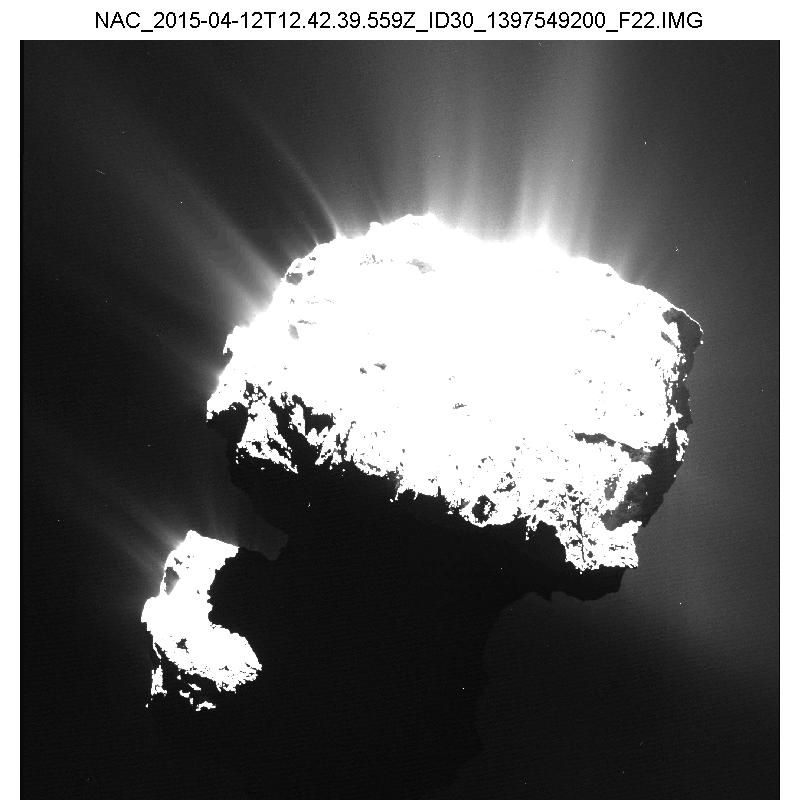}
\includegraphics[width=3.3cm]{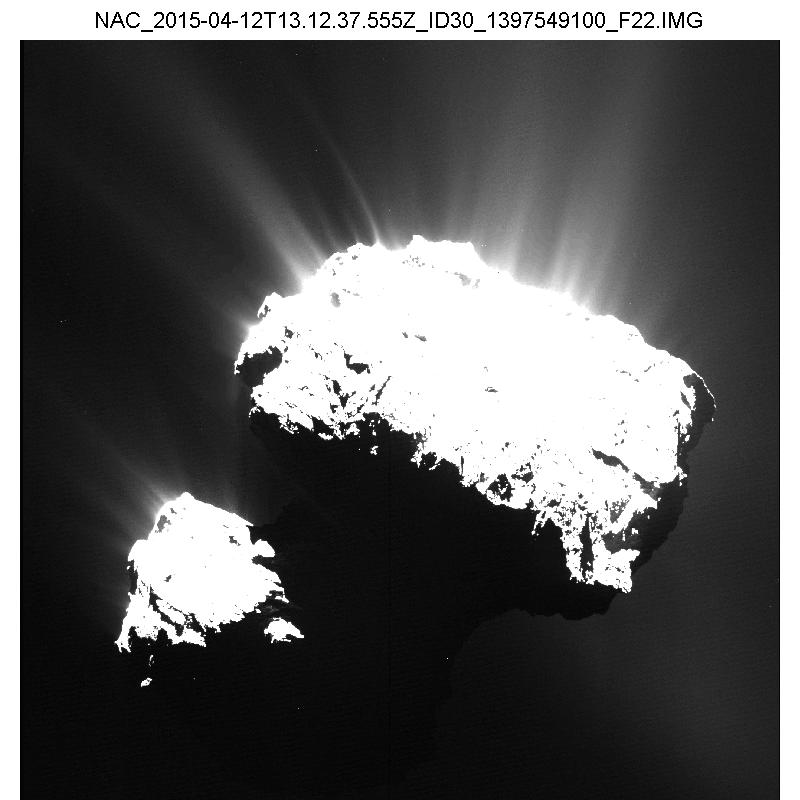}
\includegraphics[width=3.3cm]{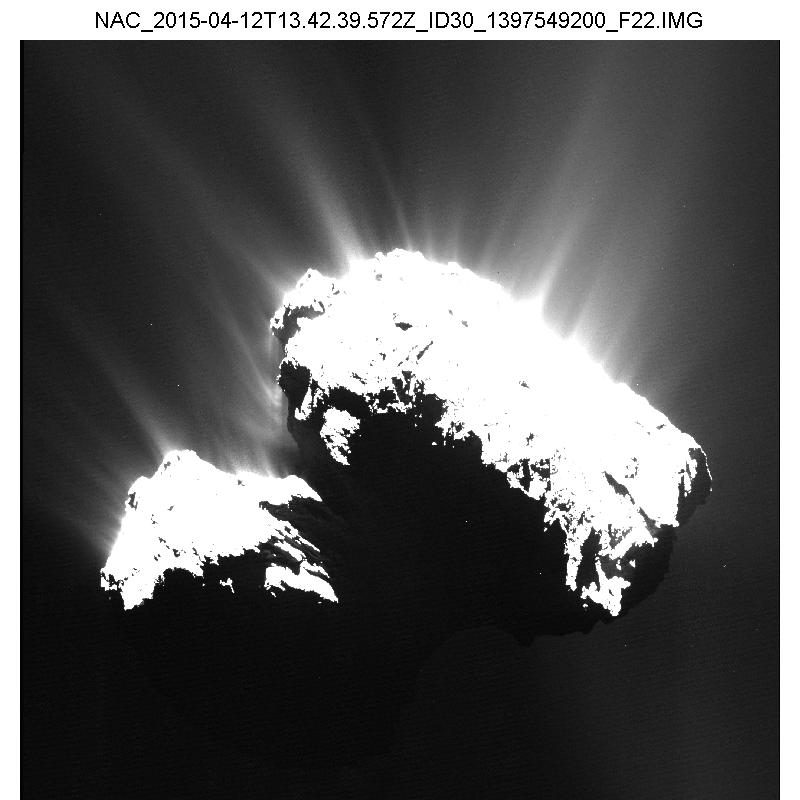}\\

\includegraphics[width=3.3cm]{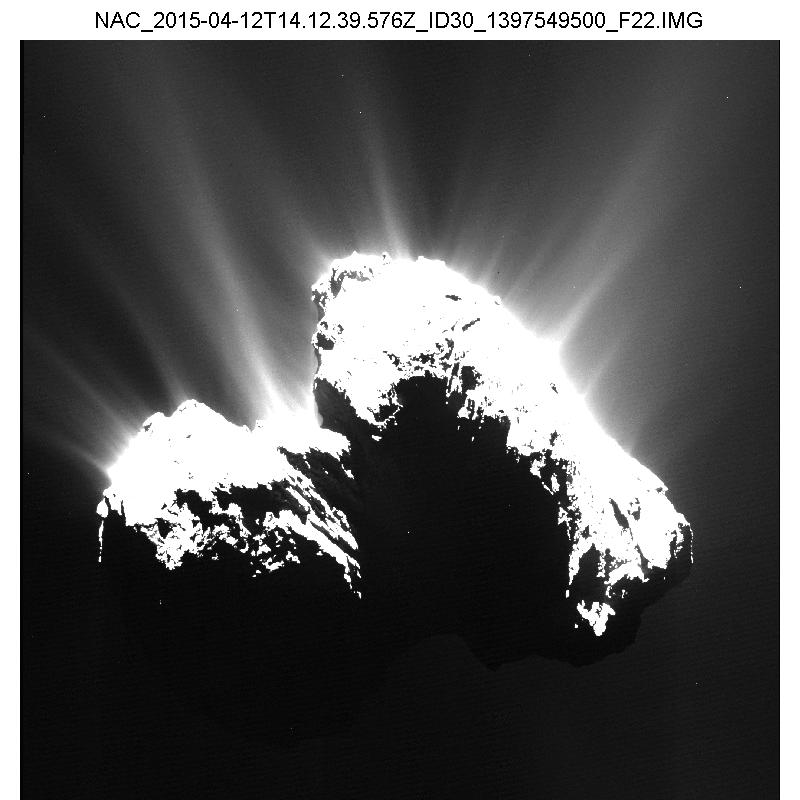}
\includegraphics[width=3.3cm]{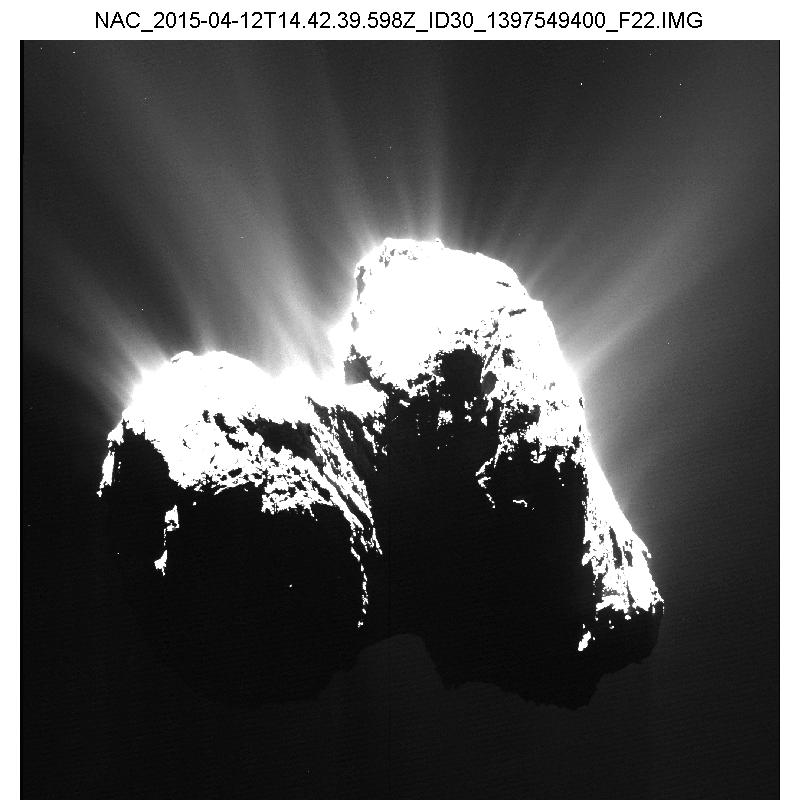}
\includegraphics[width=3.3cm]{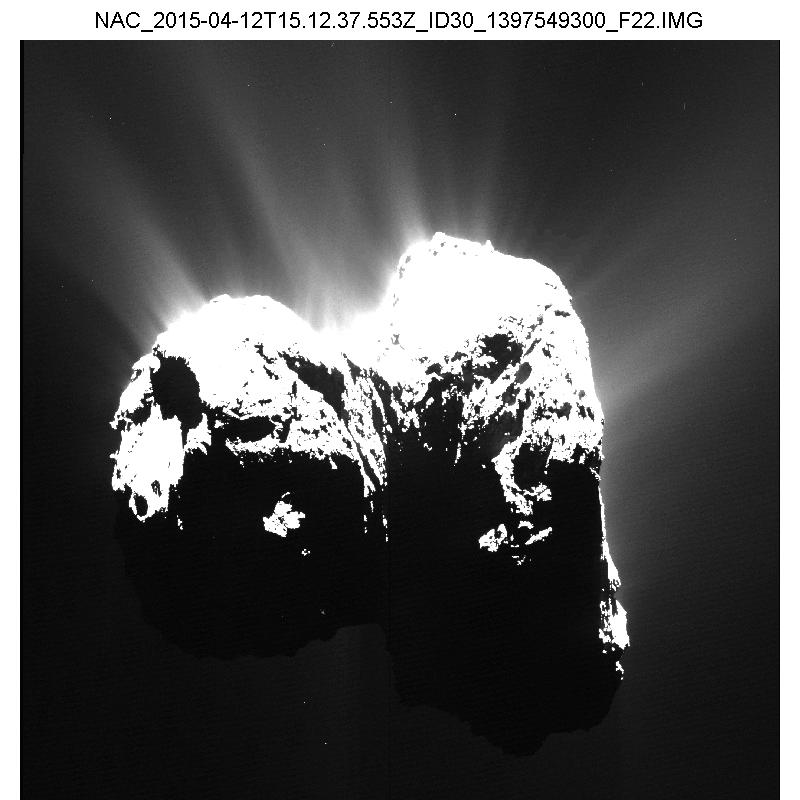}
\includegraphics[width=3.3cm]{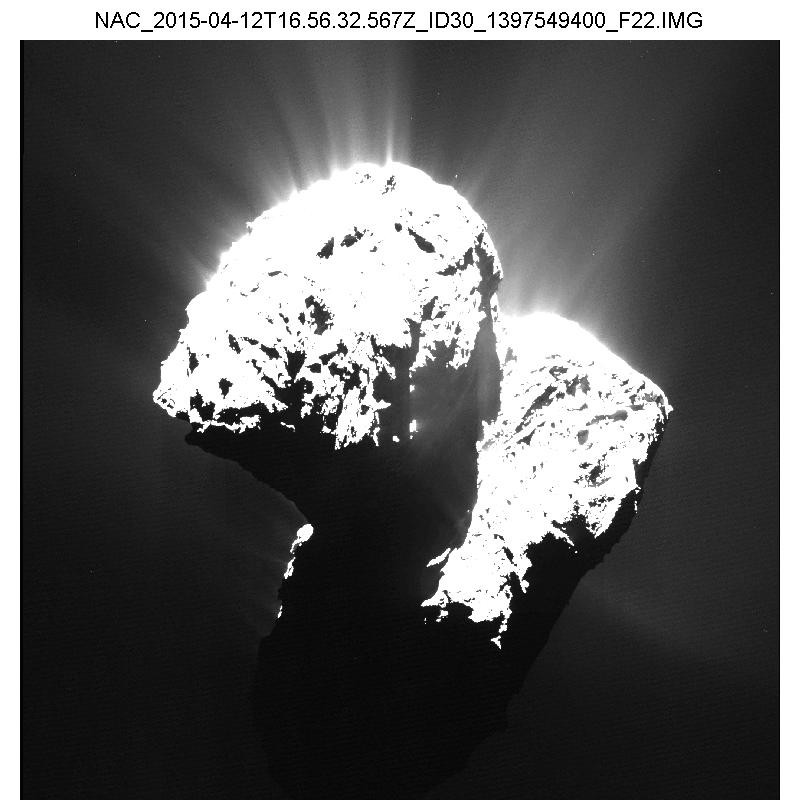}
\includegraphics[width=3.3cm]{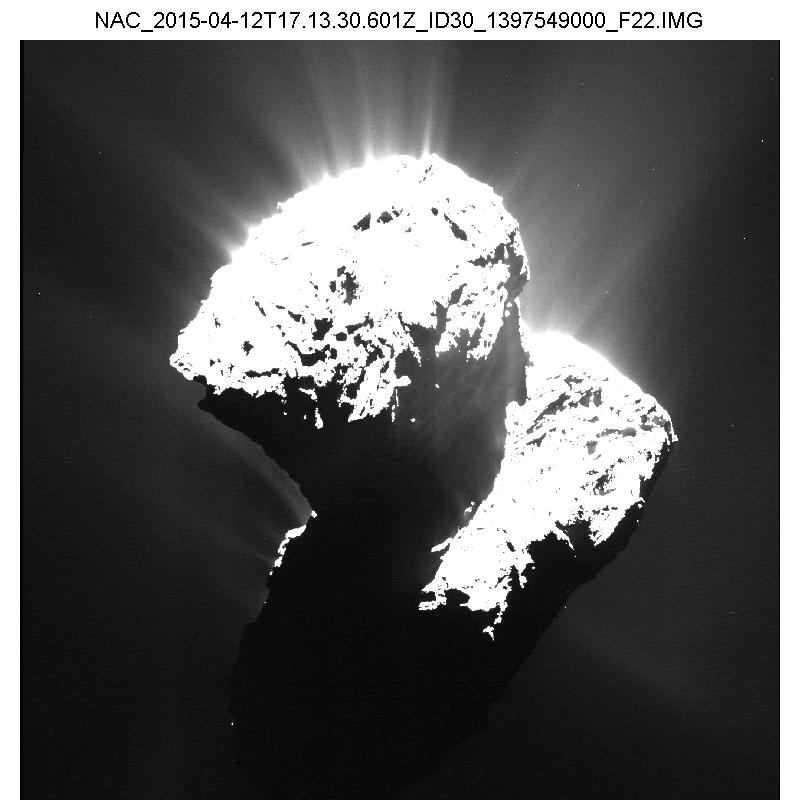}\\

\includegraphics[width=3.3cm]{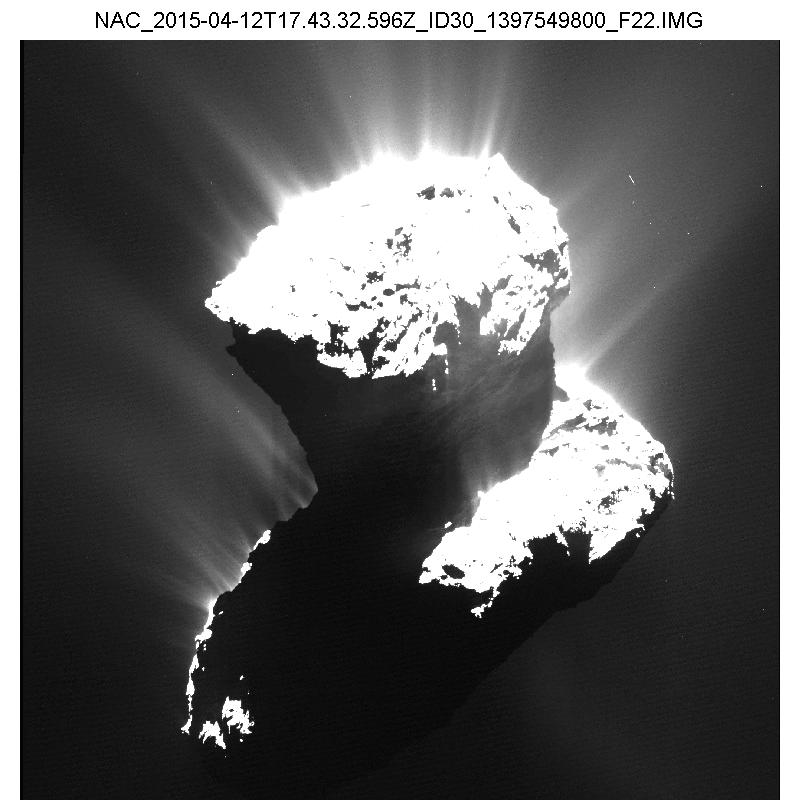}
\includegraphics[width=3.3cm]{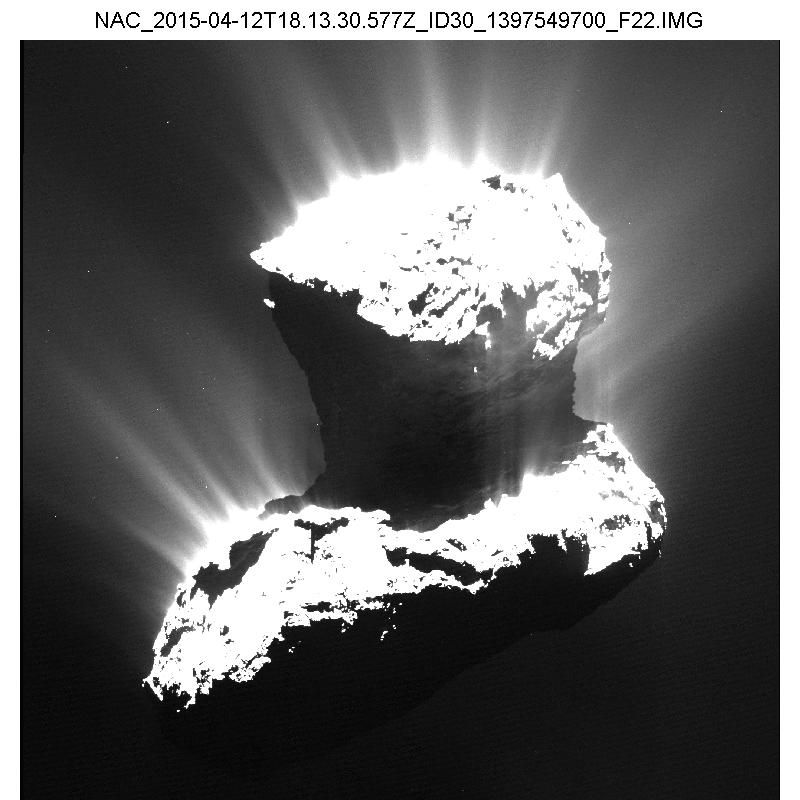}
\includegraphics[width=3.3cm]{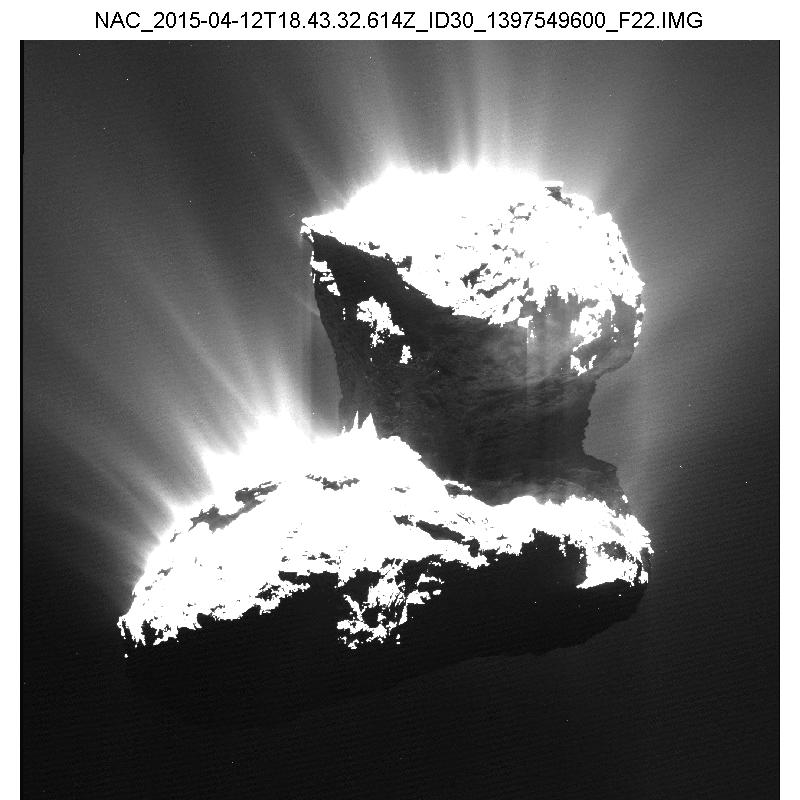}
\includegraphics[width=3.3cm]{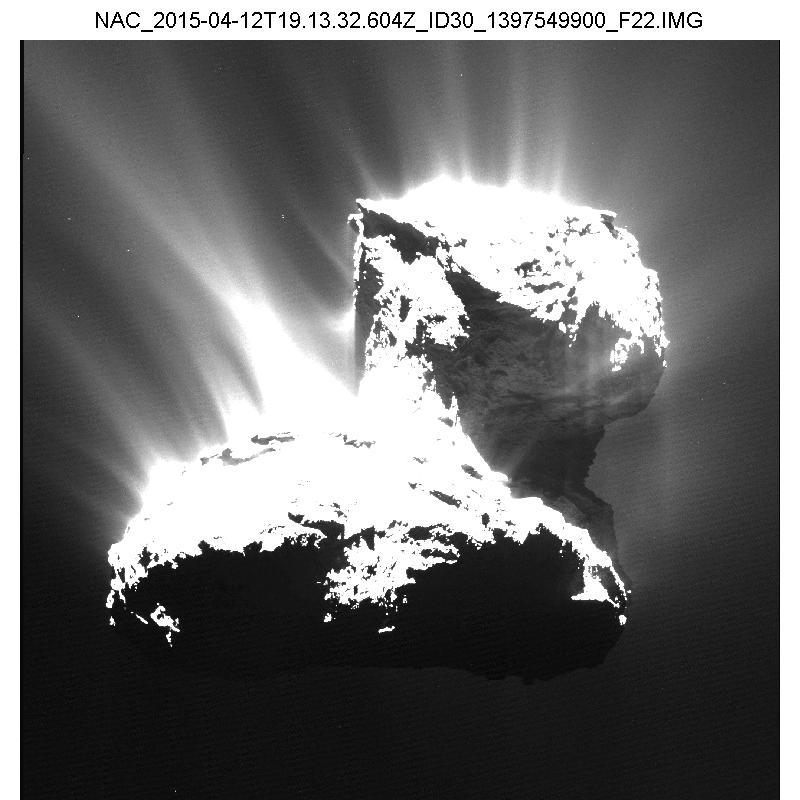}
\includegraphics[width=3.3cm]{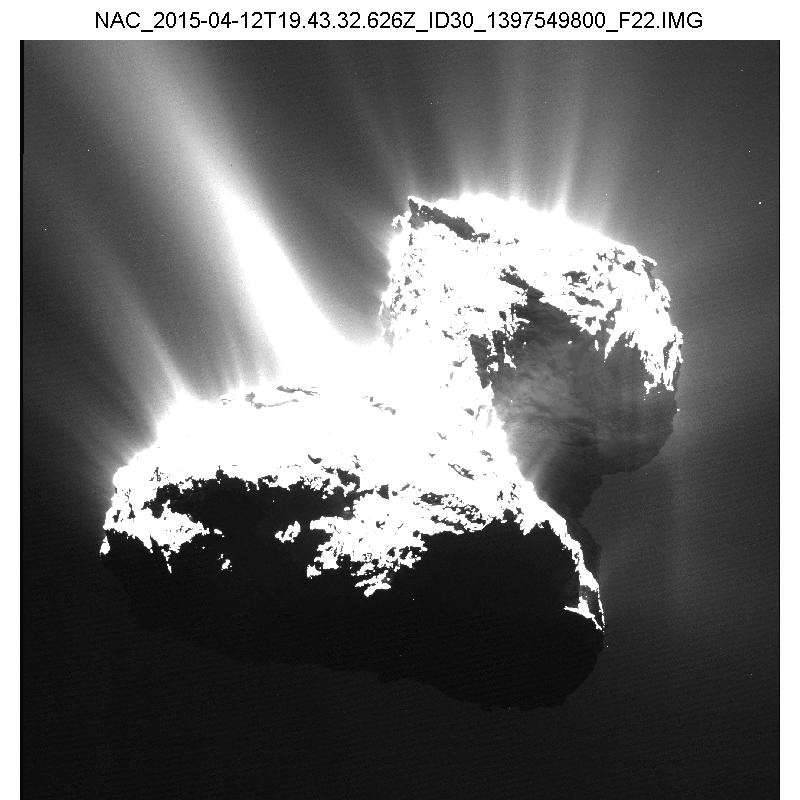}\\

\includegraphics[width=3.3cm]{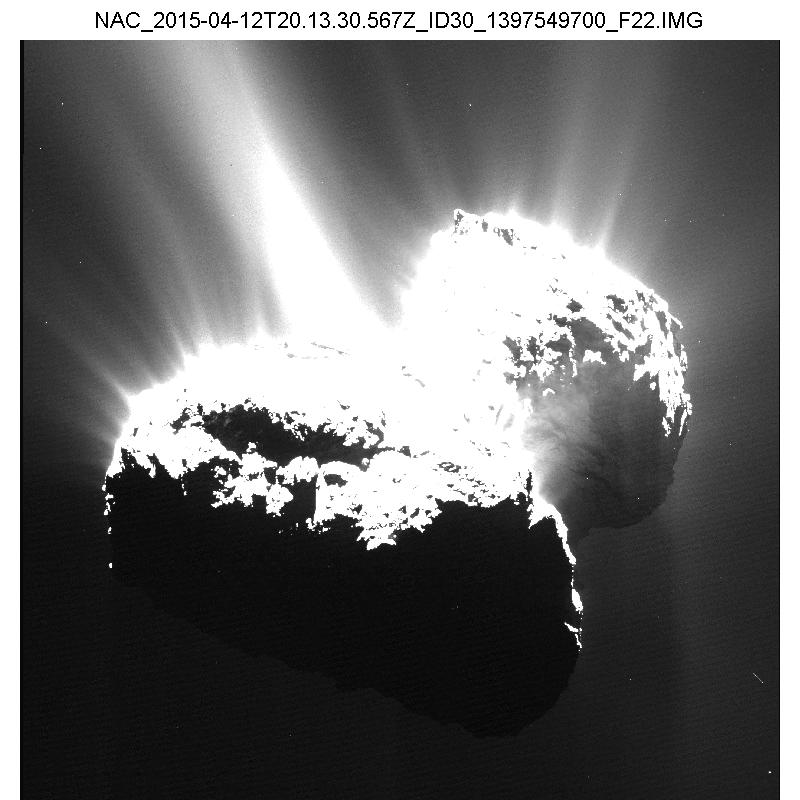}
\includegraphics[width=3.3cm]{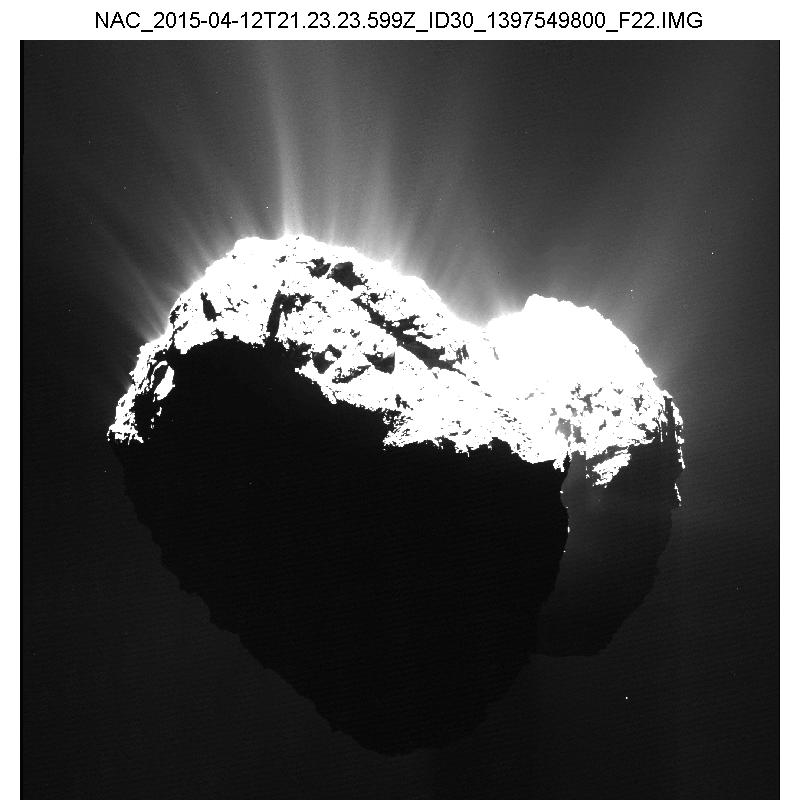}
\includegraphics[width=3.3cm]{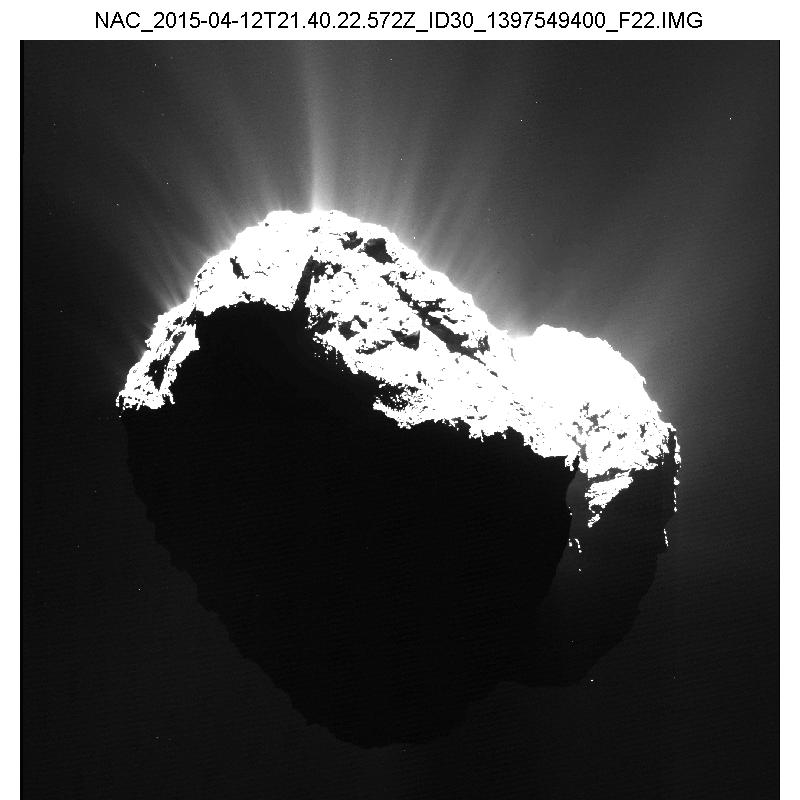}
\includegraphics[width=3.3cm]{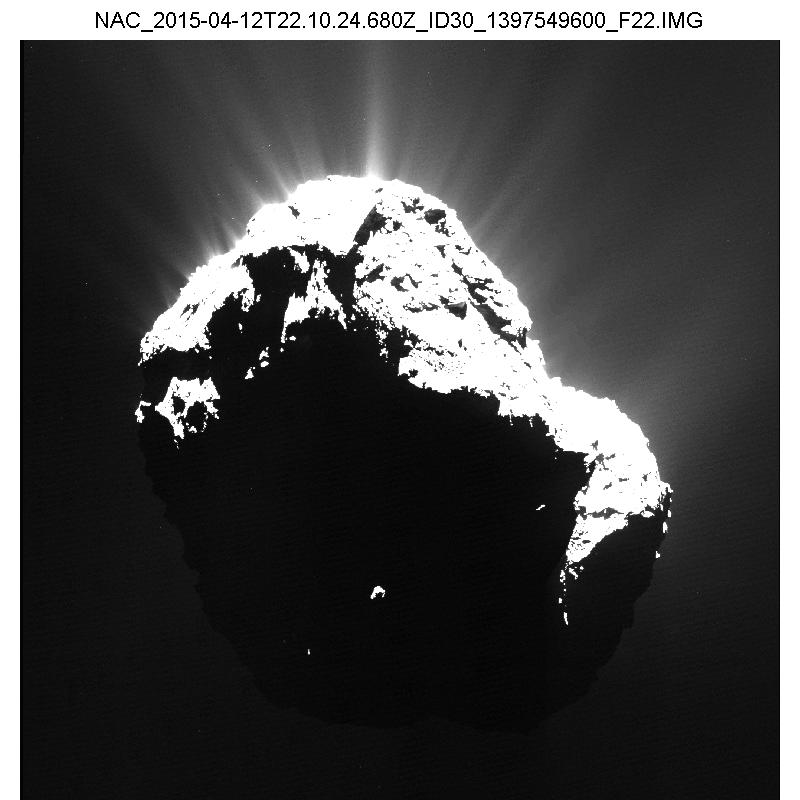}
\includegraphics[width=3.3cm]{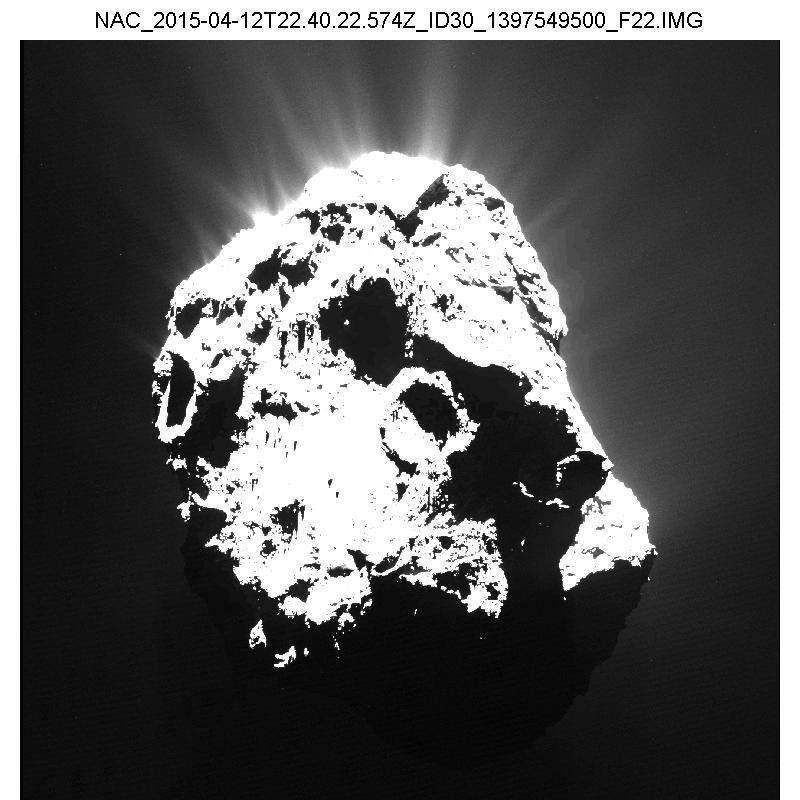}\\

\includegraphics[width=3.3cm]{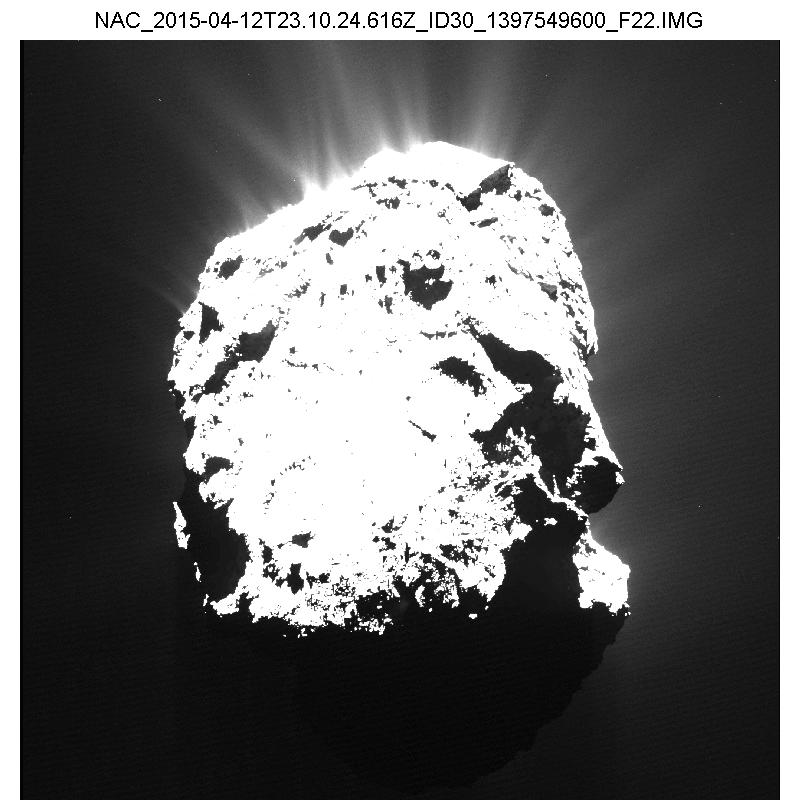}
\includegraphics[width=3.3cm]{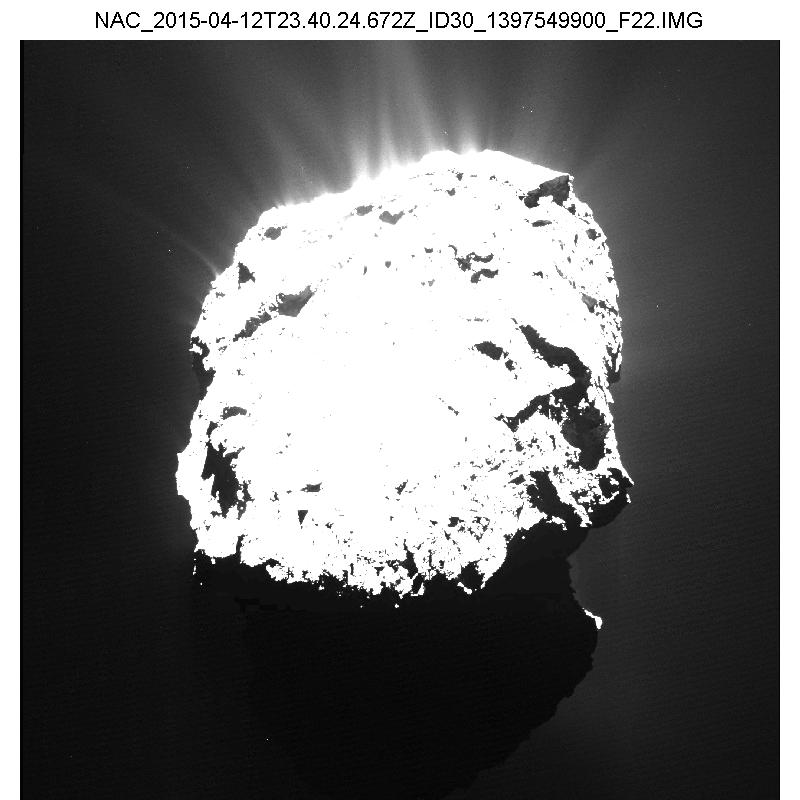}
\includegraphics[width=3.3cm]{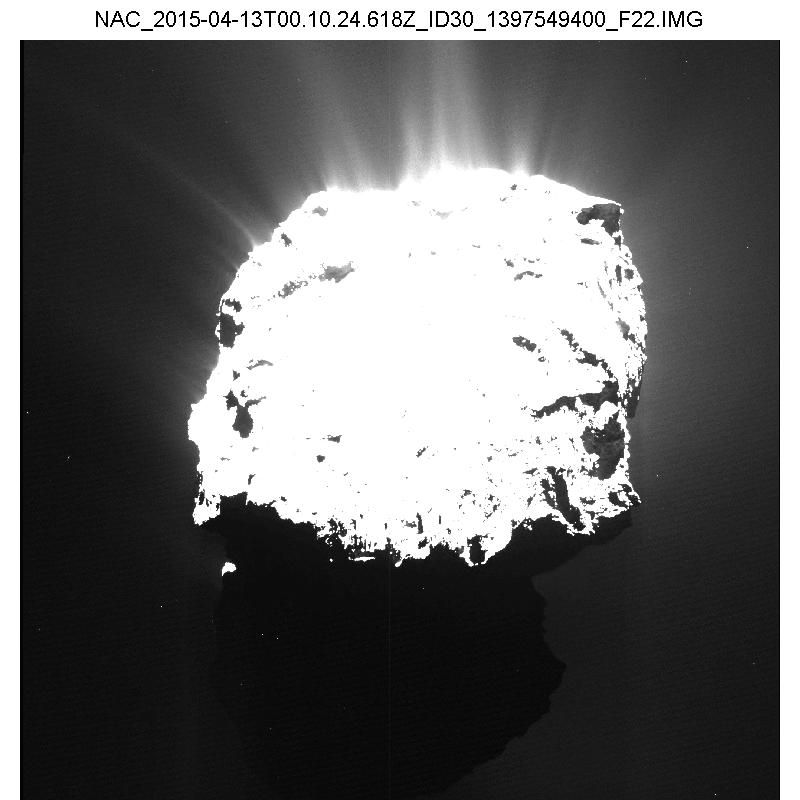}
\includegraphics[width=3.3cm]{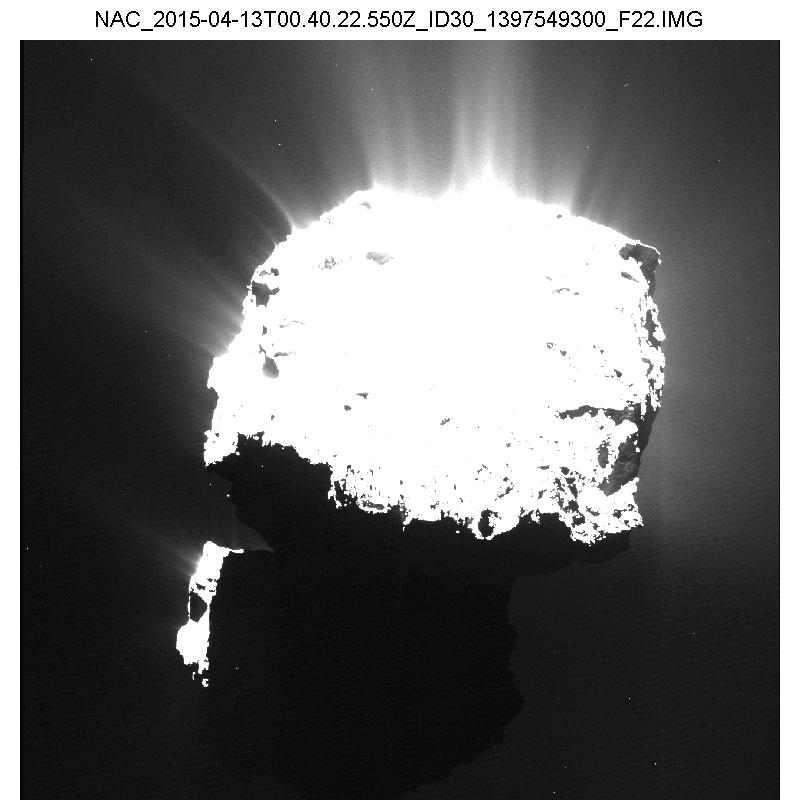}
\includegraphics[width=3.3cm]{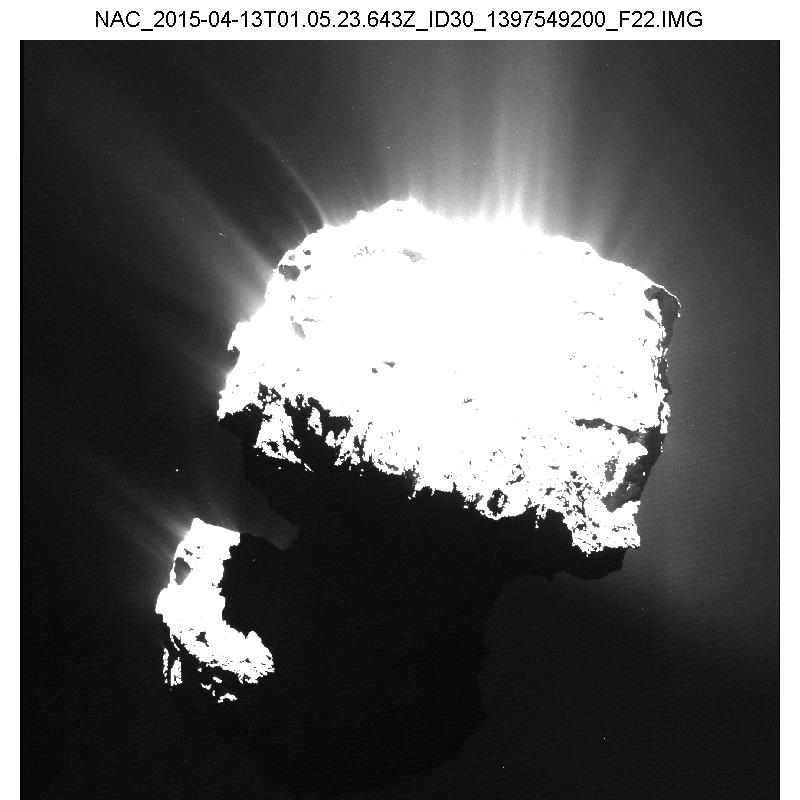}

\caption{25 NAC images from a sequence acquired on 12 April 2015, from a distance of 149 km and a sub-spacecraft latitude of -48 \degree. The sequence shows the diurnal variation of activity over a full period. Every view is separated from the previous one by about 30 minutes (15 \degree of nucleus rotation). There is a four hours gap between the last two observations. Sun and comet North pole are pointing up in these images. Brightness levels are stretched linearly to emphasize the 5\% least bright pixel values.}
\label{fig:STP051_RE}
\end{figure*}

\begin{figure*}[h!]
\centering
\includegraphics[width=14cm]{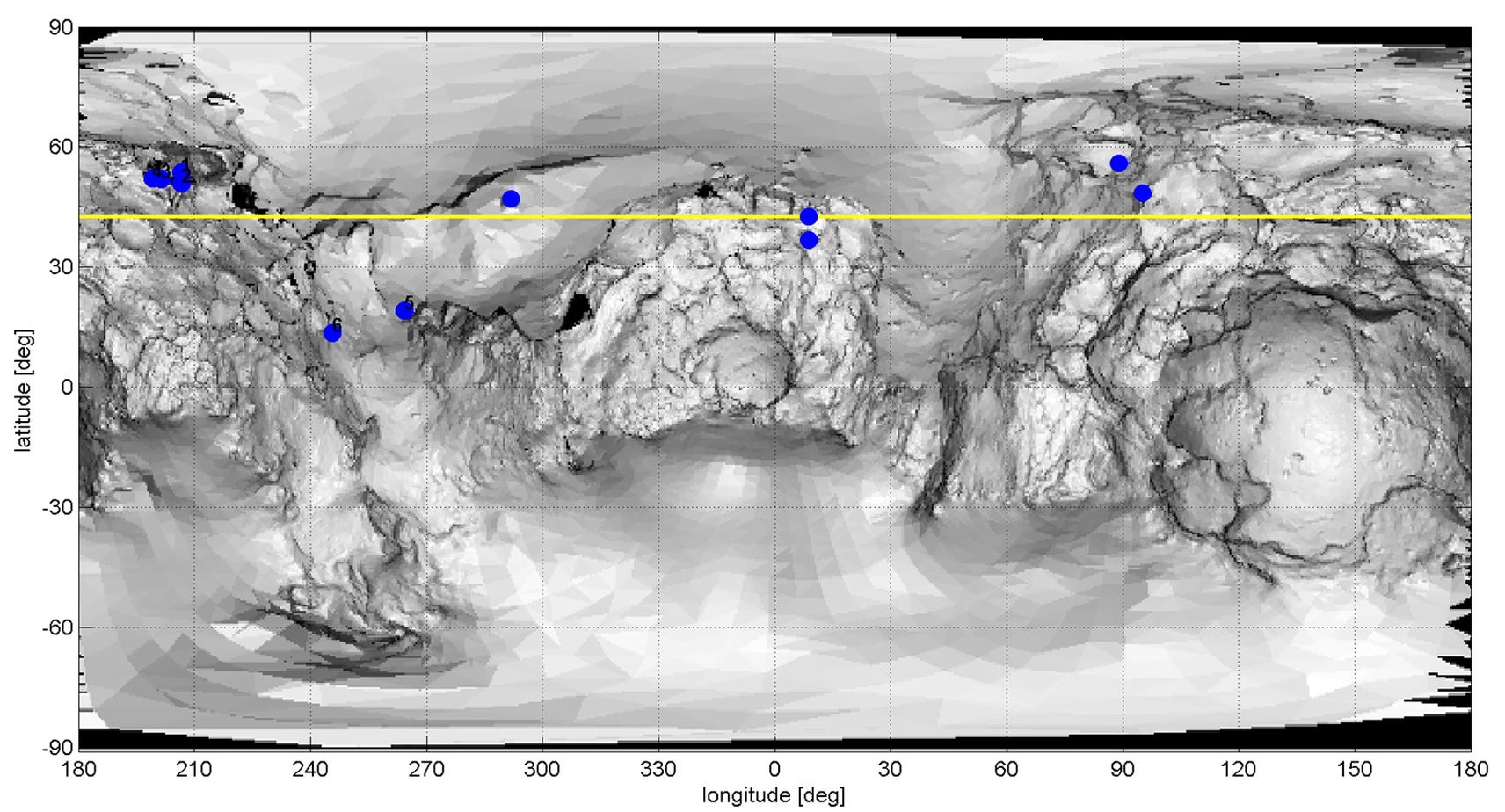}\\
\includegraphics[width=14cm]{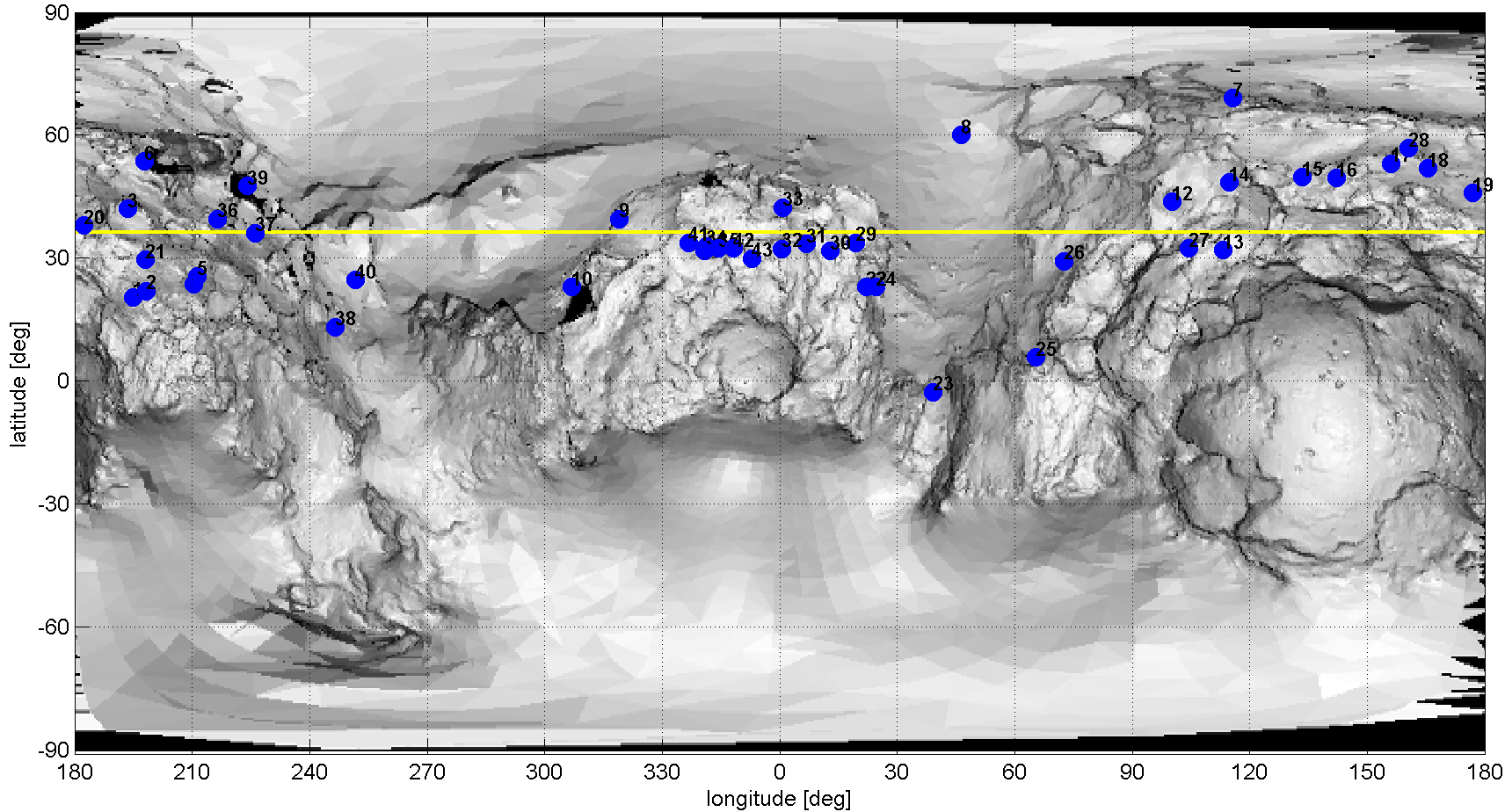}\\
\includegraphics[width=14cm]{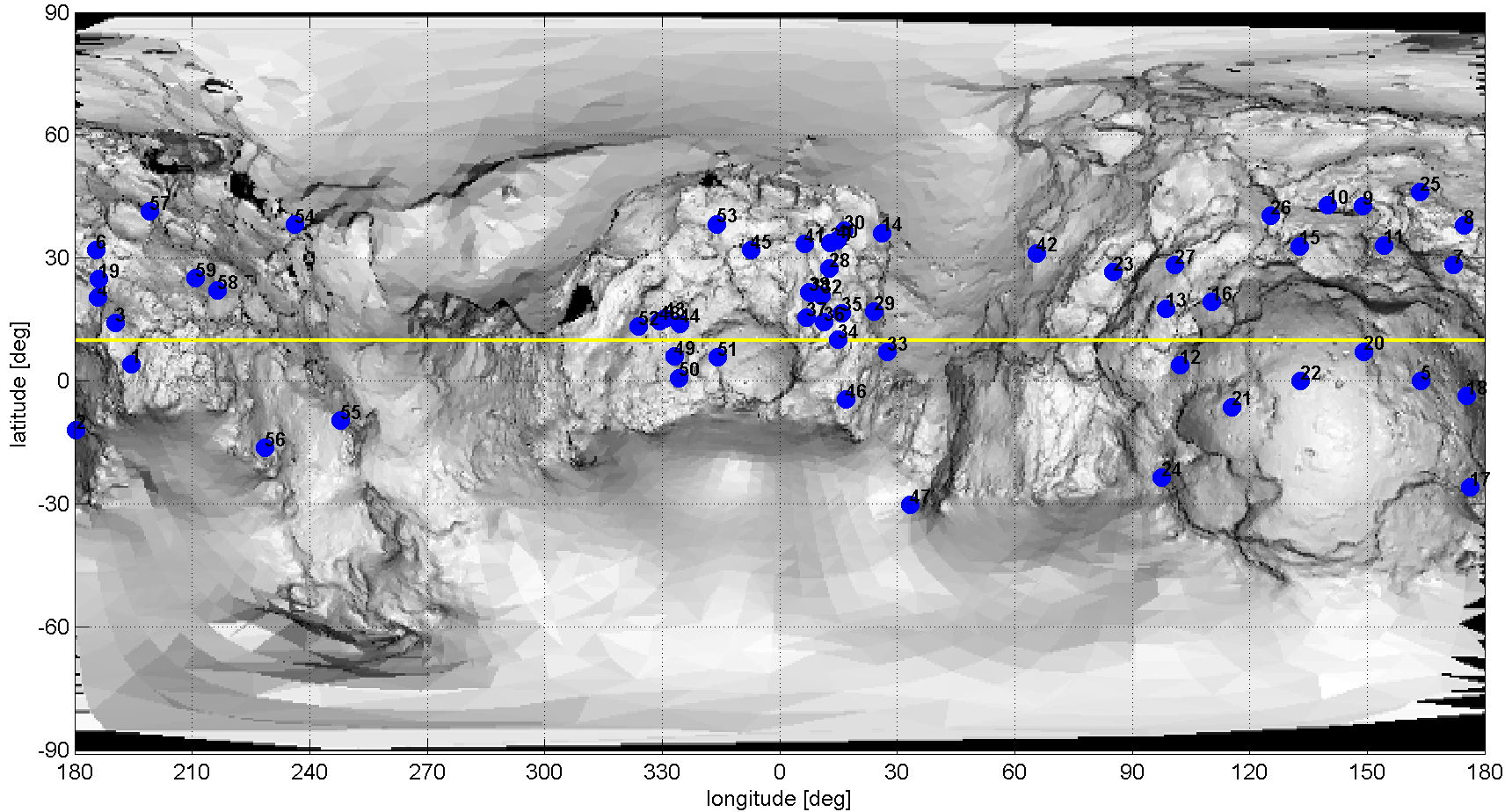}
\caption{Maps of active sources obtained at three different epochs. From top to bottom: September 2014, November 2014, April 2015. The map is shaded according surface slopes, accounting for gravity and centrifugal force. White is flat, black is a vertical wall. The yellow line marks the sub-solar latitude. Note that these maps are not an exhaustive catalog of all sources, see Section \ref{sec:results}}
\label{fig:maps}
\end{figure*}

This simple modeling approach outlines that we need to overcome a certain cohesive force to lift up dust particles. For instance of $1.96 \times 10^{-7}~N$ for particles of $100~\mu m$. This corresponds typically to a sublimation temperature larger than 205 K. Figure \ref{fig:lift_model} shows how this minimum necessary temperature evolves with respect to the tensile strength of the surface. Within the assumptions described above it can be fitted by a power law:
\begin{equation}
T_{min} = 181 \times Y^{0.0334} \quad (K)
\end{equation}

Note that this function describes only a lower limit for the gas flow necessary to lift up dust particles. One must be aware of a few limitations of our model:
\begin{itemize}
\item{We do not know the tensile strength of the surface upper layers, we only have estimates for layers a few tens of meters thick. The actual cohesion of the top surface may be much higher, especially in the presence of organic materials which can act as a glue between the grains, or because of the sintering due to the many thermal cycles that took place on the surface.}
\item{We used Stokes' equation to calculate the drag force. If the mean free path of the gas is much greater than the size of the particle, we must instead consider Epstein's model, which effectively reduces the efficiency of the drag force \citep{blum2006}. In practice, this is equivalent in replacing $v_r^2$ by $v_r v_{gas}$ in equation \ref{eq:drag}. Because in our case $v_{dust} << v_{gas}$, both regimes lead to similar drag forces.}
\item{In our model drag force and cohesion vary with the surface of the particle, i.e. the square of its radius. Therefore the ratio of drag force and cohesion is independent of the particle size.}
\item{The pressure we calculate is valid only very close to the source. As gas expands in the vacuum, the pressure drops rapidly, reducing the drag force away from the source}
\item{In addition, there may be an existing ambient pressure which may reduce the efficiency of the newly created gas flow, although it will allow a stronger collimation.}
\end{itemize}
Therefore we expect the real flux necessary to lift up the dust to be even larger than the limit we calculated earlier. Which means either the surface needs to reach even higher temperature, or other mechanisms will focus the gas flow and help it provide a much stronger drag force.
 
The limiting factor to reach such relatively high sublimation temperature is the available thermal energy. The major contributor in the energy balance of cometary surface crusts is solar irradiation. Due to the low thermal inertia of these crusts and the nearly instantaneous thermal re-emission, comparably high sublimation rates in deeper crust layers are hardly reached. Morphologic structures, such as cracks, fractures, or holes are more likely to enhance the sublimation rate at deeper depths of the crust, as they prevent large amounts of heat being re-emitted to space. These structures act as heat traps when being under favorable illumination conditions, and generating higher average temperatures and thus higher sublimation rates. They will, however, be heated only for a short time and therefore sustaining jets from a given area will require a fracture field with different fracture orientations with respect to the Sun. This is indeed what we observe in the pits in Seth region \citep{vincent2015b}: the large jet arising from a pits is composed of small features sequentially activated as different fractured areas of the walls become illuminated (typically for a couple of hours only).

The narrow geometry of fractures not only describes a heat trap, but from the point of view of gas dynamics is also similar to a pipe or a nozzle. Once released, the gas cannot expand in all directions, but its flow is constrained by the geometry of its surroundings. Depending on the temperature of the fracture walls we can have a slight cooling or warming of the gas, but overall it is likely that the fracture will accelerate the flow, hence increasing the effective drag force. Such mechanism is described for instance in \citep{yeoh2015}. We expect rather a warming when the walls are inert; sublimation at the fracture wall will lead to slightly colder gas temperatures.

Of course, this description is highly speculative as we cannot constrain the real fracture geometries from the current data, but the fact that we see enhanced gas/dust flows over these fractured areas makes this process plausible. Constraining this effect further will require detailed theoretical and experimental modeling. We refer the reader to \cite{hoefner2015} for advanced numerical studies of thermophysical conditions in cometary fractures, and implications on activity.

\subsection{A general mechanism ? Comparison with other comets}
When compared at similar spatial scales, the Jupiter Family comets visited by spacecrafts are quite similar. They have comparable size, albedo, large scale morphology. Activity occurs at similar levels and even outbursts are comparable in their amount of ejected material \citep{belton2013, tubiana2015}. Therefore why the erosion mechanism presented in previous sections would not take place on other comets as well ? There are a few evidence that this is indeed the case.

Comets 81P/Wild 2 and 9P/Tempel 1 have both been observed by space missions at similar resolutions (10~m/px). Both missions followed the comets' activity during their encounter, and trace back jets to regions of the surface \citep{sekanina2004, farnham2007, farnham2013}. Although the lower resolution and limited time coverage did not allow for a jet reconstruction as precise as the one we obtained with 67P, many jets were consistently linked to rough terrains. In particular for 81P, many jets were seen arising from within or at the edge of local depressions, quite similar to what we see in Seth region on 67P. Pits on 81P are typically shallower than on 67P, which has been interpreted as evidence for erosion and infill (see discussion in \cite{vincent2015b}). On 9P, many jets were linked to a ragged area bordering a large smooth terrain, and in particular to a the scarped edge of this terrain \cite{farnham2007}. This area, however, has been imaged in two consecutive orbits by the missions Deep Impact (in 2005) and Stardust-NExT (in 2011). Both flybys happened around perihelion passage and imaged the same area of the comet with similar illumination, thus providing the best conditions to detect surface changes. Indeed, \cite{thomas2013} report a significant retreat of the cliff, up to 50~m in some areas, as well as the lateral expansion of surface depressions in the vicinity and putatively linked this evolution to the ongoing activity. Although some uncertainty uncertainty remains as to where the source of these jets really lies (smooth patch itself ? cliff ? talus ?), we know for a fact that erosion took place in this area between 2005 and 2010 (Deep Impact and Stardust encounters) even if this region was not anymore a source of jets in the second encounter \citep{farnham2013}.

In retrospect, it seems the findings on 81P and 9P are similar to what we observe on 67P at higher resolution. The different scale of changes is likely due to the difference in insolation for both comets. Even if the smooth patch of comet 9P was illuminated only for a fraction of the rotation and at shallow angle, the receding edge would have been facing the Sun directly for a short while, receiving about $600~W.m^{-2}$ of solar energy.

On the other hand, cliffs in 67P's Seth region are in polar winter at perihelion and are not much illuminated below an heliocentric distance of 3 AU,  thus receiving an irradiance of at most $150~W.m^{-2}$. In addition to that, these areas experience strong shadowing due to the fact that they are located close to the bottom of the largest concavity of the nucleus, and therefore are illuminated only for a few hours instead of continuously for Tempel 1. This leads to a maximum erosion of 1~m only \citep{keller2015} and therefore strong changes will only occur through sudden events (like a block falling from the scarp) rather than continuous erosion.

\section{Conclusions}
The high resolution observations of dust jets arising from the nucleus of comet 67P from August 2014 to May 2015 led us to propose a new scenario explaining both the formation of cometary jets and the presence of erosional features seen on many comets. Our findings can be summarized as follows:

\begin{itemize}
\item{Seasonal and diurnal variations of activity in the Northern hemisphere of comet 67P/Churyumov-Gerasimenko are well correlated with solar illumination, hence linked to volatile sources within the diurnal thermal skin depth of the surface.}
\item{We have presented observational evidence that Northern jets arise in majority from rough terrains rather than smooth areas, more specifically from fractured walls.}
\item{We interpret this activity as a combination of meter scale jets accelerated in fractures, and decameter scale flows arising from the mast wasted brighter debris at the cliff foot.}
\item{Our findings imply that erosion of the comet surface takes place preferentially along these active cliffs, and therefore they should be showing the most significant changes, some of them already observed.}
\item{Activity driven erosion of cliffs occurs in many places on 67P, and may have been observed on 9P and 81P. we propose that this is a general process taking place on all comets.}
\end{itemize}

The process described above is very general and describes one type of activity which explains most active sources of the Northern hemisphere of 67P. It is important to keep in mind that other types of activity may arise, and our classification of active sources is still a work in progress. As the comet approaches perihelion we keep monitoring the activity and will investigate how jets and sources behave at closer heliocentric distance. We do expect to observe other types of activity, for instance previously insulated area may receive enough solar energy to allow the sublimation of volatiles without the need for specific terrain morphology.

\begin{acknowledgements}
OSIRIS was built by a consortium led by the Max Planck Institut f\"ur Sonnensystemforschung, G\"ottingen, Germany, in collaboration with CISAS, University of Padova, Italy, the Laboratoire d'Astrophysique de Marseille, France, the Instituto de Astrofisica de Andalucia, CSIC, Granada, Spain, the Scientific Support Office of the European Space Agency, Noordwijk, The Netherlands, the Instituto Nacional de Tecnica Aeroespacial, Madrid, Spain, the Universidad Politecnica de Madrid, Spain, the Department of Physics and Astronomy of Uppsala University, Sweden, and the Institut f\"ur Datentechnik und Kommunikationsnetze der Technischen Universit\"at Braunschweig, Germany. 
The support of the national funding agencies of Germany (DLR), France (CNES), Italy (ASI), Spain (MINECO), Sweden (SNSB), and the ESA Technical Directorate is gratefully acknowledged.
We thank the Rosetta Science Ground Segment at ESAC, the Rosetta Mission Operations Centre at ESOC and the Rosetta Project at ESTEC for their outstanding work enabling the science return of the Rosetta Mission.

~\\
We thank the anonymous reviewers for their constructive criticism and support of our findings. 
\end{acknowledgements}

\bibliographystyle{aa}
\bibliography{references}


%
%
%

\end{document}